\documentclass[%
superscriptaddress,
twocolumn,
nofootinbib,
amsmath,amssymb,
aps,
prd,
]{revtex4-1}

\usepackage[colorlinks]{hyperref}
\usepackage{tabularx}
\usepackage{graphicx}
\usepackage{amssymb,amsmath,bm,tensor,braket}
\usepackage[dvipsnames]{xcolor}
\usepackage[varg]{txfonts}
\usepackage{enumerate}
\usepackage{mathtools}
\usepackage{tensor}
\usepackage[capitalize]{cleveref}
\usepackage[utf8]{inputenc}
\usepackage[normalem]{ulem}
\usepackage{dcolumn}%
\usepackage{soul}
\usepackage{subfigure}
\usepackage[normalem]{ulem}
\usepackage{aas_macros}

\usepackage{comment}
\usepackage{siunitx}
\usepackage{mathrsfs}
\usepackage[T1]{fontenc}

\newcommand{\dd}{\mathrm{d}}

\newcommand{\ee}{\mathrm{e}}

\begin{document}
	
	\preprint{APS/123-QED}
	
	\title{Spectral instability of black holes: relating the frequency domain to the time domain}%
	
	\author{Yiqiu Yang}\email{yueyang1812@gmail.com}
	\affiliation{Department of Physics, School of Physics, Peking University, Beijing 100871, China}%
	\affiliation{Kavli Institute for Astronomy and Astrophysics, Peking University, Beijing 100871, China}
        \affiliation{William H. Miller III Department of Physics and Astronomy, Johns Hopkins University, Baltimore, Maryland 21218, USA}%

	\author{Zhan-Feng Mai}\email{zhanfeng.mai@gmail.com}
	\affiliation{Kavli Institute for Astronomy and Astrophysics, Peking University, Beijing 100871, China}%
        \affiliation{Guangxi Key Laboratory for Relativistic Astrophysics, School of Physical Science and Technology, Guangxi University, Nanning 530004, China}%
	
	\author{Run-Qiu Yang}\email{aqiu@tju.edu.cn}%
	\affiliation{Center for Joint Quantum Studies and Department of Physics, School of Science, Tianjin University, Tianjin 300350, China}
	
	\author{Lijing Shao}\email{lshao@pku.edu.cn}
	\affiliation{Kavli Institute for Astronomy and Astrophysics, Peking University, Beijing 100871, China}%
	\affiliation{National Astronomical Observatories, Chinese Academy of Sciences, Beijing 100012, China}
	
	\author{Emanuele Berti}\email{berti@jhu.edu}
	\affiliation{William H. Miller III Department of Physics and Astronomy, Johns Hopkins University, Baltimore, Maryland 21218, USA}
	
	\date{\today}%

	\begin{abstract}
		Recent work has shown that the quasinormal mode spectrum of black holes is unstable under small perturbations (of order $\epsilon$) of the radial potential, while the early time-domain ringdown waveform is only marginally affected.
		In this paper we provide further insight into the apparent tension between the frequency-domain and the time-domain descriptions by analyzing the scattering properties of the problem. 
		In the frequency domain, we study analytically the solutions corresponding to the perturbed potential. We show that there are two qualitatively different classes of instabilities, and that both Schwarzschild and Kerr black holes are affected by what we call a ``Type~II'' instability, i.e., an exponential migration of the mode frequencies away from their unperturbed value as the perturbing ``bump'' moves away from the peak of the unperturbed potential.
		In the time domain, we elucidate the effect of the spectral instability in terms of the causal structure of the Green's function. 
		By using an equivalent scattering problem we confirm analytically (and show numerically) that the deviation from the unperturbed waveform in the early ringdown stage is proportional to $\epsilon$ when $\epsilon\lesssim10^{-2}$.
	\end{abstract}
	
	\maketitle
	
	\allowdisplaybreaks
	
	\section{Introduction}
	\label{sec:introduction}

	Classic results in perturbation theory show that black holes (BHs) have a characteristic spectrum, like atoms in quantum mechanics~\cite{Chandrasekhar:1985kt}. The BH spectrum is characterized by a set of discrete quasinormal modes (QNMs)~\cite{1999LRR.....2....2K,1999CQGra..16R.159N,2009CQGra..26p3001B} with complex frequencies $\omega_{n}$, which contain information of the geometric structure of the background spacetime~\cite{1971PhRvL..27.1466D,1972ApJ...172L..95G,2009PhRvD..79f4016C}. 
	When the imaginary part of the frequencies $\Im \omega_{n}<0$ (assuming Fourier modes of the form $\ee^{-i\omega t}$), the QNMs correspond to damped oscillations, and the spacetime is dynamically stable~\cite{1973ApJ...185..635T,1973ApJ...185..649P}.
	
	In addition to stability considerations, QNMs are also useful to extract information from compact objects in gravitational wave (GW) astronomy~\cite{2018gwv..book.....M,2018GReGr..50...49B}. Since the detection of GW150914~\cite{LIGOScientific:2016aoc}, BH binary mergers have played an important role in both astrophysics and tests of general relativity~\cite{LIGOScientific:2016lio,LIGOScientific:2021sio}.
        Their evolution can be roughly divided in three phases: the early inspiral, a strong-field merger, and the so-called ringdown, where the GW signal is characterized by a superposition of QNMs
        that can be used to infer the merger remnant's mass and spin~\cite{Echeverria:1989hg,2007PhRvD..76j4044B} and to verify that it is indeed a Kerr BH, as predicted by general relativity~\cite{1980ApJ...239..292D,2004CQGra..21..787D,2006PhRvD..73f4030B,2015CQGra..32x3001B,2019PhRvL.123k1102I,2022PhRvL.129k1102C,2023PhRvL.131v1402C,2024PhRvD.109b4058G,2024PhRvD.109d3027W}. The observational program of testing the Kerr nature of the remnant with QNMs is usually called  ``BH spectroscopy''. Arguably the most robust tests will come from observations of the fundamental QNM (the one with the smallest imaginary part and longest damping time) from multipolar components with different angular indices $(\ell,\,m)$, while the relevance of higher overtones in waveform modeling and data analysis is debated (see e.g. Refs.~\cite{2018PhRvD..97d4048B,2019PhRvX...9d1060G,2023PhRvD.108j4020B,2023PhRvD.108d4032N,2023PhRvD.108l3018W}).

	An important assumption in these QNM-based tests of the nature of BHs is that the QNM spectrum must not be sensitive to small perturbations, such as those that may be caused by their astrophysical environment. 
	However it has long been known that BHs are affected by a spectral instability: even small changes in the effective potential describing the radial dependence of the perturbations can lead to dramatic changes in the QNM spectrum~\cite{1996PhRvD..53.4397N,1999JMP....40..980N,2020PhRvD.101j4009D}. 
	This instability has been studied in various contexts, including ``dirty BHs'' (i.e., BHs surrounded by matter~\cite{1997PhRvL..78.2894L,2014PhRvD..89j4059B,2021PhRvD.103l4013B,2022PhRvD.105f1501C,2022PhRvL.129x1103C}) and exotic compact objects~\cite{2019LRR....22....4C,2016PhRvD..94h4031C}. %
	Recent work applying pseudospectral methods to BH perturbation theory formally showed that the spectral instability affects spherically symmetric, asymptotically flat BHs ~\cite{2021PhRvX..11c1003J,2021PhRvD.104h4091D,2022PhRvL.128u1102J,2023PhRvD.107f4045Y} as well as horizonless compact objects~\cite{2023PhRvD.107f4012B} and asymptotically (anti-)de Sitter BHs~\cite{2023PhRvD.108j4002S,2023PhRvD.108j4027C,2024JHEP...05..202C,2024PhRvD.109f4068B,2024PhRvD.109d4023D} (see e.g. Ref.~\cite{2023arXiv230816227D} for a review).
        While the instability seems to be stronger for higher overtones, the fundamental mode can also be destabilized~ \cite{2022PhRvL.128k1103C}. Fortunately for the BH spectroscopy program, the spectral instability does not have a major impact on the prompt ringdown signal~\cite{2022PhRvD.106h4011B} and it has been shown that certain quantities related to scattering properties (e.g., the greybody factors) are not affected by spectral instabilities~\cite{2023PhRvD.107d4012K,2023PhRvD.107d4012K,2023PhRvL.131k1401T,2024arXiv240601692R,2024arXiv240604525O}.
        
         Our current understanding of BH spectral instabilities has some important limitations. First of all, the spectral instability of rotating BHs (the most relevant case for astrophysics) is still poorly understood. Secondly, with few exceptions, most studies of the spectral instability were performed in the frequency domain.
         Numerical work in the time domain~\cite{2022PhRvD.106h4011B} has clarified that small spectral ``bumps'' of amplitude $\epsilon$ introduce small deviations (of order $\epsilon$) from the unperturbed prompt ringdown.
	However the time-domain waveform does not simply consist of a superposition of QNMs~\cite{1986PhRvD..34..384L,1999JMP....40..980N}, and the analytic structure of the Green's function is not completely understood~\cite{2013PhRvD..87f4010C,2019PhRvD.100j4037C}. For these reasons, with the exception of an interesting analysis highlighting the role of the phase shift of the $S$-matrix~\cite{2023PhRvD.107d4012K}, there is limited analytical insight into the relation between the spectral instability and the prompt ringdown phase. 
	Last but not least, the physical mechanisms that could trigger the instability are unclear (see Ref.~\cite{2024PhRvD.110b4016C} for a recent effort in this direction). 
	
	In this paper we address the first two problems: the spectral instability of Kerr BHs, and the relation between the spectral instability and the apparent stability of the prompt ringdown phase. We use basic techniques of complex analysis in the frequency domain to prove that Kerr BHs are affected by the same spectral instability as Schwarzschild BHs: an exponential migration of the mode frequencies away from their unperturbed value as the perturbing ``bump'' moves away from the peak of the unperturbed potential (this is what we will call a ``Type~II'' instability below). Then we use standard Green's function techniques in the time domain. We define an ``equivalent scattering problem'' to clarify the influence of the (frequency-domain) spectral instability on the time-domain waveform. In particular, we show that sufficiently small changes of the unperturbed potential induce small changes of the prompt ringdown waveform, while the new (unstable) modes will only appear in the late time signal. This analysis clarifies the reasons behind the behavior observed in Refs.~\cite{2022PhRvD.106h4011B,2023PhRvD.107d4012K}.
	
	The paper is organized as follows. In Sec.~\ref{sec:frequency domain analysis} we study the scattering problem for the perturbed potential in the frequency domain. We define an appropriate ``susceptibility'', $\chi_n$, and we use it to classify different kinds of spectral instabilities. In Sec.~\ref{sec:BH instability} we apply the method to nonrotating and rotating BHs, and we show that the ``Type~II'' instability is quite generic. In Sec.~\ref{sec:t-domain} we revisit the problem in the time domain using Green's function techniques. We show that the perturbations of the original waveform can be understood as several wavepackets with different causal structures, and we demonstrate the perturbative stability of the prompt ringdown phase. In Sec.~\ref{sec:summary} we summarize our main results. Following common practice in BH perturbation theory, throughout the paper we use geometrical units ($G=c=1$) and we set the BH mass $M=1$.	

	\section{FREQUENCY-DOMAIN ANALYSIS}
	\label{sec:frequency domain analysis}

	\subsection{Perturbation equations in the frequency domain}
	\label{sub:definition}
	
	The linearization of the Einstein equations on a (Schwarzschild or Kerr) BH background leads to a set of partial differential equations. The angular dependence of these equations can be separated using appropriate spherical harmonic functions~\cite{1957PhRv..108.1063R,1970PhRvL..24..737Z,1969PhDT........13Z,1972PhRvL..29.1114T}. A Fourier transform in time leads to a single equation for some radial variable $x$ that admits discrete eigenvalues corresponding to the QNMs.

	In general, the radial equation can be expressed in the form
	\begin{equation}\label{f-domain}
		-\frac{\mathrm{d}^{2}}{\mathrm{~d} x^{2}} \tilde{\Psi}+\left[-\omega^{2}+V(x)\right] \tilde{\Psi}=0, \quad x \in \mathbb{R}, \quad \omega \neq 0		\, ,
	\end{equation}
	where $\omega$ is the Fourier frequency, $\tilde{\Psi}(\omega,x)$ denotes the Fourier transform of $ \Psi(t,x) $, and $V(x)$ is some effective potential. For a spherically symmetric background, the function $\Psi$ is related to the coefficients of spin-2 tensor spherical harmonics.

        We consider a class of short-ranged potentials which decay faster than $O(1/x)$ when $|x|\rightarrow\infty$. For the cases of interest here (including Kerr BHs) this condition is satisfied. For example, the effective potential as a function of the tortoise coordinate $x$ for Schwarzschild BHs is either the Regge-Wheeler (RW) potential $V^{RW}_{\ell}(x)$ (for odd-parity or ``axial'' perturbations) or the Zerilli potential $V^{Z}_{\ell}(x)$  (for even-parity or ``polar'' perturbations). Both of these are short-ranged potentials with a peak close to the location of the photon ring for all values of the angular harmonic index $\ell$ (see e.g. Ref.~\cite{2009PhRvD..79f4016C}), decaying as $x^{-2}$ at infinity and exponentially near the horizon. As we will briefly review in Sec.~\ref{sub:Kerr} below, the perturbation equations in the Kerr background can also be expressed in this form by using the Sasaki-Nakamura (SN) formalism. Therefore most of our considerations below apply to both spherically symmetric and rotating BHs.
	
	Short-ranged effective potentials admit wavelike solutions, sometimes called Jost solutions (see e.g., Ref.~\cite{Chandrasekhar:1985kt}), both at the horizon ($x\rightarrow-\infty$) and at infinity ($x\rightarrow \infty$). Let us define the two Jost solutions, $\tilde{\Psi}=\phi_{\omega}(x)$ and $\tilde{\Psi}=\psi_{\omega}(x)$, as those satisfying the conditions
	\begin{equation}
		\begin{aligned}
			\phi_{\omega}(x)=&\mathrm{e}^{-i \omega x}\left[1+O\left(\dfrac{1}{x}\right)\right], \quad &&x \rightarrow-\infty ,\\
			\psi_{\omega}(x)=&\mathrm{e}^{i \omega x}\left[1+O\left(\dfrac{1}{x}\right)\right], \quad &&x \rightarrow \infty .
		\end{aligned}
	\end{equation}
	As long as $\omega \neq 0$, we then have four fundamental solutions $\left\{\phi_{\omega}, \phi_{-\omega}, \psi_{\omega}, \psi_{-\omega}\right\}$. Physically, the wavefunction  $\phi_{\omega}$ corresponds to purely outgoing waves at $x\rightarrow-\infty$; $\phi_{-\omega}$ corresponds to purely ingoing waves at $x\rightarrow-\infty$; $\psi_{\omega}$ corresponds to purely outgoing waves at $x\rightarrow+\infty$; and finally,  $\psi_{-\omega}$ corresponds to purely ingoing waves at $x\rightarrow+\infty$. Any of these solutions can be expressed as a linear combination of any other two. In particular, we can consider the combination
	\begin{equation}\label{jost}
		\phi_{\omega}(x)=a(\omega) \psi_{-\omega}(x)+b(\omega) \psi_{\omega}(x) \, .
	\end{equation}

	By considering the asymptotic behavior of $ \phi_{\omega} $ in the limit $x\rightarrow+\infty$ we see that a purely ingoing wave $ \ee^{-i\omega x} $ at $x\rightarrow-\infty$ is caused by the transmission of a purely ingoing wave with amplitude $ a(\omega) $, which also leads to a reflected wave with amplitude $b(\omega)$. We can thus regard $ 1/a(\omega) $ as the transmission coefficient and $ b(\omega)/a(\omega) $ as the reflection coefficient. Note that the relation between the Jost solutions is uniquely determined by the effective potential, and therefore also the scattering coefficients are uniquely determined by $V(x)$.

	The functions $a(\omega)$ and $b(\omega)$ are defined for real frequencies $\omega$, and can be analytically continued in the complex plane.

        By definition, the QNMs $\omega_{n}$ are the complex roots of the analytically continued function $ a(\omega) $, i.e.,
	\begin{equation}\label{QNM condition}
		a\left(\omega_{n}\right)=0, \quad \omega_{n} \in \mathbb{C} .
	\end{equation}
	
	To investigate the instability of QNMs, let us consider a perturbation of the original effective potential
        \begin{equation}
          V(x)=V_{0}(x) + \epsilon V_{b}(x)\,,
        \end{equation}
        where $\epsilon\ll1$ and $V_{b}(x)$ represents a small perturbation of $V_{0}(x)$. We impose $\max\limits_{x\in \mathbb{R}}|V_{b}(x)|=1$ for convenience. 
        We focus on a large class of ``bump'' potentials that have been studied extensively to investigate the instability of the fundamental mode (but see Ref.~\cite{2024PhRvD.110b4016C} for a survey of other possibilities).
        A ``bump'' potential $V_{b}(x)$ satisfies the following two conditions: (i) it is short-ranged, and (ii) its maximum is sufficiently far away from the maximum of the original potential $V_{0}(x)$. For example, $V_{b}(x)=\mathrm{exp}\big[-(x-d)^2 \big]$ is a Gaussian ``bump'' potential centered at $x=d$, if $d$ is large enough. 
	A large distance cutoff of the original potential (such as those discussed in Refs.~\cite{1996PhRvD..53.4397N,2021PhRvX..11c1003J}) is also a ``bump'' potential according to this definition, in the sense that $V_{b}(x)$ is zero below the cutoff and it is equal to (minus) the original potential above the cutoff.
		
	The $\epsilon$-dependent Jost solutions for the modified potential will be denoted by a superscript $(\epsilon)$, while we will use a superscript $ (0) $ for the solutions of the original potential. For example, the equivalent of Eq.~(\ref{jost}) for the perturbed problem is $ \phi_{\omega}^{(\epsilon)}=a^{(\epsilon)}(\omega)\psi_{-\omega}(x)+b^{(\epsilon)}(\omega)\psi_{\omega}(x) $. The BH QNMs are now solutions $\omega^{(\epsilon)}_{n}=\omega_{n}+\delta \omega_{n}$ of $a^{(\epsilon)}(\omega)=0$, where $\omega_{n}$ is the $n$-th solution of $a^{(0)}(\omega)=0$.
        To understand the behavior of $\delta \omega_{n}$, we need to study the solution $\phi_{\omega}^{(\epsilon)}(x)$ of Eq.~\eqref{f-domain} under the perturbed effective potential, and the behavior of the coefficient $a^{(\epsilon)}(\omega)$ in the complex plane. This will allow us to study the migration of the QNMs, defined as the zeros of $ a^{(\epsilon)}(\omega) $.		
	
	\subsection{Wavelike solutions of the perturbed problem}
	\label{sub:solution}

	We will now find an approximate solution for $ \phi^{(\epsilon)}_{\omega} $. A more rigorous derivation is provided in Appendix~\ref{appendix:scattering problem}.	
	The analogy between the radial perturbation equation~(\ref{f-domain}) and a one-dimensional Schr\"{o}dinger equation is well known (see e.g. Ref.~\cite{Chandrasekhar:1985kt}). 
	For the moment we will assume $\omega$ to be real, and later we will perform an analytic continuation in the complex plane to find the complex zeros, i.e., the QNMs.
	
\begin{figure}
	\centering
	\includegraphics[width=1.0\linewidth]{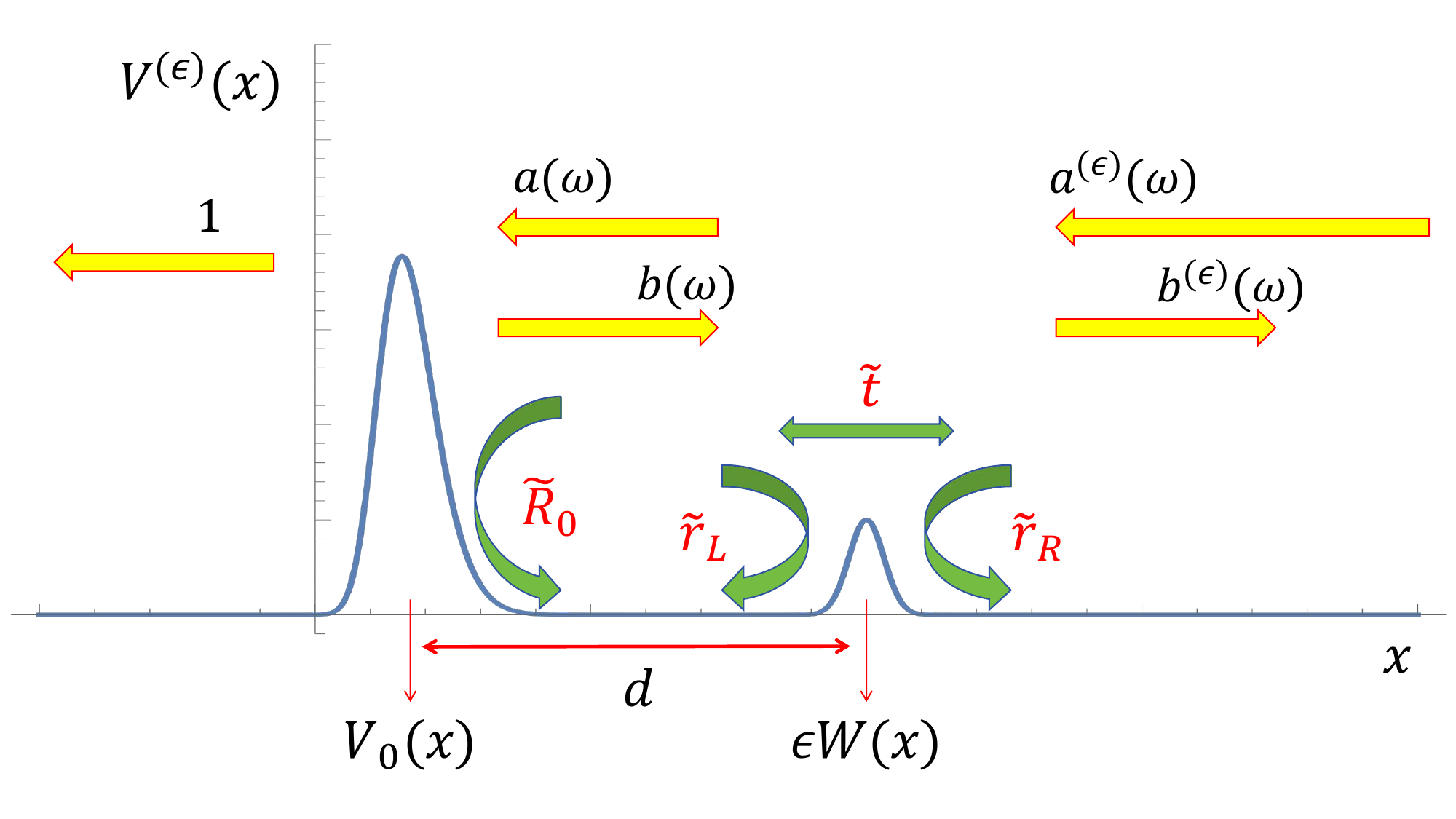}
	\caption{Illustration of the one-dimensional scattering problem. Yellow arrows correspond to the ingoing or outgoing waves for the Jost solution $\phi_{\omega}^{(\epsilon)}$ in different regions. Green arrows show the definitions of the transmission and reflection coefficients.}
	\label{fig:scattering}
\end{figure}

	As we discussed in Sec.~\ref{sub:definition}, we can interpret the Jost solution $\phi^{(\epsilon)}_{\omega}(x)$ as an incoming wave from $x=+\infty$ with amplitude $a^{(\epsilon)}(\omega)$, while the reflected and transmitted waves have amplitudes $b^{(\epsilon)}(\omega)$ and $1$, respectively. However now the potential consists of two ``peaks'' (see Fig.~\ref{fig:scattering}), so the incoming wave experiences multiple reflections and transmissions, leading to a modification of the scattering coefficients.
	Let us assume the bump potential to be located around $x=d>0$, and let us denote $W(x)=V_{b}(x+d)$ for convenience (the $d<0$ case is a trivial extension).
	We will denote the transmission coefficient of $W(x)$ by $\tilde{t}=\tilde{t}(\omega)$; 
	the reflection coefficients of $W(x)$ from the left and from the right by $\tilde{r}_{L}=\tilde{r}_{L}(\omega)$ and $\tilde{r}_{R}=\tilde{r}_{R}(\omega)$, respectively;
        and the reflection coefficient of the original effective potential $V_{0}(x)$ by $\tilde{R}_{0}=b^{(0)}(\omega)/a^{(0)}(\omega)$.
        Sometimes we will omit the $\omega$-dependence for simplicity.
        The definition of these scattering coefficients for $\phi^{(\epsilon)}_{\omega}(x)$ is illustrated in Fig.~\ref{fig:scattering}.

	If we trace the transmitted wave backwards into the intermediate region between $V_{0}(x)$ and the bump $W(x)$, the amplitude of the ingoing and outgoing waves should remain the same as in the unperturbed problem: as the bump potential only becomes important at a large distance, to produce a transmitted wave of unit amplitude near the horizon, the wavefunction in the intermediate region should not change. We can thus expect that an ingoing wave with amplitude $a^{(0)}(\omega)$ must be produced by the incoming wave from $+\infty$ in the intermediate region. However the ingoing wave in the intermediate region is a superposition of many components. At first we have the transmitted wave $\tilde{t}a^{(\epsilon)}(\omega)$. This transmitted wave is then reflected to the right by the unperturbed potential and reflected again to the left by the ``bump'', adding another factor $\tilde{r}_{L}\tilde{R}_{0}$ as well as a phase factor of $\mathrm{e}^{2i\omega d}$, because it must propagate over an additional distance of $2d$. 
	We should also sum the components resulting from waves that are reflected more times and travel further, and their sum should equal $a^{(0)}(\omega)$ (see Refs.~\cite{2017PhRvD..96h4002M,2019JCAP...12..020H} for similar considerations). This yields
	\begin{equation}\label{aeps}
		\begin{aligned}
			a^{(\epsilon)}(\omega)& =\dfrac{a^{(0)}(\omega)}{\tilde{t}+\tilde{t} \tilde{R}_0 \tilde{r}_L \ee^{2 i \omega d}+\tilde{t}\left(\tilde{R}_0 \tilde{r}_L \ee^{2 i \omega d}\right)^2+\cdots}\\
			& =\frac{1}{\tilde{t}}(1-\tilde{R}_0 \tilde{r}_L \ee^{2 i \omega d})a^{(0)}(\omega) \\
			& =\frac{1}{\tilde{t}}\left[a^{(0)}(\omega)-b^{(0)}(\omega) \tilde{r}_L \ee^{2 i \omega d}\right] \, . \\
		\end{aligned}
	\end{equation}
	Using similar arguments, we get an expression for $b^{(\epsilon)}(\omega)$,
	\begin{equation}\label{beps}
		\begin{aligned}		
			b^{(\epsilon)}(\omega)&= \left[\tilde{r}_R+\tilde{t}^{2} \tilde{R}_0 e^{2i \omega d}\left(1+\tilde{R}_0 \tilde{r}_L e^{2 i \omega d}+\cdots\right)\right]a^{(\epsilon)}(\omega)\ee^{-2i\omega d}\\
			&=b^{(0)}(\omega) \frac{\tilde{t}^2-\tilde{r}_R \tilde{r}_L}{\tilde{t}}+a^{(0)}(\omega) \frac{\tilde{r}_{R}}{\tilde{t}} \ee^{-2 i \omega d} \, .  \\
		\end{aligned}
	\end{equation}
        The factor $|\tilde{R}_{0}\tilde{r}_{L}\ee^{-2i\omega d}|$ is always smaller than 1 for real $\omega$. For complex $\omega$ this will not always be the case. We can solve the equation assuming $\omega$ to be real, and then perform an analytic continuation of Eq.~(\ref{aeps}) and Eq.~(\ref{beps}).

Note that these different waves can be superposed because in the frequency domain we deal with a stationary problem. In the frequency domain, the instability can be seen as arising from adding together waves that are misaligned in the time domain, and requiring them to cancel out through mutual interference. By considering their finite propagation speed, one can partially ``cure'' this instability (see Sec.~\ref{sec:t-domain} below).

The solutions for $a^{(\epsilon)}(\omega)$ and $b^{(\epsilon)}(\omega)$ in Eq.~(\ref{aeps}) and Eq.~(\ref{beps}) are still based on an approximation, as we neglect the asymptotic behavior of both the unperturbed potential and the bump potential in the intermediate region, treating $\phi_{\omega}^{(\epsilon)}(x)$ as noninteracting waves. As long as the bump potential is far away from the original effective potential and both are short-ranged, this approximation captures the most significant features of the problem. We will revisit this issue in Sec.~\ref{sub:more on approxiamtion} below.
		
	\subsection{Susceptibility and instability in the frequency domain}
	\label{sub:kai}

	To quantify the modification in the QNM frequencies due to the perturbation, let us define the susceptibility of the $n$-th QNM as
	\begin{equation}\label{kai}
	\chi_{n}\equiv \lim _{\epsilon \rightarrow 0} \frac{\delta \omega_{n}}{\epsilon} \, .
	\end{equation}
	Here we assume both $\epsilon$ and $\delta\omega_{n}$ to be small enough that $a^{(0)}(\omega^{(\epsilon)}_{n})\approx a^{(0)}(\omega_{n})+a^{(0)\prime}(\omega_{n})\delta\omega_{n}$, where a prime denotes a derivative with respect to $\omega$. 
	Similarly, we assume the difference between $a^{(\epsilon)}(\omega)$ and $ a^{(0)}(\omega) $ to be of order $\epsilon$.
        By imposing $a^{(0)}(\omega_{n})=a^{(\epsilon)}(\omega^{(\epsilon)}_{n})\simeq
        a^{(\epsilon)}(\omega_{n})+a^{(\epsilon)\prime}(\omega_{n})\delta\omega_{n}=0$, and approximating $a^{(\epsilon)\prime}(\omega_{n})\simeq a^{(0)\prime}(\omega_{n})$, we find
	\begin{equation}\label{susceptibility}
		\begin{aligned}
			\chi_{n}&=-\lim _{\epsilon \rightarrow 0}\left[ \dfrac{1}{\epsilon}\dfrac{a^{(\epsilon)}(\omega_{n})-a^{(0)}(\omega_{n})}{a^{(0)\prime}(\omega_{n})}\right]\\
			&=\lim _{\epsilon \rightarrow 0}\left(\dfrac{1}{\epsilon}\frac{\tilde{r}_L}{\tilde{t}}\right) \dfrac{ b^{(0)}(\omega_{n})}{a^{(0)\prime}(\omega_{n})}\ee^{2 i \omega_{n} d}\,,
		\end{aligned}
	\end{equation}
	where $\tilde{t},  \tilde{r}_{L} $ are all evaluated at $\omega_{n}$.
	
        This leads to the following definition:

        \noindent
        \textbf{\em Type~I instability}.
        A ``Type~I instability'' of the QNM with frequency $\omega_{n}$ under the effect of the perturbative potential $W(x)$ occurs when $\chi_{n}$ turns out to be singular. 
        \par
        There are two different possibilities for a Type~I instability. In the first case, the position of the $n$-th QNM frequency in the complex plane changes discontinuously, so the gap between $\omega^{(\epsilon)}_{n}$ and $\omega_{n}$ is finite as $\epsilon\to 0$. In the second case, $\omega^{(\epsilon)}_{n}$ is continuous at $\epsilon=0$, but the corresponding susceptibility is infinite. Under a Type~I  instability, the structure of the QNM spectrum can change drastically even for small $\epsilon$. From Eq.~(\ref{susceptibility}), a Type~I instability can happen in two ways: 
        \begin{itemize}
        \item[(I)] The coefficient $ \dfrac{\tilde{r}_{L}}{\tilde{t}} $ (related to the properties of the ``bump'') diverges. This case is not physical, because it means that a wave with frequency $\omega_{n}$ is completely reflected by the bump.
        \item[(II)]  $a^{(0)\prime}\left(\omega_{n}\right)$ is zero, in addition to the unperturbed QNM condition $ a^{(0)}(\omega_{n})=0$.
        \end{itemize}
	
        Apart from this somewhat unnatural Type~I instability, there is a second, more interesting type of instability, first pointed out by \citet{2014PhRvD..89j4059B} (see also Refs.~\cite{2022PhRvL.128k1103C,1999PhRvD..59d4034L}). Intuitively, a ``bump'' located at large distance from the unperturbed potential should not cause a dramatic distortion of the spectrum. However, we can see from Eq.~(\ref{susceptibility}) that the susceptibility depends on $ d $, and if $\Im \omega_{n}<0$, this leads to an exponential divergence with $d$. This leads to our second definition:

        \noindent        
        \textbf{\em Type~II instability}.
        A ``Type~II instability'' of the QNM with frequency $\omega_{n}$ under the effect of the perturbation $W(x)$ occurs if for an arbitrarily large value $K$ we can always find a distance $d$ such that the susceptibility $\chi_{n}(d)>K$.
        
	From Eq.~(\ref{susceptibility}) we see that the susceptibility $ \chi_{n} $ is a function of the distance parameter $d$. When $\Im \omega_{n}<0$, $\left|\chi_{n}(d)\right|$ can be arbitrarily large as $d$ increases, so we can always find a $d$ such that $\omega^{(\epsilon)}_{n}-\omega_{n}$ is comparable to $\omega_{n}$ for any arbitrarily small $\epsilon$. 
	In this case, the potential $W$ itself does not cause the instability of $\omega_{n}$, which would occur anyway as long as the bump is located far enough from the peak of the unperturbed potential.
	The condition for a Type~II instability to occur is that $\tilde{r}_{L}/\tilde{t}$ evaluated at $\omega_{n}$ must be nonzero,
        which is physically plausible (the reflection rate is nonzero, and the transmission rate is nondivergent).
	In the next section we will further explore Type~II instabilities for both nonrotating and rotating BHs. 

	\section{Spectral instability}
	\label{sec:BH instability}
	
	\subsection{First-order expression for $\chi_{n}$}
	\label{sub:first order chi}

        Before applying our formalism to the BH case, it is useful to compute the factor $ \tilde{r}_{L}/\tilde{t} $ in Eq.~(\ref{susceptibility}) at first order in $\epsilon$.
	As we show in Appendix~\ref{appendix:scattering problem}, at this order we can write the scattering coefficients as
	\begin{equation}\label{first order scattering coefficient}
		\begin{aligned}
			\dfrac{\tilde{r}_{L}(\omega)}{\tilde{t}}=& \dfrac{\ee^{-2i\omega d}}{2i\omega}\int_{-\infty}^{+\infty}\epsilon V_{b}(x)\mathrm{e}^{2i\omega x}\mathrm{d}x=\dfrac{\epsilon}{2i\omega}\widetilde{\mathrm{W}}(2\omega) \, , \\
			\tilde{t}(\omega) =& 1+\dfrac{1}{2i\omega}\int_{-\infty}^{+\infty}\epsilon V_{b}(x)\mathrm{d}x=1+\dfrac{\epsilon}{2i\omega}\widetilde{\mathrm{W}}(0) \, ,
		\end{aligned}
	\end{equation}
	where $ W(x)=V_{b}(x+d) $ and $\widetilde{\mathrm{W}}(\omega)$ denotes the spatial Fourier transform of $W(x)$, i.e.,
	$$
	\widetilde{\mathrm{W}}(\omega)=\int^{+\infty}_{-\infty}W(x)\mathrm{\ee}^{i\omega x}\mathrm{d}x\, .
	$$
	Notice that Eq.~(\ref{first order scattering coefficient}) is the first-order approximation of a series expansion in $ \epsilon/\omega $.
        Then Eq.~(\ref{susceptibility}) gives
	\begin{equation}\label{chi1}
		\chi_{n}(d)=\frac{b^{(0)}\left(\omega_{n}\right)}{2i a^{(0)\prime}\left(\omega_{n}\right) \omega_{n}} \widetilde{\mathrm{W}}\left(2 \omega_{n}\right)\ee^{2 i \omega_{n} d}  \, .
	\end{equation}
	For the case of a Gaussian bump we have $ \widetilde{W}(\omega) =\sqrt{\pi}\ee^{-\omega^{2}/4}$,  and therefore $ \chi_{n}(d) $ grows exponentially with $d$. 

        \begin{figure*}
          \centering
          \includegraphics[width=1.0\linewidth]{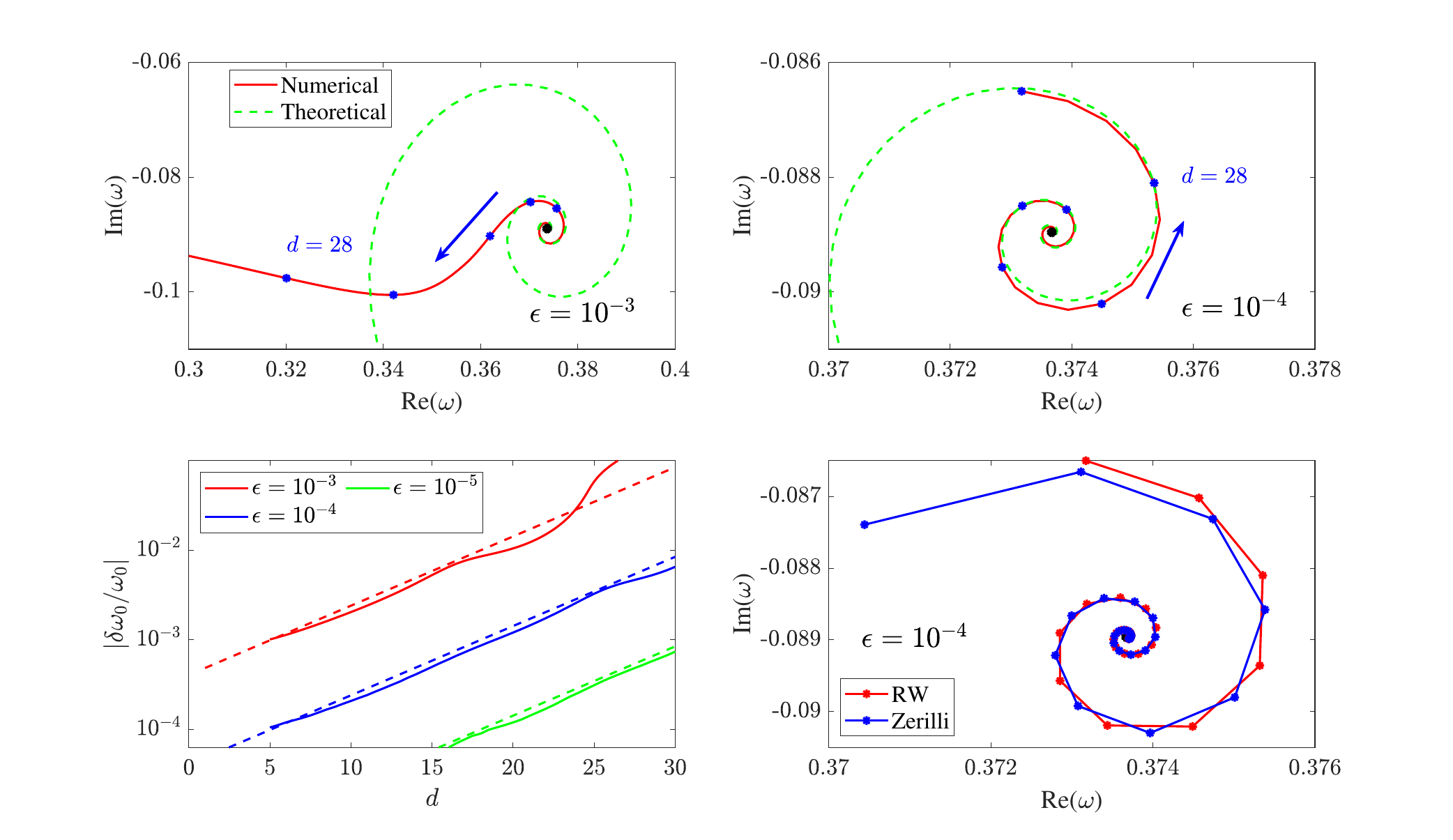}
          \caption{QNM frequencies computed by solving numerically the eigenvalue problem for a Gaussian bump, $ W(x)=\mathrm{exp}(-x^{2}) $, and compared with theoretical predictions. Top left: QNM frequencies for $\epsilon=10^{-3}$ and for an unperturbed RW potential. The red line (numerical calculation) and the green dashed line (theoretical prediction) are in good agreement for small deviations from the unperturbed mode. Blue dots refer to the perturbed mode at specific values of $ d $, ranging from 20 to 28 in steps of 2. The blue arrow shows the direction of increasing $ d $. Top right: same as the top left panel, but for $\epsilon=10^{-4}$. Blue dots now refer to values of $ d $ ranging from 20 to 30 in steps of 2. Bottom left: logarithmic plot of the relative deviation from the unperturbed fundamental mode as a function of $d$ for selected values of $\epsilon$ computed numerically (solid lines) versus the theoretical prediction (dashed lines). Bottom right: comparison of the perturbed fundamental mode frequencies for the RW potential and for the Zerilli potential, both with the same Gaussian bump ($\epsilon=10^{-4}$). In both cases, $ d $ ranges from 5 to 30 in steps of 1. }
          \label{fig:shootting}
        \end{figure*}
	
	If we consider an ``infrared cutoff'' of the effective potential~\cite{1996PhRvD..53.4397N,1999JMP....40..980N,2021PhRvX..11c1003J}, i.e., we truncate the right-hand side of the RW potential, we can set $ \epsilon V_{b}=- \ell (\ell+1)/x^{2} $ for $ x>d $ and sufficiently large $ d $. Substituting into Eq.~(\ref{first order scattering coefficient}) we find
	\begin{equation}
		\begin{aligned}
			(\tilde{r}_{L})_{\rm cutoff}&=-\ell(\ell+1)\ee^{2i\omega 	d}\int^{+\infty}_{0}\dfrac{\ee^{2i\omega x}}{(d+x)^{2}}\mathrm{d}x\\
			&=\dfrac{\ell(\ell+1)}{2i\omega}\dfrac{\ee^{2i\omega d}}{d^{2}}\left[1+{\cal O} \left(\dfrac{1}{\omega d}\right) \right]\,.
		\end{aligned}
	\end{equation}
	The dependence of $ \chi_{n}\propto \tilde{r}_{L} $ is slightly more complicated than for a Gaussian bump, but the characteristic exponential factor yielding QNM instability is still present.
	
	\subsection{Instability of Schwarzschild black holes}
	\label{sub:Schwarzschild}

        Now that we have computed the QNM susceptibility for a Gaussian bump and for a potential with a cutoff, it is trivial to apply the formulas to a Schwarzschild BH (see Refs.~\cite{2021PhRvX..11c1003J,2022PhRvL.128k1103C} for previous work).
        Our main goal is to describe some universal properties of Type~II instabilities that will also apply to the Kerr case to be discussed below.
	
	For concreteness, we apply Eq.~(\ref{chi1}) to the RW potential perturbed by a Gaussian bump, i.e., $ W(x)=\mathrm{exp}(-x^{2}) $. We compute numerically the new modes using a simple shooting method for selected values of $ \epsilon $. The corresponding migration trajectories for the fundamental mode are shown in Fig.~\ref{fig:shootting}. In the bottom-right panel we also compute the QNM frequencies for the Zerilli potential to show that isospectrality is broken by the perturbing potential.

	\noindent
	\textbf{\em Exponential instability.} Equation~(\ref{chi1}) implies that as the location of the bump $ W(x) $ moves away from the maximum of the unperturbed potential, $ \delta\omega_{n} $ increases exponentially,
	\begin{equation}
	\delta\omega_{n}=\epsilon\chi_{n}\propto\epsilon|\ee^{2i\omega_{n}d}|=\epsilon\ee^{-2 \Im \omega_{n}d} \,.
	\end{equation}
	The top two panels and the bottom-left panel of Fig.~\ref{fig:shootting} illustrate this for the fundamental mode $ \omega_{0} $ of the RW potential. The perturbed mode frequency $ \omega^{(\epsilon)}_{0} $ moves along a spiral as $ d $ increases. However, in the top-right panel ($\epsilon=10^{-4}$) we observe that when $d$ is large enough the mode frequency no longer moves along a spiral, but rather it moves towards the real axis. The same happens in the top-left panel for $\epsilon=10^{-3}$. This is because Eq.~(\ref{susceptibility}) assumes small deviations $\delta\omega_{n}$, so that we can expand $a^{(0)}(\omega)$ near $\omega_{n}$. When $d$ increases and $\chi_{n}$ increases exponentially, the perturbative analysis is no longer valid. The bottom-left panel of Fig.~\ref{fig:shootting} shows indeed that, once $ |\delta \omega_{0}|\gtrsim10^{-3} $, its growth with $d$ is no longer described by a simple exponential. In this sense, the condition $\delta\omega_{n}/\omega_{n}\approx10^{-2}$ is a rough criterion to determine the critical value of $ d_{c} $ at which the QNM frequency $\omega_{n}$ is highly destabilized. Because $\delta\omega_{n}\propto \epsilon\ee^{-2 \Im \omega_{n}d} $, this criterion also leads to a relation between $\epsilon$ and $ d_{c} $: as $\epsilon$ increases, the critical value of $ d $ for $\omega_{n}$ to destabilize decreases logarithmically. This explains the behavior observed numerically by \citet{2022PhRvL.128k1103C}: the critical line for the destabilization of the fundamental mode is a straight line in the $d$-$\log \epsilon$ plane.

	\noindent	
	\textbf{\em Overtone instability.} By definition of the overtone number, the factor $ |2 \Im \omega_{n}| $ increases with $ n $. Therefore, the QNM overtones will deviate more and more from their unperturbed values at a fixed $ d $, i.e., higher overtones are more unstable under small perturbations. For any $ d $, there will be a critical but finite $ n_{m} $ such that the deviation can no longer be described using a perturbative description when $ n>n_{m} $, so that higher overtones are always unstable.
        Our treatment cannot establish whether small perturbations near the light ring destabilize higher overtones, because of the assumptions we made in Sec.~\ref{sub:definition} on the location of the ``bump'' potential. We will get back to this point in Sec.~\ref{sub:more on approxiamtion}.

        \noindent
	\textbf{\em Isospectrality breaking.} Isospectrality (the fact that the axial RW potential and the polar Zerilli potential have the same QNM spectrum) is a special property of Schwarzschild BHs, and it is broken if we perturb the effective potential. From Eq.~(\ref{kai}) we see that the susceptibility is proportional to $ b^{(0)}(\omega) $. Although the RW and Zerilli potentials yield the same transmission coefficient, their reflection coefficients differ by a phase factor (see e.g. Chapter~27 of Ref.~\cite{Chandrasekhar:1985kt}), therefore the perturbed spectrum will be different for any $\epsilon\neq 0$.	

	\subsection{Instability of Kerr black holes}
	\label{sub:Kerr}

	An important restriction that we have imposed so far on the unperturbed potential is that it must be short-ranged. This condition must be satisfied in order for the solution to behave as a ``plane wave'', $\mathrm{e}^{\pm i\omega x}$, in the asymptotic regions.

        The condition is not satisfied for the Teukolsky equation~\cite{1973ApJ...185..635T} describing perturbations around Kerr BHs. Asymptotic solutions of the Teukolsky equation are not pure waves, but rather they are proportional to $r^{\pm s}$ when $r\rightarrow+\infty$ and to $\Delta^{\pm s/2}$ when $r\rightarrow r_{+}$, where $s$ is the spin of the perturbing field and $\Delta=r^2-2Mr+a^2$ (here $M$ and $a$ are the BH mass and spin parameters, respectively).
        To apply our method we can transform the Teukolsky equation into an equivalent form such that the potential is short-ranged, and the wave equation contains a second derivative term but no first derivative term. The Sasaki-Nakamura (SN) formalism~\cite{1982PThPh..67.1788S} casts the perturbation equations in this form. Within the SN formalism we can then apply our method, and this leads to the prediction that QNMs must be unstable also for Kerr BHs.
	
	The SN equation has the form
	\begin{equation}\label{SN}
		\frac{d^2 \psi}{d r_*^2}+\left(\frac{\mathcal{F}^{\prime}}{2}-\frac{\mathcal{F}^2}{4}-\mathcal{U}\right) \psi = 0 \, , 
	\end{equation}
	where we follow the notation of Ref.~\cite{2013PhRvD..88d4018Z}. Here $\psi$ is the wave function, and the tortoise coordinate $r_{*}$ is defined by the condition $\dd r /\dd r_{*}=(r^2+a^2)/\Delta$. With the identification $r_{*}\rightarrow x$, the equation has the same form as Eq.~(\ref{f-domain}). For completeness, the other functions are defined in Appendix~\ref{appendix:SN}.

	The potential $V_{\rm{SN}}=\left(\mathcal{F}^{\prime}/2-\mathcal{F}^2/4-\mathcal{U}\right)$ is short-ranged, and therefore it admits asymptotic wavelike solutions. If we add a bump potential to the left-hand side of Eq.~(\ref{SN}), the arguments we made above in the Schwarzschild case will apply. For a real frequency $\omega$, Eq.~(\ref{SN}) admits solutions such that
	\begin{equation*}
		\left\{
		\begin{aligned}
			\psi(r_{\star})\rightarrow&\mathrm{e}^{-i\omega r_{\star}},\ &r_{\star}\rightarrow-\infty \,, \\
			\psi(r_{\star})\rightarrow&a_{\rm{SN}}(\omega)\mathrm{e}^{-i\omega r_{\star}}+b_{\rm{SN}}(\omega)\mathrm{e}^{i\omega r_{\star}},\ &r_{\star}\rightarrow+\infty \,.
		\end{aligned}
		\right.
	\end{equation*}
	We now add a bump potential $W(r_{\star})$ to $ V_{\rm{SN}}$, following the steps in Secs.~\ref{sub:solution} and~\ref{sub:kai}. We can then introduce an analogous definition of the susceptibility for the Kerr spacetime,
	\begin{equation}\label{kerr susceptibility}
		\begin{aligned}
			\chi_{n}^{\rm Kerr}&=\lim _{\epsilon \rightarrow 0}\left(\dfrac{1}{\epsilon}\frac{\tilde{r}_L}{\tilde{t}}\right) \dfrac{ b^{(0)}_{\rm{SN}}(\omega_{n})}{a^{(0)\prime}_{\rm{SN}}(\omega_{n})}\ee^{2 i \omega_{n} d}\\
			&=\frac{b_{\rm{SN}}^{(0)}\left(\omega_{n}\right)}{2i a_{\rm{SN}}^{(0)\prime}\left(\omega_{n}\right) \omega_{n}} \widetilde{\mathrm{W}}\left(2 \omega_{n}\right)e^{2 i \omega_{n} d}  \,.
		\end{aligned}
	\end{equation}
	This equation gives conditions for the spectral instability of Kerr BHs, as the arguments given in Sec.~\ref{sub:kai} still apply.
	
	It is worth noting that our argument is only valid within the SN formalism, so the scattering coefficients in Eq.~(\ref{kerr susceptibility}) must be computed within the SN (rather than the Teukolsky) formalism. 
	The ``bump'' potential $W(r_{\star})$ also refers to the SN formalism; it should be added directly to Eq.~(\ref{SN}), and its physical meaning is not as clear. However, it is reasonable to expect that a short-ranged bump potential in the SN formalism can be related to a similar short-ranged potential in the Teukolsky formalism.
        A numerical calculation of specific examples is an interesting topic for future work.
	
	\subsection{Some comments on approximations and limits of validity}
	\label{sub:more on approxiamtion}

	So far we have considered a bump potential, $ V_{b} $, required to be (i) short-ranged, and (ii) sufficiently far away from the unperturbed potential. These two restrictions justify the approximations made in Sec.~\ref{sec:frequency domain analysis}.
	
	The first assumption seems natural: a bump potential that decays like $1/r$ (or more slowly) cannot be regarded as a ``small'' modification, as it will change the asymptotic behavior of the whole potential, and thus the asymptotic behavior of the Jost solutions. 
	The second assumption is not as natural and it may be broken in specific scenarios, but it is necessary for our proof of Type~II instability (recall Sec.~\ref{sub:solution}, where we derived the solution by considering the wave amplitude in the intermediate region). Therefore our proof only applies to bumps that are sufficiently far away from the light ring.
	
	We can relax assumption (ii) by considering the first-order approximation from the beginning. In Appendix~\ref{appendix:first order approximation} we show that the susceptibility can be rewritten as
	\begin{equation}\label{first order chi zhengwen}
		\chi_{n}=\dfrac{1}{2 i \omega a^{(0)\prime}(\omega)} \int_{-\infty}^{\infty} \psi_{\omega}(x) V_{b}(x) \phi_{\omega}(x) \mathrm{d} x \, , 
	\end{equation}
	where $ \psi_{\omega} $ and $ \phi_{\omega} $ are the Jost solutions defined in Sec.~\ref{sub:definition}. If we further impose that $ V_{b} $ is mainly localized in the asymptotic region of $ V_{0} $, where the Jost solutions can be replaced by their asymptotic expression, we recover Eq.~(\ref{chi1}), but now this is not necessary. From Eq.~\eqref{first order chi zhengwen} we see that as the bump potential moves towards the light ring, the susceptibility also changes continuously. Equation~(\ref{chi1}) implies that as $ d $ decreases (for fixed $ \omega_{n} $) the susceptibility decreases monotonically. Eventually $ d $ becomes so small that assumption (ii) breaks down, and then the susceptibility will change continuously as described by Eq.~(\ref{first order chi zhengwen}). 
	While the susceptibility may not continue to decrease monotonically, we expect its magnitude not to be large, because (at least for first few overtones) Eq.~(\ref{chi1}) predicts $ \chi_{n} $ to decrease with $d$. Moreover, the Jost solutions $ \psi_{\omega} $ and $ \phi_{\omega} $ are not divergent in the region around the unperturbed potential, thus the integral will not be very large. 
	It is unclear (while plausible) whether high QNM overtones will be destabilized by small changes around the light ring, but the above arguments suggest that the first few modes will not be sensitive to these changes. This seems plausible: QNMs are related to the broad properties of the light ring, and small modifications should not drastically affect them.

	\section{Time-domain analysis}
	\label{sec:t-domain}

        Working in the frequency domain, we have proved that the BH QNM instability caused by a bump is a very general phenomenon. However for detection purposes it is more relevant to understand whether, and to what extent, this instability affects the waveform in the time domain~\cite{2022PhRvD.106h4011B}.

        In this section we will consider, for simplicity, the wave equation describing perturbations of Schwarzschild BHs,
	\begin{equation}\label{master equation}
		\left(\frac{\partial^{2}}{\partial t^{2}}-\frac{\partial^{2}}{\partial x^{2}}\right) \Psi+V(x)\Psi=0\,,
	\end{equation}
	with appropriate initial conditions for the wave function $ \Psi(t=0,x) $ and its first derivative $ \partial_{t}\Psi(t=0,x) $. 
	
	In Sec.~\ref{sub:Green function} we briefly review the standard Green's function treatment. 
	In Sec.~\ref{sub:difference internal} we investigate ``internal'' initial data (localized at distances smaller than the location of the bump)
        and the causal structure of the Green's function, showing that the wavepackets related to perturbed QNMs are delayed with respect to the wavepackets carrying the original ringdown signal.
	In Sec.~\ref{sub:equivalent scattering problem}, combining theoretical arguments and numerical results, we show that the deviation of the perturbed waveform from the original waveform in the early ringdown phase, which is the most relevant for detection purposes, is proportional to $ \epsilon $ as long as $\epsilon\lesssim10^{-2}$, so the early ringdown waveform is not affected by the spectral instability.
        In Sec.~\ref{sub:difference external} we further consider ``external'' initial data (localized at distances larger than the location of the bump) with similar conclusions.
        This analysis confirms and extends the main findings of Ref.~\cite{2022PhRvD.106h4011B}.
	
        \begin{figure*}
          \centering
          \includegraphics[width=1.0\linewidth]{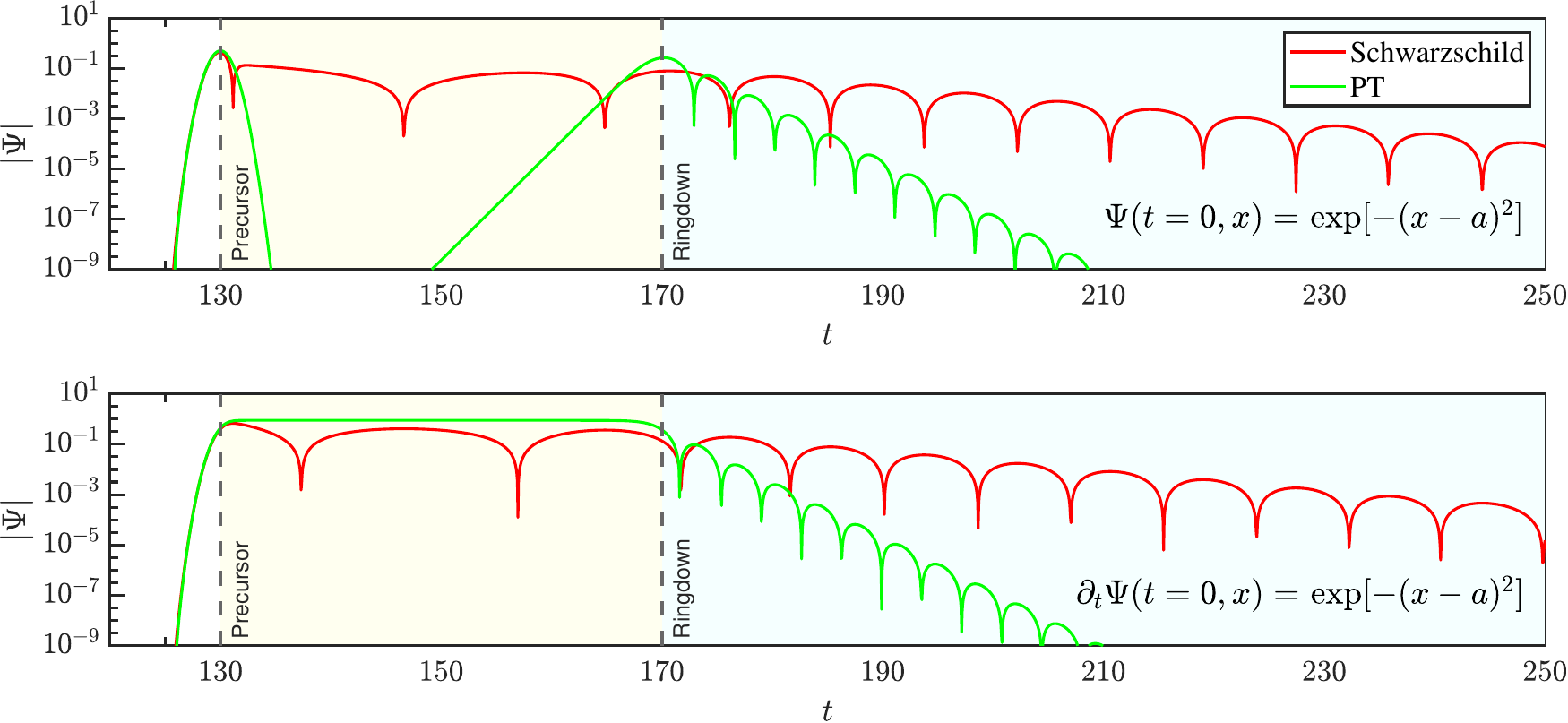}
          \caption{Illustrative time-domain waveform for axial (RW) perturbations of a Schwarzschild BH (in red) and for the PT potential (in green). The region shaded in yellow is the epoch between the arrival of the prompt wavepacket (precursor) and the beginning of the ringdown phase (shaded in blue).
            The top panel refers to Gaussian initial conditions on the wavefunction: $ \Psi(t=0,x)=\mathrm{exp}\left[-(x-a)^{2}\right]$ and $\partial_{t}\Psi(t=0,x)=0$ with $ a=20 $. 
            The bottom panel refers to Gaussian initial condition on its derivative: $ \Psi(t=0,x)=0$ and $\partial_{t}\Psi(t=0,x)=\mathrm{exp}\left[-(x-a)^{2}\right]$ with $ a=20 $.
            In both cases the observer is located at $ x=150 $. }
          \label{fig:schwarzchild-pt}
        \end{figure*}

	\subsection{Green's function}
	\label{sub:Green function}

	The use of Green's function techniques is standard in the study of BH perturbations in the time domain~\cite{1986PhRvD..34..384L,Andersson:1995zk}, but we briefly review it here to establish notation. 
	
	We will mainly work using Laplace transforms and define a new frequency variable $s=-i\omega$. Let $\hat{G}\left(s, x, x^{\prime}\right)$ be any Green's function of Eq.~(\ref{f-domain}), i.e., any solution of the equation
	\begin{equation}
		\left[\partial_x^2-s^2-V(x)\right] \hat{G}\left(s, x, x^{\prime}\right)=\delta\left(x-x^{\prime}\right)  \, .
	\end{equation}
	Imposing QNM boundary conditions, the Green's function can be constructed by writing
	\begin{equation*}
		\begin{aligned}
			\hat{G}\left(s, x, x^{\prime}\right)=\frac{1}{\mathcal{W}(s)} & {\left[\theta\left(x-x^{\prime}\right) \hat{\phi}_{-}\left(s, x^{\prime}\right) \hat{\phi}_{+}(s, x)\right]} \\
			& \left.+\theta\left(x^{\prime}-x\right) \hat{\phi}_{-}(s, x) \hat{\phi}_{+}\left(s, x^{\prime}\right)\right] \, , 
		\end{aligned}
	\end{equation*}
	where $\mathcal{W}(s)$ is the Wronskian of two solutions of the homogeneous equation, $\hat{\phi}_{-}(s,x)$ and $\hat{\phi}_{+}(s,x)$, with the following asymptotic behavior:
	\begin{equation*}
		\hat{\phi}_{-}(s,x\rightarrow -\infty)\rightarrow \ee^{+sx},\ \ \hat{\phi}_{+}(s,x\rightarrow +\infty)\rightarrow \ee^{-sx} \, .
	\end{equation*}
	Note that $\hat{\phi}_{-}$ is just the Jost solution $\phi_{\omega}$ defined in Sec.~\ref{sub:definition}. For an observer at future infinity, the Green's function then reduces to
	\begin{equation}\label{s-Green}
		\hat{G}\left(s, x, x^{\prime}\right)=\frac{1}{\mathcal{W}(s)} \hat{\phi}_{-}\left(s, x^{\prime}\right) \ee^{-sx} .
	\end{equation}
	We will mostly follow the conventions of Sec.~\ref{sub:definition}. We denote $\hat{\phi}_{-}(s,x)\rightarrow A(s)e^{sx}+B(s)e^{-sx}$ when $x\rightarrow+\infty$, so that $A(s)=a(\omega),B(s)=b(\omega)$ in the new notation, and the Wronskian is given by $ \mathcal{W}(s)=2sA(s) $.
	The time-domain Green's function can be obtained by an inverse Laplace transform:
	\begin{equation}\label{inverse}
		\begin{aligned}
			G\left(t, x, x^{\prime}\right) =& \mathscr{L}^{-1}\left[\hat{G}(s,x,x^{\prime})\right]\\
			=&\int_{s_{0}-i\infty}^{s_{0}+i\infty}\mathrm{d}s ~\ee^{s(t-x)}\frac{1}{2sA(s)} \hat{\phi}_{-}\left(s, x^{\prime}\right) \, .
		\end{aligned}
	\end{equation}
	The solution of the perturbation equations in terms of the initial conditions is
	\begin{equation}\label{initial integral}
		\begin{aligned}
			\phi(t, x)=-\int_{-\infty}^{\infty} \dd x^{\prime}\left[\partial_t G\left(t, x, x^{\prime}\right) \phi(t=0, x^{\prime})\right.\\
			\left.+G\left(t, x, x^{\prime}\right) \partial_t \phi(t=0, x^{\prime})\right] \, .
		\end{aligned}
	\end{equation}
	It is useful to review some properties of the Green's function for later convenience. Assume that the initial conditions are localized in the asymptotic region, $x^{\prime}\gg1$. We have $ \hat{\phi}^{(0)}_{-}\left(s,x^{\prime}\right)=A^{(0)}(s) \ee^{s x^{\prime}}+B^{(0)}(s) \ee^{-s x^{\prime}} $, so Eq.~(\ref{inverse}) reduces to
	\begin{equation}\label{original Green}
			G\left(t, x, x^{\prime}\right)=\int_{s_{0}-i\infty}^{s_{0}+i\infty}\mathrm{d}s\ \frac{1}{2s} \left(\ee^{s (t-x+x^{\prime})}+\dfrac{B^{(0)}(s)}{A^{(0)}(s)} \ee^{s (t-x-x^{\prime})}\right) \, . 
	\end{equation}
	The inverse Laplace transform $ F(s) $ of a function $ f(t) $  satisfies the ``retardation theorem'',
	\begin{equation*}
		\mathscr{L}[f(t-\tau) \mathcal{\theta}(t-\tau)]=\ee^{-\tau s} F(s) \, .
	\end{equation*}
	The calculation of the inverse Laplace transform for Schwarzschild BHs involves a contour integral along a semicircle on the left side of the complex plane with a branch cut along the negative real axis, including the contribution from residues at the poles away from the real axis~\cite{1986PhRvD..34..384L}.
	
        We can interpret the two terms with different exponential factors in Eq.~\eqref{original Green} as two waves with different propagation histories (see Ref.~\cite{2019JCAP...12..020H} for similar considerations in a different context). The first term represents a wavepacket propagating directly from the initial position to the observer and arriving at $ t=x-x^{\prime} $. 
        The second term represents a wavepacket that first propagates inward, is reflected by the potential barrier, and then propagates to the observer, arriving at $ t=x+x^{\prime} $. 
        It is this second term that  produces QNM ringing due to its poles. In other words, the wave reflected by the potential barrier at the location of the light ring is responsible for triggering the excitation of QNMs that then propagate to the observer.

        This intuitive picture is complicated by the fact that QNM eigenfunctions do not form a complete set. In general, as illustrated in Fig.~\ref{fig:schwarzchild-pt}, there are three main contributions to the Green's function~\cite{1986PhRvD..34..384L,2018gwv..book.....M}:

        \noindent
        \textbf{\em (i) Precursor.} This contribution comes mainly from integration along the large semicircle with $|s|\rightarrow\infty$, with contributions from the integration along the branch cut and from the (divergent) sum over QNMs~\cite{2012PhRvD..86b4021C}. 
        Physically, the precursor represents both the ``geometric optics'' component of the waveform (prompt light-cone propagation of the high-frequency part of the initial data from its initial location to the observer) and its dispersion. 
        The ``geometric optics'' part starts at $t\approx x-x^{\prime}$ and dominates the prompt signal.
        For BHs (red curves in Fig.~\ref{fig:schwarzchild-pt}), dispersion occurs during propagation because of the power-law decay of the effective potential. 
        These dispersion effects are not present for the P\"oschl-Teller (PT) potential $V_{\rm PT}={\rm sech}(x)^{2}$ (green curves in Fig.~\ref{fig:schwarzchild-pt}), which decays exponentially.
		
        \noindent
        \textbf{\em (ii) QNMs.} The QNM contribution comes from the poles of the function $A^{(0)}(s)$ on the left-half plane. As we have seen, the wave first propagates inwards and it is reflected by the potential peak before propagating towards the observer, so this stage starts at $t\approx x+x^{\prime}$.
		
        \noindent
        \textbf{\em (iii) Power-law tail.} The integration along the branch cut gives power-law tails in the time domain if the potential has a power-law decay (tails are not present for the PT potential).
        In the present work we are not interested in the late-time waveform, so this part is not shown in Fig.~\ref{fig:schwarzchild-pt}.
			
	\subsection{Difference Green's functions for ``internal'' initial data}
	\label{sub:difference internal}

	To understand how spectral instability affects the time-domain waveform, let us define a ``difference Green's function'' computed by subtracting the unperturbed Green's function from the perturbed Green's function:
	\begin{equation}\label{difference G}
		\begin{aligned}		
			& \Delta \hat{G}\left(s, x, x^{\prime}\right)=\frac{\ee^{-s x}}{2 s A^{(\epsilon)}(s)} \hat{\phi}_{-}^{(\epsilon)}\left(s, x^{\prime}\right)-\frac{\ee^{-s x}}{2 s A^{(0)}(s)} \hat{\phi}^{(0)}_{-}\left(s, x^{\prime}\right) \, .
		\end{aligned}
	\end{equation}
	To proceed, we need to specify the location of our initial data (namely, $ x^{\prime} $). We will distinguish between the case $0<x^{\prime}<d$ (``internal'' initial data, discussed here) and the case $x^{\prime}>d$ (``external'' initial data, discussed in Sec.~\ref{sub:difference external} below). 
	For convenience we choose $ x^{\prime} $ to be sufficiently far away from the origin that $\hat{\phi}_{-}(x^{\prime})$ can be treated as a wavelike function. 
	
	For internal initial data, we can neglect the difference between $\hat{\phi}^{(\epsilon)}_{-}\left(s, x^{\prime}\right) $ and $\hat{\phi}^{(0)}_{-}\left(s,x^{\prime}\right)$ (an approximation we have already used in Sec.~\ref{sub:solution}, and justified more rigorously in Appendix~\ref{appendix:scattering problem}):
	\begin{equation}
		\hat{\phi}^{(\epsilon)}_{-}\left(s, x^{\prime}\right) \approx\hat{\phi}^{(0)}_{-}\left(s,x^{\prime}\right)=A^{(0)}(s) e^{-s x}+B^{(0)}(s) \ee^{s x} \, .
	\end{equation}
	By substituting into Eq.~(\ref{difference G}) and using Eq.~(\ref{aeps}) we find
	\begin{equation}\label{four waves}
		\begin{aligned}
			\Delta \hat{G}_{\rm int}\left(s, x, x^{\prime}\right)&= \frac{A^{(0)}(s)-A^{(\epsilon)}(s)}{2 s A^{(0)}(s)A^{(\epsilon)}(s)} \hat{\phi}^{(0)}_{-}\left(s, x^{\prime}\right)\ee^{-s x}\\
			&= \frac{1}{2 s} \frac{A^{(0)}(s)}{A^{(\epsilon)}(s)} \frac{\tilde{t}-1}{\tilde{t}} \ee^{-s\left(x-x^{\prime}\right)} \\
			&+\frac{1}{2 s}\frac{B^{(0)}(s)}{A^{(\epsilon)}(s)} \frac{\tilde{t}-1}{\bar{t}} \ee^{-s\left(x+x^{\prime}\right)} \\
			&+\frac{1}{2 s} \frac{B^{(0)}(s)}{A^{(\epsilon)}(s)} \frac{\tilde{r}_L}{\tilde{t}} \ee^{-s\left(x-x^{\prime}+2 d\right)} \\
			&+\frac{1}{2 s} \frac{B^{(0)}(s)}{A^{(\epsilon)}(s)} \tilde{R}_0 \frac{\tilde{r}_L}{\tilde{t}} \ee^{-s\left(x+x^{\prime}+2 d\right)} \, .
		\end{aligned}
	\end{equation}
\begin{figure}
	\centering
	\includegraphics[width=1.0\linewidth]{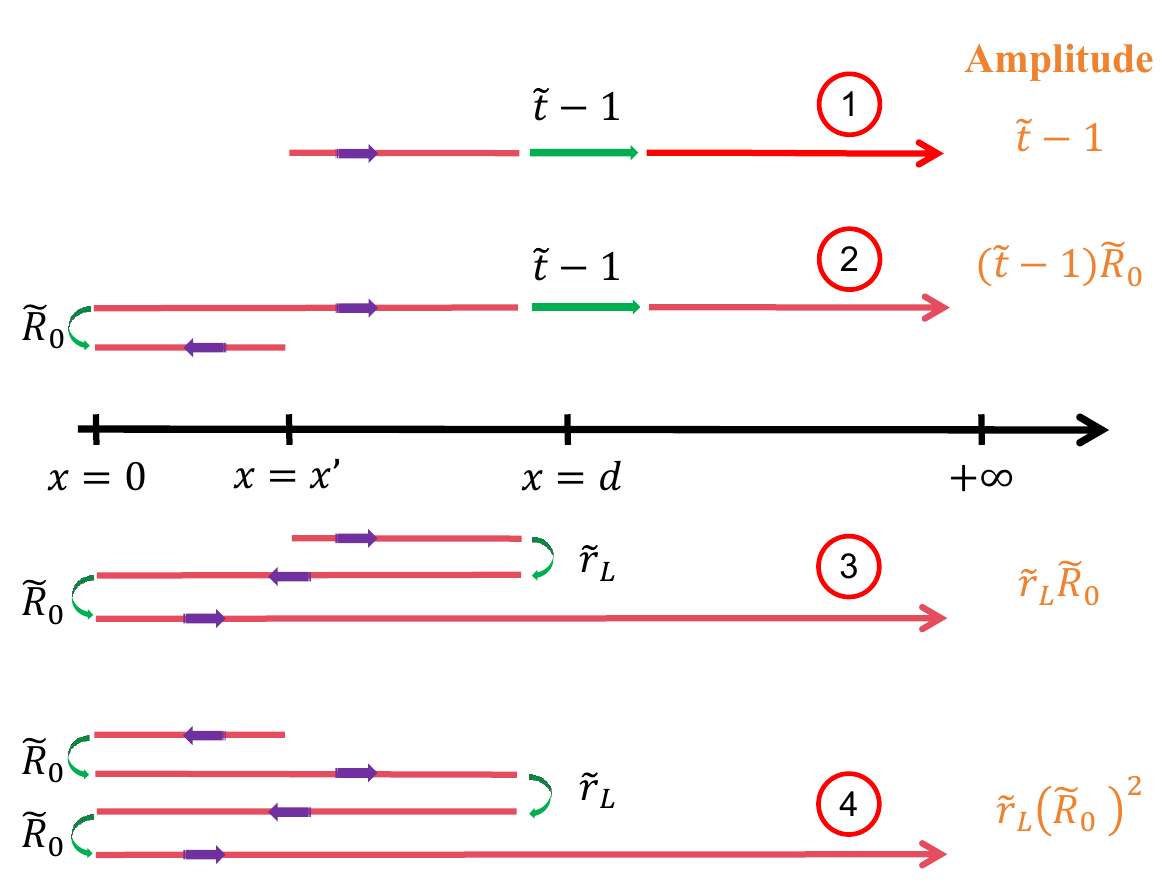}
	\caption{Interpretation of the four components of the difference Green's function (\ref{four waves}), representing four different waves with different arrival times. Green arrows represent transmission or reflection on the unperturbed potential or on the bump. The amplitude of the four components at the observer's location is shown in orange on the right.}
	\label{fig:four waves}
\end{figure}
Equation~(\ref{four waves}) consists of four terms. The discussion of Sec.~\ref{sub:Green function} implies that the difference Green's function consists of four waves with different arrival times whose interpretation is very natural, as shown in Fig.~\ref{fig:four waves} and discussed below:
	\begin{itemize}
		\item[(1)] The first wave (arriving at $t=x-x^{\prime}$) is generated when the outgoing precursor hits the bump from the left, and its amplitude decreases after transmission. 
		Since we are considering the difference between the perturbed and unperturbed signal, this transmission event generates a new component of the wave with amplitude equal to the part ``blocked'' by the bump. 
		\item[(2)] The second wave (arriving at $t=x+x^{\prime}$) is generated when the ingoing part of initial wavepacket (the one carrying the QNM signal to the observer) is transmitted through the bump.
		\item[(3)] The third wave (arriving at $t=x-x^{\prime}+2d$) is the counterpart of the first wave that is first reflected ``to the left'' by the bump potential, then reflected ``to the right'' by the unperturbed potential, and finally transmitted through the bump.
		\item[(4)] The fourth wave (arriving at $t=x+x^{\prime}+2d$) is, similarly, the counterpart of the second wave that is reflected by the bump, reflected again by the unperturbed potential, and finally transmitted through the bump.
	\end{itemize}
        In our proof of instability in the frequency domain, adding waves with different histories was enough to prove the generic instability of QNMs induced by the bump.
        The time-domain analysis of the difference Green's function reveals a more intricate causal structure due to retardation, with different classes of waves separated in time from each other. 
        This causal structure will be used below to explain why the spectral instability does not significantly affect the early ringdown signal.
	
	It is interesting to analyze the poles of Eq.~(\ref{four waves}). The first term (except for $s=0$) has a pole only when $A^{(\epsilon)}(s)=0$, i.e., at the perturbed QNMs. 
        Note that the QNMs of the bump potential (which are poles of both $\tilde{r}_{L}$ and $\tilde{t}$) will not contribute, because $\tilde{t}$ appears in both the numerator and denominator.
        In other words, the first wave rings down (if the oscillation can be observed at all) at the frequency of the perturbed QNMs, not at the frequency of the unperturbed QNMs or of the QNMs associated with the bump. 
        The same argument applies to all other components except for the fourth wave, which also contains the poles of $\tilde{R}_{0}$, so it rings down at both the unperturbed and perturbed QNM frequencies.

	Rather than taking the inverse Laplace transform of each component of Eq.~(\ref{four waves}), which requires a careful treatment to cancel infinities~\cite{1986PhRvD..34..384L,2012PhRvD..86b4021C}, we will now develop some insight into the behavior of the Green's function by considering an equivalent scattering problem.
        
	\subsection{Equivalent scattering problem} 
	\label{sub:equivalent scattering problem}

        The waveform cannot be derived by simply taking the QNM part of the four waves appearing in the difference Green's function: other contributions (especially the precursor part) must be taken into consideration.
	
\begin{figure}
		\centering
		\includegraphics[width=1.0\linewidth]{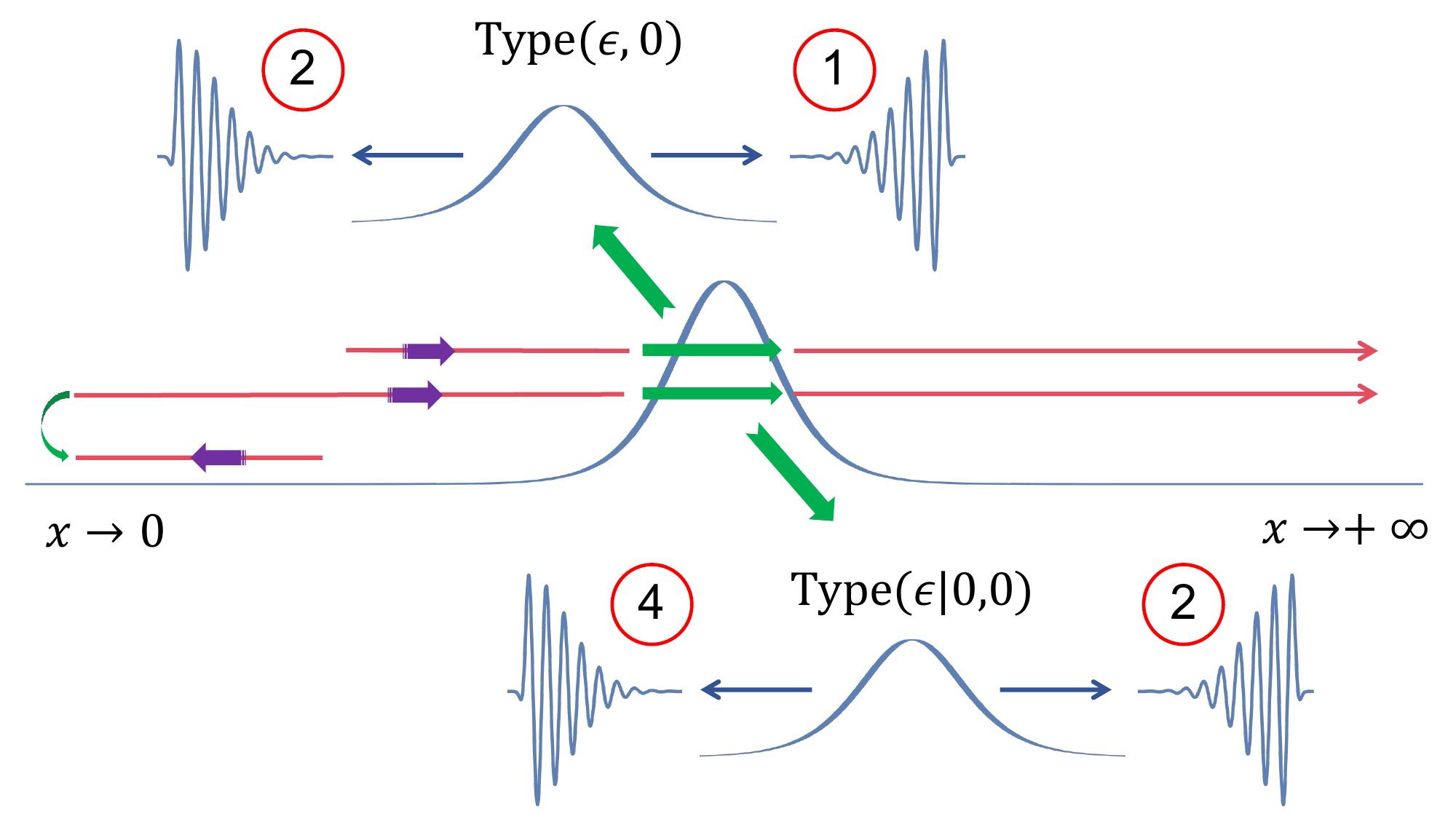}
		\caption{Schematic diagram of the equivalent scattering problem discussed in Sec.~\ref{sub:equivalent scattering problem}. The first and third wave can be seen as the result of a Type $(\epsilon,0)$ scattering problem, with initial conditions generated by the transmission of the original wavepacket through the bump. Similarly, the second and fourth wave result from a Type $(\epsilon|0,0)$ scattering problem, with initial conditions generated by the transmission of the wavepacket which carries the original QNM signal.}
		\label{fig:equivalent-problem}
	\end{figure}

	However there is an alternative and more intuitive way to treat the problem, illustrated in Fig.~\ref{fig:equivalent-problem}. As we discussed above, when the free propagating wavepacket meets the bump it generates a transmitted wave with a lower amplitude compared to the original wave.
	If we now regard this as an ``initial wavepacket'', the first and third components of Fig.~\ref{fig:four waves} can be seen as being generated by this ``initial wavepacket''.
        Similar considerations apply to the second and fourth components: the initial data propagating to the left, reflected by the original effective potential and then transmitted by the bump can be seen as generating another equivalent ``initial wavepacket'', which leads to the second and fourth waves.

	To express this idea mathematically, let us define a special Green's function $G_{(a)}^{(b)}$:
	\begin{equation}
		G_{(a)}^{(b)}\left(s, x, x^{\prime}\right)=\frac{\ee^{-s x}}{2 s A^{(a)}(s)}\hat{\phi}^{(b)}(s,x^{\prime}) \, ,
	\end{equation}
	where $a,\,b$ refer to the corresponding function on the right-hand side. For example, $a=\epsilon$ means $ A^{(a)}(s)= A^{(\epsilon)}(s)$, and $b=0$ means $\hat{\phi}^{(b)}(s,x^{\prime})= \hat{\phi}^{(0)}(s,x^{\prime})$. The meaning of this Green's function is clear: it is the Green's function of an artificial scattering problem, with pole structure given by $A^{(a)}(s)$ and wave solution given by $\hat{\phi}^{(b)}(s,x)$. For example, $G^{(0)}_{(0)}$ is just the Green's function of the original scattering problem. In general, we will call this a ``Type $(a,\,b)$'' scattering problem. The advantage of these definitions should be clear soon.
	
	Let us now add the first and third waves together, and the second and fourth waves together in pairs, as each pair has a similar origin. We then expand $\tilde{t}$ and $\tilde{r}_{L}$ to first order in $\epsilon$ (as in Sec.~\ref{sub:kai}). For example, for the first and third waves we have
	\begin{widetext}
		\begin{equation}
			\begin{aligned}
				\frac{\ee^{-s\left(x-x^{\prime}\right)}}{2 s A^{(\epsilon)}(s)}\left[A^{(0)}(s)\left(1-\dfrac{1}{\tilde{t}}\right)+B^{(0)}(s) \frac{\tilde{r}_L}{\tilde{t}} \ee^{-2 s d}\right]&=\frac{\ee^{-s\left(x-x^{\prime}\right)}}{2 s A^{(\epsilon)}(s)}\left(-\frac{\epsilon}{2 s}\right)\left[A^{(0)}(s) \int_{-\infty}^{+\infty} V_b(z) \dd z+B^{(0)}(s) \int_{-\infty}^{+\infty} V_b(z) \ee^{-2 s z} \dd z\right] \\
				& =\frac{\ee^{-s\left(x-x^{\prime}\right)}}{2 s A^{(\epsilon)}(s)}\left(-\frac{\epsilon}{2 s}\right) \int_{-\infty}^{+\infty}\left(A^{(0)}(s) \ee^{s z}+B^{(0)} \ee^{-s z}\right) V_b(z) \ee^{-s z} \dd  z\\
				&=-\frac{\epsilon}{2 s} \int_{-\infty}^{+\infty} \dd z V_b(z)\ee^{-s(z-d)}   G_{(\epsilon)}^{(0)}(s, x-x^{\prime}+d, z) \, , 
			\end{aligned}
			\end{equation}
	\end{widetext}
	where we have used the integral expression of the scattering coefficients at first order and the fact that the bump $ V_{b}(z) $ is located in a small region around $ z\approx d $, so that we can make the approximation $\ee^{-s(z-d)}\simeq 1$. 
	The other term can be calculated similarly. By performing the inverse Laplace transform, we obtain a neat form for the time-domain difference Green's function in the case of internal initial data, namely:
	\begin{widetext}
	\begin{equation}\label{internal initial condition equivalent}
		\begin{aligned}
			\Delta G_{\rm int}\left(t, x, x^{\prime}\right)=-\int_0^t d \tau\int_{-\infty}^{+\infty} d z\left[\frac{\epsilon}{2} V_b(z)\right] \left[G_{(\epsilon)}^{(0)}(\tau, x-x^{\prime}+d, z)+G_{(\epsilon|0)}^{(0)}\left(\tau, x+x^{\prime}+d, z\right)\right] \, ,
		\end{aligned}
	\end{equation}
	\end{widetext}
	where 
	$$
	G^{(0)}_{(\epsilon|0)}(s,x,x^{\prime})=\dfrac{B^{(0)}(s)\ee^{-sx}}{2sA^{(\epsilon)}(s)A^{(0)}(s)}\hat{\phi}^{(0)}(s,x^{\prime}) \,.
	$$
	This is an equivalent scattering problem with QNMs given by both $A^{(\epsilon)}(s)=0$ and $A^{(0)}(s)=0$, which explains why we denote it as $ G^{(0)}_{(\epsilon|0)} $.

        The interpretation of this result is straightforward. The integrand of Eq.~\eqref{internal initial condition equivalent} is the sum of two Green's functions: one represents a Type $(\epsilon,0)$ scattering problem, and the other a Type $(\epsilon|0,0)$ problem. The integral with respect to $z$ is analogous to the first line of Eq.~(\ref{initial integral}), which means that we can regard it as the integral of a Green's function with initial condition $\partial_{t}\Psi(t=0,z)\rightarrow\frac{\epsilon}{2}V_{b}(z)$.

        For the Type $ (\epsilon,0) $ scattering problem, this is physically equivalent to the following situation: the ``observer'' is at $x-x^{\prime}+d$, the initial data at $z\approx d$, and the effective potential corresponds to a Green's function of Type $(\epsilon,0)$. 
        Then we can expect that the initial data will generate an ingoing wavepacket and an outgoing wavepacket. One of these meets the observer at $t\approx x-x^{\prime}$, carrying the precursor wave and its dispersion tail, while the other propagates inward, is reflected, and then gets to the observer location at $t\approx x-x^{\prime}+2d$, carrying information about the perturbed QNMs. 

        For the Type $ (\epsilon|0,0) $ case, similarly, we have the ``observer'' at $x+x^{\prime}+d$, the initial data at $z\approx d$, and the effective potential corresponds to a Green's function of Type $(\epsilon|0,0)$. The precursor will arrive at $t\approx x+x^{\prime}$, and the perturbed QNMs at $ t\approx x+x^{\prime}+2d $. 
	
	\begin{figure*}
		\centering
		\includegraphics[width=1.0\linewidth]{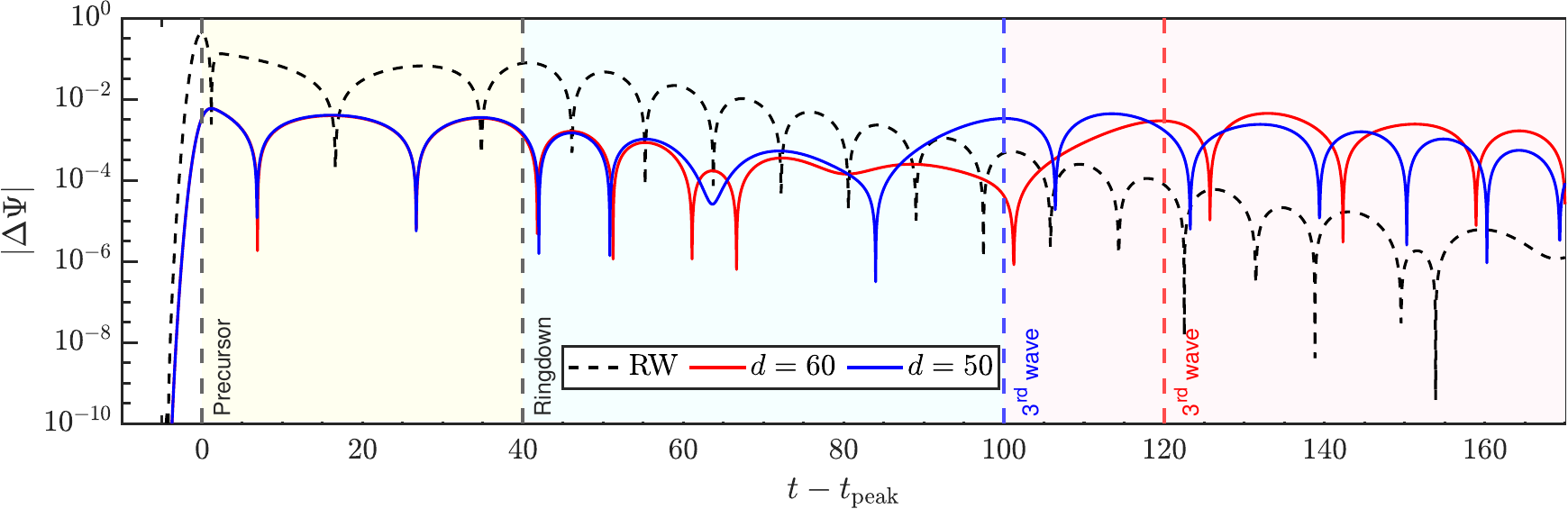}
		\includegraphics[width=1.0\linewidth]{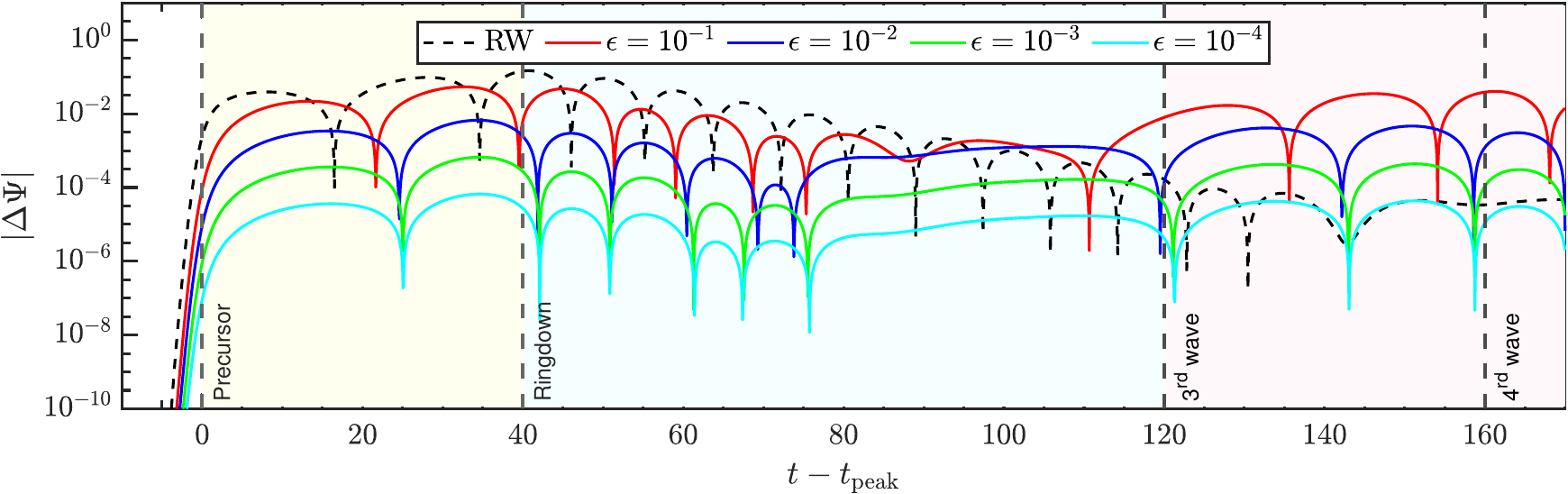}
		\includegraphics[width=1.0\linewidth]{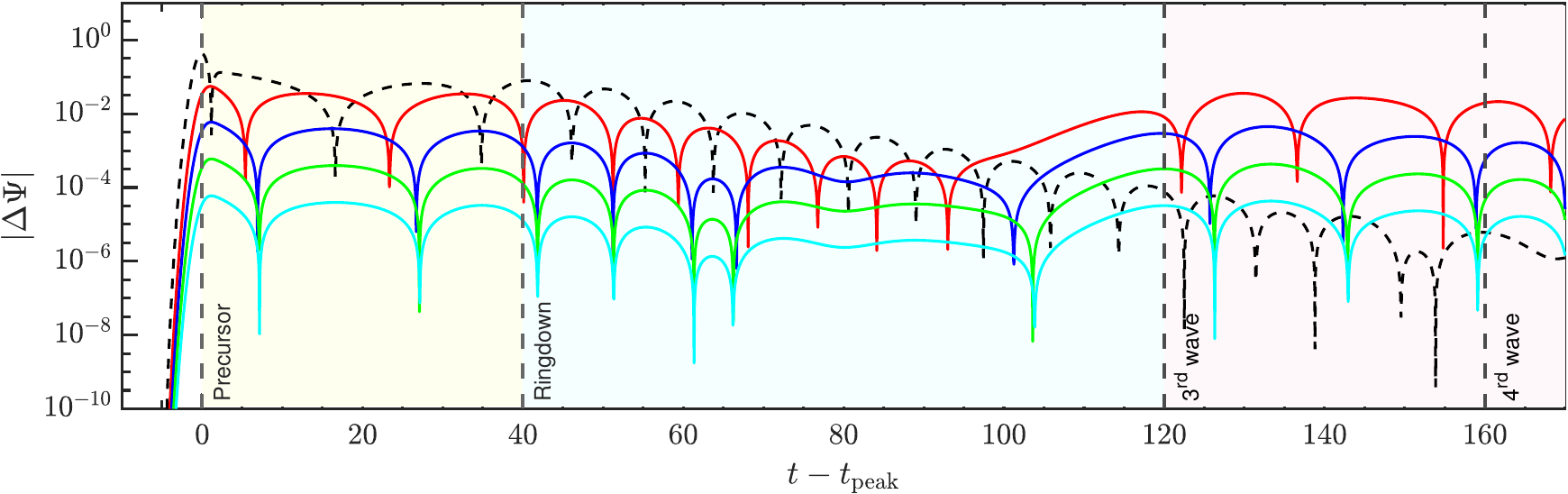}
		\caption{Numerical waveform obtained by integrating Eq.~(\ref{master equation}) in time with a fourth-order Runge-Kutta method. In all the panels, the unperturbed potential is the RW potential, and the perturbation is a Gaussian bump $V_{b}(x)=\epsilon\,\mathrm{exp}\left[-(x-d)^{2}\right]$. The observer is located at $x=150$, and we have chosen the origin of the time axis to be at the peak of the original signal, $t_{\rm peak}\approx x-x^{\prime}=130$. In the pink-shaded region, the waveform is dominated by the perturbed QNMs. Top panel: the difference between the perturbed and unperturbed waveforms when $\epsilon=10^{-2}$ with initial data $\Psi(t=0,x)=\mathrm{exp}\left[-(x-20)^{2}\right]$ and $\partial_{t}\Psi(t=0,x)=0$. The red and blue lines correspond to $d=60$ and $d=50$, respectively. The black dashed line shows the original waveform for comparison. Vertical lines mark the arrival time of different waves: the black vertical lines represent the arrival of the precursor and the beginning of the ringdown; red and blue vertical lines mark the arrival of the third wave, which happens at slightly different times for different values of $d$ (see Fig.~\ref{fig:four waves}). Middle panel: same as the top panel for a bump located at $d=60$. Here we consider purely ingoing  initial data, $\Psi(t=0,x)=\mathrm{exp}\left[-(x-20)^{2}\right]$ and $\partial_{t}\Psi(t=0,x)=\partial_{x}\Psi(t=0,x)$, with different colors corresponding to different values of $\epsilon$. Bottom panel: same as the middle panel, except that the initial data is now a Gaussian in $\Psi$: $\Psi(t=0,x)=\mathrm{exp}\left[-(x-20)^{2}\right]$, $\partial_{t}\Psi(t=0,x)=0$.}
		\label{fig:internal initial condition}
	\end{figure*}

        \begin{figure*}
          \centering
          \includegraphics[width=1.0\linewidth]{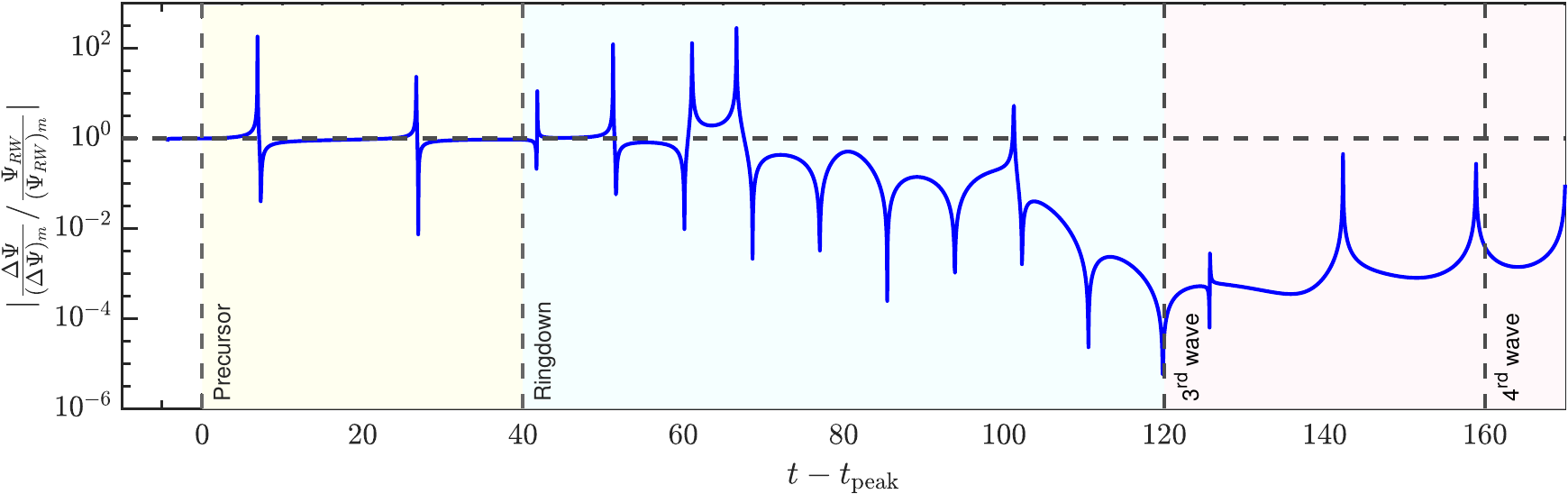}
          \caption{Difference between the $\epsilon=10^{-2}$ case shown in the middle panel of Fig.~\ref{fig:internal initial condition} and the unperturbed signal shown in the bottom panel of Fig.~\ref{fig:schwarzchild-pt}. Each of the two signals is normalized so that the maximum amplitude is set to unity. The horizontal black dashed line corresponds to the two waveforms being equal. Sharp cusps at early times are caused by numerical artifacts around zero crossings ($|\Psi|=0$).}
          \label{fig:difference between Sch and new}
        \end{figure*}

        \begin{figure*}
          \centering
          \includegraphics[width=1.0\linewidth]{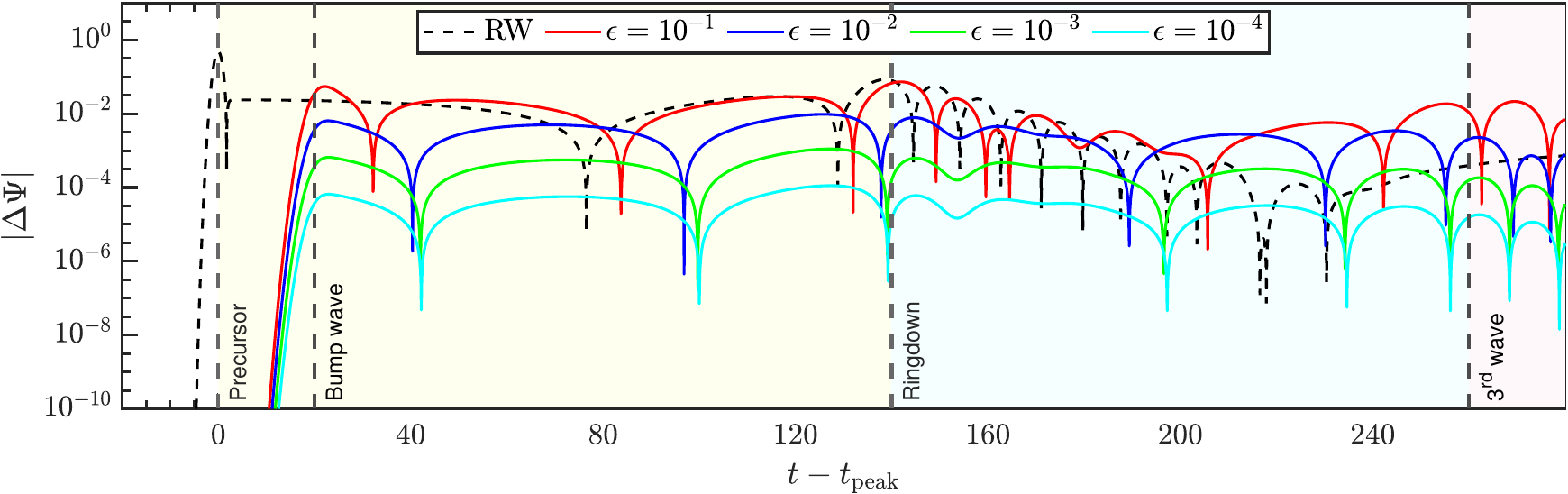}
          \caption{Numerical waveform for a RW potential with a Gaussian bump $ V_{b}=\mathrm{exp}\left[-(x-d)^{2}\right] $ located at $ d=60 $, and ``external'' initial data $\Psi=\mathrm{exp}\left[-(x-70)^2\right]$. The observer is at $x=150$, and we have set $t_{\rm peak}\approx 80$. The vertical lines and colored regions have the same meaning as in Fig.~\ref{fig:internal initial condition}.}
          \label{fig:external initial condition}
      \end{figure*}

      In short, we expect a superposition of two signals similar to the Schwarzschild BH signal in the bottom panel of Fig.~\ref{fig:schwarzchild-pt}. The first and second waves of Fig.~\ref{fig:four waves} are both dominated by the precursor and its dispersion tail, although they correspond to different Green's functions. 
        If the properties of these waves are not dramatically different from the original scattering problem, we may expect the difference between the perturbed signal and the original signal to be proportional to $\epsilon$ at the early ringdown stage. Therefore, a sufficiently small bump will not significantly affect the early ringdown signal.

	We verify these analytical arguments by numerical integration of Eq.~(\ref{master equation}). In Fig.~\ref{fig:internal initial condition} we show numerical calculations of the difference between the unperturbed waveform and the one corresponding to a Gaussian bump. We have checked that other bumps, such as PT, give similar results. Here we comment on some interesting features of these results.

	\noindent
	\textbf{\em Separation of different waves.} The top panel of Fig.~\ref{fig:internal initial condition} shows that the first wave comes at the same time as the original signal (black dashed line), carrying the precursor of the first term of Eq.~(\ref{internal initial condition equivalent}). The blue and red lines overlap, so this feature is rather insensitive to the value of $d$. 
	The second wave is first observed when the ringdown starts at $t-t_{\rm peak}\approx 2x^{\prime}=40$, but the first wave is still dominant. 
	The third wave, which rings down at the perturbed QNM frequencies, is observed at the expected time ($t-t_{\rm peak}\approx 100$ for $ d=50 $, and $ t-t_{\rm peak}\approx 120 $ for $ d\approx 60 $), while the fourth wave is hardly visible. 
	
	To better illustrate the presence of the second term in Eq.~(\ref{internal initial condition equivalent}), in the middle panel of Fig.~\ref{fig:internal initial condition} we repeat the analysis for purely ingoing initial data: $\partial_{t}\Psi(t=0,x)=\partial_{x}\Psi(t=0,x)$. 
	Purely ingoing initial data greatly suppresses the magnitude of the first and third waves (although we cannot fully eliminate them due to dispersion).
        Now the second wave can be observed around the ringdown starting time $ t-t_{\rm peak}\approx40 $, where we observe some amplification. The fourth wave is elusive, although there is a maximum in the waveform amplitude for small bumps ($\epsilon\lesssim 10^{-2}$) around $t-t_{\rm peak}\approx 150$.
	This is consistent with expectations: the first and third waves are represented by the first term of Eq.~(\ref{internal initial condition equivalent}), and they are generated by an equivalent ``initial wavepacket'' when the outgoing part of the initial data meets the bump.

        \noindent
	\textbf{\em Perturbative stability of the early ringdown phase.} As expected, the first and second waves carry the precursor of the two equivalent scattering problems discussed in Eq.~(\ref{internal initial condition equivalent}). 
	The middle and bottom panels of Fig.~\ref{fig:internal initial condition} show that as long as $\epsilon\lesssim10^{-2}$, these waves are proportional to $\epsilon$  (the $\epsilon=10^{-1}$ case is different).
	This is most evident in the bottom panel: for small $\epsilon$, the difference between the perturbed and unperturbed signals is proportional to $\epsilon$ until $t-t_{\rm peak}\approx90$, where either the third wave starts taking over or we observe effects that are not captured by the first-order expansion in $\epsilon$.
        The middle panel is dominated by the second and fourth waves, but once again, the difference between the perturbed and unperturbed signals is proportional to $\epsilon$ for $\epsilon\lesssim10^{-2}$ in the early ringdown phase (roughly, when $t\lesssim x-x^{\prime}+2d$). 

	Another remarkable fact is shown in Fig.~\ref{fig:difference between Sch and new}. If we consider one of the perturbed waveforms (say, for $\epsilon=10^{-2}$) and compare it with the unperturbed Schwarzschild signal (lower panel of Fig.~\ref{fig:schwarzchild-pt}) using the same normalization, we see that the difference Green's function is almost identical to the original Schwarzschild BH Green's function before $t\approx180$, when the ringdown phase starts. 
	
	The expansion used to derive Eq.~(\ref{internal initial condition equivalent}) breaks down for large perturbations ($\epsilon=10^{-1}$). More work is needed to understand the Green's function in this nonperturbative regime. 

        \noindent
	\textbf{\em New modes}. New QNMs dominate the signal only after $t\approx x- x^{\prime}+2d$ (see Ref.~\cite{2014PhRvD..89j4059B}). At this time the wave amplitude becomes comparable to the oscillation amplitude of the early ringdown modes. 
	The excitation of the new QNMs can be compared with the excitation of the unperturbed QNMs using Eq.~(\ref{four waves}). 
	The fourth and third waves are nearly indistinguishable, being close in time and comparable in amplitude.

	\subsection{Difference Green's function for ``external'' initial data}
	\label{sub:difference external}

	We have shown that the early ringdown waveform is stable by considering the time evolution of ``internal'' initial data (i.e., when the initial wavepacket is located between the unperturbed potential peak and the bump). Here we generalize the arguments of Sec.~\ref{sub:difference internal} to ``external'' initial data, located on the right side of the bump.
        
        For $x^{\prime}>d$ we  substitute $\phi^{(\epsilon)}_{-}\approx A^{(\epsilon)}(s) \ee^{-s x}+B^{(\epsilon)}(s) \ee^{s x}$ into Eq.~(\ref{difference G}). We find
	\begin{equation}\label{three waves}
		\begin{aligned}
			\Delta \hat{G}_{\rm ext}\left(s, x, x^{\prime}\right)&\approx \frac{1}{2 s} \tilde{r}_{R} \ee^{-s\left(x+x^{\prime}-2d\right)} \\
			&+\frac{1}{2 s}\frac{B^{(0)}(s)}{A^{(\epsilon)}(s)} \frac{\tilde{t}-1}{\bar{t}} \ee^{-s\left(x+x^{\prime}\right)} \\
			&+\frac{1}{2 s} \frac{B^{(0)}(s)}{A^{(\epsilon)}(s)} \tilde{R}_0 \frac{\tilde{r}_L}{\tilde{t}} \ee^{-s\left(x+x^{\prime}+2 d\right)} \, .
		\end{aligned}
	\end{equation}
	This can be interpreted as three waves arriving at different times. The second and third waves of Eq.~(\ref{three waves}) appear also in Eq.~(\ref{four waves}). The main difference is the first term, which represents a wave reflected by the bump, whose poles are the QNMs of the bump potential itself. Proceeding as in Sec.~\ref{sub:difference internal}, we can rewrite Eq.~(\ref{three waves}) in the form
	\begin{equation}\label{external initial condition equivalent}
		\begin{aligned}
			\Delta G_{\rm ext}&\left(t, x, x^{\prime}\right)=G_{\rm bump}\left(t, x-d, x^{\prime}-d\right)\\
			&-\int_0^t \mathrm{d} \tau\int_{-\infty}^{+\infty}\mathrm{d} z\left[\frac{\epsilon}{2} V_b(z)\right]  G_{(\epsilon,0)}^{(0)}(\tau, x-x^{\prime}+d, z) \, , 
		\end{aligned}
	\end{equation}
	where $G_{\rm bump}(t,x-d,x^{\prime}-d)=\mathscr{L}^{-1}\left[\dfrac{\tilde{r}_{R}}{2s}\mathrm{e}^{-s(x+x^{\prime})}\right]$ is the Green's function describing a purely ingoing wavepacket initially located at $x^{\prime}-d$, reflected by the bump located at $x=0$, and arriving at the observer's location at $t=x+x^{\prime}-2d$. 
	In general, this Green's function contains precursor, QNM ringing and power-law tail contributions. While in principle this first wave could display QNM oscillations related to the bump potential, the precursor (which  usually has a larger amplitude than the QNM ringing) will also contribute.

        This is what we observe in the numerical integrations shown in Fig.~\ref{fig:external initial condition}.	
	Since the dominant contribution from Eq.~(\ref{external initial condition equivalent}) is the same as the second term of Eq.~(\ref{internal initial condition equivalent}), our analytical and numerical results lead to the same conclusion: the deviation between the perturbed and unperturbed signals in the early ringdown stage is proportional to $\epsilon$, as long as $\epsilon$ is sufficiently small. 
        New QNMs only appear after $ t\approx x+x^{\prime}+2d $, with amplitudes comparable to the unperturbed QNMs.
	
	\section{Conclusions}
	\label{sec:summary}

	In this paper we used Green's function arguments to prove that the spectral instability of BH QNMs in the presence of a perturbing ``bump'' is generic, and that it has only a small influence on the early ringdown waveform. 
	
	Working in the frequency domain, we exploited the analogy with the scattering problem for a one-dimensional Schr\"{o}dinger equation to find approximate wavefunctions in the presence of a bump located around $x=d$, which we assumed to be short-ranged and sufficiently far away from the peak of the unperturbed potential. We proved that a universal ``Type~II'' instability (i.e., an exponential migration of the modes away from their unperturbed value as $d$ grows) must be present for both nonrotating (Schwarzschild) and rotating (Kerr) BHs. This instability can be quantified using Eq.~(\ref{chi1}).
	
	We then studied the causal structure of the Green's function in the time domain. We found that the wavepacket carrying the ``new'' QNMs is separated from the wavepacket carrying the original ringdown signal, and that the ``new'' QNMs will only appear at late times. Using both analytical and numerical arguments, we confirmed that the deviation between the perturbed and unperturbed waveforms in the early ringdown stage (which is most interesting observationally) is proportional to the magnitude of the bump $\epsilon$ as long as $\epsilon\lesssim10^{-2}$.
       
        Our analysis leads to more physical insight into the apparent tension between the frequency-domain and time-domain descriptions.
	In the frequency domain, we found the solution corresponding to the perturbed potential by adding all the waves transmitted and reflected back and forth in the intermediate cavity. The Type~II instability is caused by the phase factor due to the extra propagation of the reflected waves.
        A frequency-domain analysis usually deals with stationary solutions. However the QNM problem is not self-adjoint, energy is not conserved, and this ultimately leads to a spectral instability.
        In the time domain, this apparent problem is somehow cured: the inverse Laplace transform shows that the phase factors appear in the causal structure of the Green's function, revealing that the different waves are separated in time, and that the wavepacket carrying the new QNMs generated by this instability will only be observed at late times.
	
        We have required the bump potential to be short-ranged and localized sufficiently far away from the unperturbed potential.
	Because of this assumption, our method is not applicable to small changes near the peak of the effective potential. However, as we discussed in Sec.~\ref{sub:more on approxiamtion}, this restriction can be relaxed, and it is reasonable to expect that the ``susceptibility'' $\chi_n(d)$ defined in Eq.~\eqref{kai} will be a continuous function as $d$ decreases. 
	Then small perturbations around the light ring should not trigger serious instabilities at least for the first few overtones, although their influence on higher overtones should be investigated more carefully (see e.g. Refs.~\cite{2022PhRvD.106h4011B,2022arXiv220900679K,2024PhRvD.110b4016C}).
	
	From the observational point of view, our conclusions are only applicable within linear perturbation theory. In our analysis, the robustness of the early ringdown phase is guaranteed by the scattering nature of the perturbation problem. 
	The full nonlinear description of the ``precursor'' signal preceding the ringdown in a binary BH merger remains an important open problem (see e.g. Ref.~\cite{2023PhRvD.108j4020B}). 

        \noindent
        \textbf{\em Note Added.} After the completion of this work, we have become aware of a similar study by Ianniccari {\em et al.}, and we have coordinated the arXiv submission.
        Our results, when overlap is possible, agree with theirs. Our findings are also consistent with a recent preprint by Motohashi~\cite{2024arXiv240715191M}.
        
        \acknowledgments

        This work was supported by the National SKA Program of China (2020SKA0120300), the National Natural Science Foundation of China (11991053, 12375051), the Beijing Natural Science Foundation (1242018), the Max Planck Partner Group Program funded by the Max Planck Society, and the High-performance Computing Platform of Peking University. Y.Y. is supported by the Principal's Fund for the Undergraduate Student Research Study at Peking University. E.B. is supported by NSF Grants No. AST-2307146, PHY-2207502, PHY-090003 and PHY-20043, by NASA Grants No. 20-LPS20-0011 and 21-ATP21-0010, by the John Templeton Foundation Grant 62840, by the Simons Foundation, and by the Italian Ministry of Foreign Affairs and International Cooperation Grant No.~PGR01167. This work was carried out at the Advanced Research Computing at Hopkins (ARCH) core facility (\url{rockfish.jhu.edu}), which is supported by the NSF Grant No.~OAC-1920103. The authors also acknowledge the Texas Advanced Computing Center (TACC) at The University of Texas at Austin for providing {HPC, visualization, database, or grid} resources that have contributed to the research results reported within this paper \cite{10.1145/3311790.3396656}.
	
	\appendix
	\section{Mathematical treatment of the scattering problem}
	\label{appendix:scattering problem}
	In this appendix we present a more rigorous derivation of the results presented in Sec.~\ref{sec:frequency domain analysis}.
	For a one-dimensional scattering problem, the wave equation in the frequency domain reads
	\begin{equation}\label{scattering}
		\begin{aligned}
			& {\left[\partial_x^2+\omega^2-V(x)\right] \widetilde{\Psi}(x)=0} \, . 
		\end{aligned}
	\end{equation}
	We look for a set of Jost solution with the following asymptotic behavior:
	\begin{equation}\label{jost-}
		\phi_\omega(x) \rightarrow \left\{
		\begin{aligned}
			&\ee^{-i\omega x}, &&x\rightarrow-\infty \,, \\
			& \frac{1}{\tilde{T}_1(\omega)} \ee^{-i \omega x}+\frac{\tilde{R}_1(\omega)}{\tilde{T}_{1}(\omega)} \ee^{i \omega x}, &&x \rightarrow +\infty  \,,
		\end{aligned}
	\right.
	\end{equation}

	\begin{equation}\label{jost+}
	\phi_\omega(x) \rightarrow \left\{
	\begin{aligned}
		& \frac{1}{\tilde{T}_2(\omega)} \ee^{i \omega x}+\frac{\tilde{R}_2(\omega)}{\tilde{T}_{2}(\omega)} \ee^{-i \omega x}, &&x \rightarrow -\infty  \,, \\
		&\ee^{i\omega x}, &&x\rightarrow+\infty \,.
	\end{aligned}
	\right.
\end{equation}
	The Jost solutions $ \phi_{-\omega} $, $ \psi_{-\omega} $ can be simply obtained by changing the sign of $\omega$. We can arbitrarily choose two linearly independent solutions of Eq.~(\ref{scattering}) among the four Jost solutions.  We can compute the (constant) Wronskian of any two solutions, for example:
	\begin{equation}
		\mathcal{W}\left[\phi_{\omega},\psi_{\omega}\right]=\phi_{\omega}\dfrac{\dd\psi_{\omega}}{\dd x}-\psi_{\omega}\dfrac{\dd\phi_{\omega}}{\dd x} \, . 
	\end{equation}
	By evaluating it explicitly in the limits $x\rightarrow\pm \infty$ we find
	\begin{equation}
		\tilde{T_{1}}(\omega)=\tilde{T_{2}}(\omega)=\tilde{T}(\omega) \, . 
	\end{equation}
	Similarly, by calculating $ \mathcal{W}\left[\phi_{-\omega},\psi_{\omega}\right] $ and $ \mathcal{W}\left[\phi_{\omega},\psi_{-\omega}\right] $ we get, respectively,
	\begin{equation}\label{left and right}
		\begin{aligned}
			\dfrac{\tilde{R}_{1}(-\omega)}{\tilde{T}(-\omega)}=-\dfrac{\tilde{R}_{2}(\omega)}{\tilde{T}(\omega)}\, , \quad \ 		\dfrac{\tilde{R}_{2}(-\omega)}{\tilde{T}(-\omega)}=-\dfrac{\tilde{R}_{1}(\omega)}{\tilde{T}(\omega)} \, . 
		\end{aligned}
	\end{equation}
	Now we try to solve for these coefficients. Equation~(\ref{scattering}) can be transformed into an integral equation. For $ \phi_{\omega} $ and $ \psi_{\omega} $, we have the relations
	\begin{equation}\label{integral equation}
          \begin{aligned}
            & \phi_{\omega}(x)=\ee^{-i \omega x}+\int_{-\infty}^x \frac{\ee^{i \omega(x-y)}-\ee^{i \omega(y-x)}}{2 i \omega} V(y) \phi_\omega(y) \dd y  \, , \\
            & \psi_\omega(x)=\ee^{i \omega x}-\int_x^{+\infty} \frac{\ee^{i \omega(x-y)}-\ee^{i \omega(y-x)}}{2 i \omega}V(y) \psi_\omega(y) \dd y \, , 
          \end{aligned} 
        \end{equation}
  which can be easily verified. By comparing Eq.~(\ref{integral equation}) with Eqs.~(\ref{jost-}) and~(\ref{jost+}), we find:
	\begin{equation}\label{scattering coefficient}
		\begin{aligned}
			1-\frac{1}{\tilde{T}(\omega)}=&\int_{-\infty}^{+\infty} \frac{\ee^{-i \omega y}}{2 i \omega} V(y) \psi_\omega(y) \dd  y=\int_{-\infty}^{+\infty} \frac{\ee^{i \omega y}}{2 i \omega} V(y) \phi_\omega(y) \dd y  \, , \\
			\frac{\tilde{R}_1(\omega)}{\tilde{T}(\omega)}=&\int_{-\infty}^{+\infty} \frac{\ee^{-i \omega y}}{2 i \omega} V(y) \phi_\omega(y) \dd y \, , \\
			\frac{\tilde{R}_2(\omega)}{\tilde{T}(\omega)}=&\int_{-\infty}^{+\infty} \frac{\ee^{i \omega y}}{2 i \omega} V(y) \psi_\omega(y) \dd y \, .
		\end{aligned}
	\end{equation}
Recalling the definitions of $ \phi^{(0)}_{\omega}(x) $ and $ \phi^{(\epsilon)}_{\omega}(x) $ from Sec.~\ref{sub:definition}, we have, respectively,
\begin{equation}\label{eps integral equation}
	\begin{aligned}
		 \phi^{(0)}_{\omega}(x)=& \ee^{-i \omega x}+\int_{-\infty}^x \frac{\ee^{i \omega(x-y)}-e^{i \omega(y-x)}}{2 i \omega} V_{0}(y) \phi^{(0)}_\omega(y) \dd y  \, , \\
		 \phi^{(\epsilon)}_{\omega}(x)=&\ee^{-i \omega x}+\int_{-\infty}^x \frac{\ee^{i \omega(x-y)}-e^{i \omega(y-x)}}{2 i \omega} V^{(\epsilon)}(x) \phi^{(\epsilon)}_\omega(y) \dd y \, , \\
	\end{aligned}
\end{equation}
where $ V^{(\epsilon)}=V_{0}+\epsilon V_{b} $. Defining $\phi^{(\epsilon)}_{\omega}(x)-\phi^{(0)}_{\omega}(x)=\delta\phi_{\omega}(x) $, we find
	\begin{equation}
		\begin{aligned}
			\delta\phi_{\omega}(x)=&\int_{-\infty}^x \frac{\ee^{i \omega(x-y)}-\ee^{i \omega(y-x)}}{2 i \omega} V_{0}(y) \delta\phi_{\omega}(y) \dd y \, , \\
			&+\int_{-\infty}^x \frac{\ee^{i \omega(x-y)}-\ee^{i \omega(y-x)}}{2 i \omega} \epsilon V_{b}(y) \phi^{(\epsilon)}_{\omega}(y) \dd y \, .
		\end{aligned}
	\end{equation}
	It can be shown that $\delta \phi_{\omega}(y)$ is proportional to $\int_{-\infty}^{y}\epsilon V_{b}(z)\dd z$, so that the first integral is related to $ V_{0}(y)\int_{-\infty}^{y}\epsilon V_{b}(z)\dd z $, a spatial ``overlap'' of the two potentials. If we assume that the two potentials are localized and that their maxima are sufficiently far away from each other, we can drop the first integral. By taking the limit $ x\rightarrow+\infty $, using Eqs.~(\ref{jost-}) and~(\ref{jost+}) we find
	\begin{equation}
		\begin{aligned}
			\left[a^{\epsilon}(\omega)-a^{(0)}(\omega)\right]\ee^{-i\omega x}+\left[b^{(\epsilon)}(\omega)-b^{(0)}(\omega)\right]\ee^{i\omega x}\\
			=\int_{-\infty}^{+\infty} \frac{e^{i \omega(x-y)}-e^{i \omega(y-x)}}{2 i \omega} \epsilon V_{b}(y) \phi^{(\epsilon)}_{\omega}(y) \dd y \, .
		\end{aligned}
	\end{equation}
	Taking the approximation further, $ \phi^{(\epsilon)}_{\omega}(y) $ can be treated as a superposition of the Jost solutions of the bump potential: on the left side of $ V_{b} $, it is the sum of an incoming wave with an amplitude proportional to $ b^{(0)}(\omega) $ plus an outgoing wave with an amplitude proportional to $ a^{(0)}(\omega) $ (up to a phase factor, which can be determined). Let us denote the reflection rate $ \tilde{R}_{2}(\omega) $ for the bump potential on the left hand side by $ \tilde{r}_{L} $, and the transmission rate for the bump by $ \tilde{t} $. Using Eq.~(\ref{scattering coefficient}) and the definition $ W(x)=V_{b}(x+d) $, we can transform the integral above into a combination of scattering coefficients. For example, for the term proportional to $ \ee^{-i\omega x} $ we find 
	\begin{equation}
		\begin{aligned}
			&\int_{-\infty}^{+\infty} -\frac{e^{i \omega y}}{2 i \omega} \epsilon V_{b}(y) \phi^{(\epsilon)}_{\omega}(y) \dd y\\
			&=\int_{-\infty}^{+\infty} -\frac{e^{i \omega (z+d)}}{2 i \omega} \epsilon W(z) \left[a^{(0)}(\omega)\ee^{-i\omega d}\phi_{\omega}(z)+b^{(0)}(\omega)\ee^{i\omega d}\phi_{-\omega}(z)\right] \dd z\\
			&=a^{(0)}(\omega)(\dfrac{1}{\tilde{t}}-1)-b^{(0)}(\omega)\dfrac{\tilde{r}_{L}}{\tilde{t}}\\
			&=a^{(\epsilon)}(\omega)-a^{(0)}(\omega) \, , 
		\end{aligned}
	\end{equation}
	that is, Eq.~(\ref{aeps}). We find Eq.~(\ref{beps}) by the same procedure.
	
	Finally, for a small bump, we compute the scattering coefficients at first order. Consider for example $ \tilde{r}_{L}/\tilde{t} $. Using the third line of Eq.~(\ref{scattering coefficient}), for small perturbing potentials we can approximate $ \psi_{\omega}(y) $ as $ \mathrm{exp}(i\omega y) $ to leading order, so that
	\begin{equation}
		\dfrac{\tilde{r}_{L}}{\tilde{t}}\approx\dfrac{1}{2i\omega}\int_{-\infty}^{+\infty}\ee^{2i\omega y}W(y)\dd y \, . 
	\end{equation}
	The other coefficients can be calculated at leading order in a similar way.

	\section{Sasaki-Nakamura formalism}
	\label{appendix:SN}
	
	For completeness, here we briefly review the SN formalism, following  \citet{2013PhRvD..88d4018Z}. The Teukolsky equation for gravitational perturbations is
	\begin{equation}\label{Teu}
		\Delta^{2}\dfrac{\mathrm{d}}{\mathrm{d}r}\dfrac{1}{\Delta}\dfrac{\mathrm{d}}{\mathrm{d}r}R+V(r)R=0 \, , 
	\end{equation}
	where the potential $ V(r) $ for gravitational perturbations is given in Ref.~\cite{1973ApJ...185..635T}. Now we can perform the transformation (see Appendix~B of Ref.~\cite{1982PThPh..67.1788S}):
        \begin{equation}
          \begin{aligned}
            X &=\frac{\sqrt{r^2+a^2}}{\Delta}\left(\alpha(r) R+\frac{\beta(r)}{\Delta} R^{\prime}\right) \,,\\
            \alpha &=-\frac{i K}{\Delta^2} \beta+3 i K^{\prime}+\lambda+\frac{6 \Delta}{r^2} \,, \\
            \beta &=\Delta\left[-2 i K+\Delta^{\prime}-4 \frac{\Delta}{r}\right] \, , 
          \end{aligned}
	\end{equation}
	where primes stand for derivatives with respect to $r$,
	\begin{equation}
		\begin{aligned}
                  K &=\left(r^2+a^2\right) \omega-a m\,, \\
                  \Delta &=r^2-2 M r+a^2\,, \\
                  \lambda &=A_{lm}+(a\omega)^{2}-2am\omega\,, 
                \end{aligned}
	\end{equation}
        and $ A_{lm} $ is the separation constant appearing in the angular equation. Then the SN wave function $X$ satisfies
	\begin{equation}
		\frac{d^2 X}{d r_*^2}-\mathcal{F} \frac{d X}{d r_*}-\mathcal{U} X=\mathcal{S} \, , 
	\end{equation}
	where the tortoise coordinate is defined as
	\begin{equation}
		r_*=r+\frac{2 M r_{+}}{r_{+}-r_{-}} \log \left(r-r_{+}\right)-\frac{2 M r_{-}}{r_{+}-r_{-}} \log \left(r-r_{-}\right)  \, .
	\end{equation}
	The functions $\mathcal{F}$ and $\mathcal{U}$ are given by
	\begin{equation}\label{trans}
		\begin{aligned}
			\mathcal{F} & =\frac{\Delta}{r^2+a^2} F, \\ 
			F &\equiv \frac{\gamma^{\prime}}{\gamma} \,, \\
			\gamma & \equiv \alpha\left(\alpha+\frac{\beta^{\prime}}{\Delta}\right)-\frac{\beta}{\Delta}\left(\alpha^{\prime}-\frac{\beta}{\Delta^2} V\right), \\
			\mathcal{U} & =\frac{\Delta U}{\left(r^2+a^2\right)^2}+G^2+\frac{d G}{d r_*}-\frac{\Delta G F}{r^2+a^2}, \\
			G & \equiv-\frac{\Delta^{\prime}}{r^2+a^2}+\frac{r \Delta}{\left(r^2+a^2\right)^2}, \\
			U & =-V+\frac{\Delta^2}{\beta}\left[\left(2 \alpha+\frac{\beta^{\prime}}{\Delta}\right)^{\prime}-\frac{\gamma^{\prime}}{\gamma}\left(\alpha+\frac{\beta^{\prime}}{\Delta}\right)\right] .
		\end{aligned}
	\end{equation}
	We can now eliminate the first derivative by defining
	\begin{equation}
		X=\exp \left[\int \frac{\mathcal{F}}{2} d r_*\right] \psi=\psi \sqrt{\gamma} \,,
	\end{equation}
	with the result
	\begin{equation}\label{SN appendix}
		\frac{d^2 \psi}{d r_*^2}+\left(\frac{\mathcal{F}^{\prime}}{2}-\frac{\mathcal{F}^2}{4}-\mathcal{U}\right) \psi = 0 \, . 
	\end{equation}
	The potential $V_{\rm{SN}}=\left(\mathcal{F}^{\prime}/2-\mathcal{F}^2/4-\mathcal{U}\right)$ is now short-ranged and admits wavelike solutions.

        For a real frequency $\omega$, Eq.~(\ref{SN appendix}) admits the solution
	\begin{equation}
		\left\{
		\begin{aligned}
			\psi\rightarrow&\mathrm{e}^{-i\omega r_{*}},\ &r_{*}\rightarrow-\infty \,, \\
			\psi\rightarrow&A(\omega)\mathrm{e}^{-i\omega r_{*}}+B(\omega)\mathrm{e}^{i\omega r_{*}},\ &r_{*}\rightarrow+\infty \, . 
		\end{aligned}
		\right.
	\end{equation}
	If we add a bump potential $W(r_{*})$ to $ V_{\rm SN}$ on the left-hand side of Eq.~(\ref{SN appendix}), the arguments in the main text will apply.

        Assuming $A^{( \epsilon )}(\omega)=A(\omega)+\epsilon P(\omega)$, we get
	\begin{equation}
		P(\omega)=-\dfrac{1}{2i\omega}\left[A(\omega)\widetilde{\mathrm{W}}_{0}+B(\omega)\widetilde{\mathrm{W}}(2\omega)\ee^{2i\omega d} \right]  \,.
	\end{equation}
	We can compute the Kerr susceptibility, Eq.~\eqref{kerr susceptibility}, by applying the same reasoning as in Eq.~\eqref{susceptibility}, with the result
	\begin{equation}
          \chi_{n}^{\rm Kerr}=
          \lim _{\epsilon \rightarrow 0}\frac{\delta \omega_{n}}{\epsilon}=-\dfrac{P(\omega)}{A^{\prime}(\omega)} \, . 
	\end{equation}

	\section{Another derivation of the first-order approximation}
	\label{appendix:first order approximation}
        Consider the solution of Eq.~(\ref{f-domain}) for the perturbed potential $V \rightarrow \tilde{V}=V_{0}+\epsilon V_{b}$. The perturbed wave function will change from $\phi_{\omega}$ to $\tilde{\phi}_{\omega}$. At leading order in $\epsilon$, we can write
	\begin{equation}\label{new wave function}
			\tilde{\phi}_{\omega}=\phi_{\omega}+\epsilon\Delta\phi_{\omega} +\mathcal{O}\left( \epsilon^{2}\right)  \, .
	\end{equation}
	The leading-order modification of the wave function must satisfy the equation
	\begin{equation}\label{first order equation}
			-\frac{\mathrm{d}^{2}}{\mathrm{~d} x^{2}} \Delta\phi_{\omega}+\left[-\omega^{2}+V_{0}(x)\right] \Delta\phi_{\omega}=-V_{b}(x) \phi_{\omega}(x)  \, ,
	\end{equation}
	where we have assumed a short-ranged bump potential and neglected the term $ \epsilon V_{b}(x)\Delta\phi_{\omega} $, as it will not change the asymptotic behavior of the Jost solutions. 
	The function $\tilde{\phi}_{\omega}$ should still satisfy the boundary condition $\tilde{\phi}_{\omega}=\mathrm{e}^{-i \omega x}$ as $x \rightarrow-\infty$. This implies the following boundary condition for $\Delta\phi_{\omega}$:
	\begin{equation}\label{asym 0}
          \Delta\phi_{\omega}(-\infty)=0 \, .
        \end{equation}
	If both the unperturbed potential and the bump are short-ranged, we have
	\begin{equation}\label{asymp deltaphi}
			\Delta\phi_{\omega}(x)=q(\omega) \mathrm{e}^{i \omega x}\left[1+O(1/x)\right]+p(\omega) \mathrm{e}^{-i \omega x}\left[1+O(1/x)\right] 
	\end{equation}
	for some appropriate coefficients $p(\omega)$ and $q(\omega)$. Using the Jost solution $\phi_{\omega}$ defined in Eq.~\eqref{jost} of Sec.~\ref{sub:definition}, the general solution of Eq.~(\ref{first order equation}) can be found using Green's function techniques: 
	\begin{equation}\label{constant variation}
	\Delta\phi_{\omega}=\dfrac{F_{\omega}^{-}(x) \psi_{\omega}(x)-F_{\omega}^{+}(x) \psi_{-\omega}(x)}{\mathcal{W}\left[\psi_{\omega}, \psi_{-\omega}\right]} \, , 
\end{equation}
	where
	\begin{equation}\label{F}
			F_{\omega}^{ \pm}(x)\equiv -\int_{-\infty}^{x} \psi_{ \pm \omega}(s) V_{b}(s) \phi_{\omega}(s) \mathrm{d} s \, ,
	\end{equation}
	and $\mathcal{W}\left[\psi_{\tilde{\omega}}, \psi_{-\omega}\right]\equiv \psi_{\omega} \psi_{-\omega}^{\prime}-\psi_{-\omega} \psi_{\omega}^{\prime}=-2 i \omega$ is the Wronskian. Taking the asymptotic behaviors of Eq.~(\ref{asymp deltaphi}) into account, we then find that $2 i \omega p(\omega)=F_{\omega}^{+}(\infty)$, i.e., 
	\begin{equation}\label{p}
		p(\omega)=-\frac{1}{2 i \omega} \int_{-\infty}^{\infty} \psi_{\omega}(x) V_{b}(x) \phi_{\omega}(x) \mathrm{d} x \, .
	\end{equation}
	We now expand the right-hand side of Eq.~(\ref{new wave function}) for large $x$: 
	$$
	\begin{aligned}
		\tilde{\psi}_{\omega} & =[b(\omega)+\epsilon q(\omega)] \mathrm{e}^{i \omega x}(1+\cdots) \\
		& +[a(\omega)+\epsilon p(\omega)] \mathrm{e}^{-i \omega x}(1+\cdots)+\mathcal{O}\left( \epsilon^{2}\right) .
	\end{aligned}
	$$
	Imposing the QNM condition of Eq.~(\ref{QNM condition}) on $ a(\omega)+\epsilon p(\omega) $ at $ \omega=\omega^{(\epsilon)}_{n} $ then gives, at leading order in $\epsilon$,
	\begin{equation}\label{first order chi}
		\chi_{n}=\dfrac{1}{2 i \omega a^{\prime}(\omega)} \int_{-\infty}^{\infty} \psi_{\omega}(x) V_{b}(x) \phi_{\omega}(x) \mathrm{d} x \,.
	\end{equation}
	If we further impose that $ V_{b} $ is localized in the asymptotic region of $ V_{0} $, where the Jost solutions can be replaced by their asymptotic expressions, we recover Eq.~(\ref{chi1}).
        However, this assumption is not necessary.

	\bibliography{refs}

\begin{thebibliography}{75}%
\makeatletter
\providecommand \@ifxundefined [1]{%
 \@ifx{#1\undefined}
}%
\providecommand \@ifnum [1]{%
 \ifnum #1\expandafter \@firstoftwo
 \else \expandafter \@secondoftwo
 \fi
}%
\providecommand \@ifx [1]{%
 \ifx #1\expandafter \@firstoftwo
 \else \expandafter \@secondoftwo
 \fi
}%
\providecommand \natexlab [1]{#1}%
\providecommand \enquote  [1]{``#1''}%
\providecommand \bibnamefont  [1]{#1}%
\providecommand \bibfnamefont [1]{#1}%
\providecommand \citenamefont [1]{#1}%
\providecommand \href@noop [0]{\@secondoftwo}%
\providecommand \href [0]{\begingroup \@sanitize@url \@href}%
\providecommand \@href[1]{\@@startlink{#1}\@@href}%
\providecommand \@@href[1]{\endgroup#1\@@endlink}%
\providecommand \@sanitize@url [0]{\catcode `\\12\catcode `\$12\catcode
  `\&12\catcode `\#12\catcode `\^12\catcode `\_12\catcode `\%12\relax}%
\providecommand \@@startlink[1]{}%
\providecommand \@@endlink[0]{}%
\providecommand \url  [0]{\begingroup\@sanitize@url \@url }%
\providecommand \@url [1]{\endgroup\@href {#1}{\urlprefix }}%
\providecommand \urlprefix  [0]{URL }%
\providecommand \Eprint [0]{\href }%
\providecommand \doibase [0]{http://dx.doi.org/}%
\providecommand \selectlanguage [0]{\@gobble}%
\providecommand \bibinfo  [0]{\@secondoftwo}%
\providecommand \bibfield  [0]{\@secondoftwo}%
\providecommand \translation [1]{[#1]}%
\providecommand \BibitemOpen [0]{}%
\providecommand \bibitemStop [0]{}%
\providecommand \bibitemNoStop [0]{.\EOS\space}%
\providecommand \EOS [0]{\spacefactor3000\relax}%
\providecommand \BibitemShut  [1]{\csname bibitem#1\endcsname}%
\let\auto@bib@innerbib\@empty
\bibitem [{\citenamefont {Chandrasekhar}(1985)}]{Chandrasekhar:1985kt}%
  \BibitemOpen
  \bibfield  {author} {\bibinfo {author} {\bibfnamefont {S.}~\bibnamefont
  {Chandrasekhar}},\ }\href@noop {} {\emph {\bibinfo {title} {{The mathematical
  theory of black holes}}}}\ (\bibinfo {year} {1985})\BibitemShut {NoStop}%
\bibitem [{\citenamefont {{Kokkotas}}\ and\ \citenamefont
  {{Schmidt}}(1999)}]{1999LRR.....2....2K}%
  \BibitemOpen
  \bibfield  {author} {\bibinfo {author} {\bibfnamefont {K.~D.}\ \bibnamefont
  {{Kokkotas}}}\ and\ \bibinfo {author} {\bibfnamefont {B.~G.}\ \bibnamefont
  {{Schmidt}}},\ }\href {\doibase 10.12942/lrr-1999-2} {\bibfield  {journal}
  {\bibinfo  {journal} {Living Reviews in Relativity}\ }\textbf {\bibinfo
  {volume} {2}},\ \bibinfo {eid} {2} (\bibinfo {year} {1999})},\ \Eprint
  {http://arxiv.org/abs/gr-qc/9909058} {arXiv:gr-qc/9909058 [gr-qc]}
  \BibitemShut {NoStop}%
\bibitem [{\citenamefont {{Nollert}}(1999)}]{1999CQGra..16R.159N}%
  \BibitemOpen
  \bibfield  {author} {\bibinfo {author} {\bibfnamefont {H.-P.}\ \bibnamefont
  {{Nollert}}},\ }\href {\doibase 10.1088/0264-9381/16/12/201} {\bibfield
  {journal} {\bibinfo  {journal} {Classical and Quantum Gravity}\ }\textbf
  {\bibinfo {volume} {16}},\ \bibinfo {pages} {R159} (\bibinfo {year}
  {1999})}\BibitemShut {NoStop}%
\bibitem [{\citenamefont {{Berti}}\ \emph {et~al.}(2009)\citenamefont
  {{Berti}}, \citenamefont {{Cardoso}},\ and\ \citenamefont
  {{Starinets}}}]{2009CQGra..26p3001B}%
  \BibitemOpen
  \bibfield  {author} {\bibinfo {author} {\bibfnamefont {E.}~\bibnamefont
  {{Berti}}}, \bibinfo {author} {\bibfnamefont {V.}~\bibnamefont {{Cardoso}}},
  \ and\ \bibinfo {author} {\bibfnamefont {A.~O.}\ \bibnamefont
  {{Starinets}}},\ }\href {\doibase 10.1088/0264-9381/26/16/163001} {\bibfield
  {journal} {\bibinfo  {journal} {Classical and Quantum Gravity}\ }\textbf
  {\bibinfo {volume} {26}},\ \bibinfo {eid} {163001} (\bibinfo {year}
  {2009})},\ \Eprint {http://arxiv.org/abs/0905.2975} {arXiv:0905.2975 [gr-qc]}
  \BibitemShut {NoStop}%
\bibitem [{\citenamefont {{Davis}}\ \emph {et~al.}(1971)\citenamefont
  {{Davis}}, \citenamefont {{Ruffini}}, \citenamefont {{Press}},\ and\
  \citenamefont {{Price}}}]{1971PhRvL..27.1466D}%
  \BibitemOpen
  \bibfield  {author} {\bibinfo {author} {\bibfnamefont {M.}~\bibnamefont
  {{Davis}}}, \bibinfo {author} {\bibfnamefont {R.}~\bibnamefont {{Ruffini}}},
  \bibinfo {author} {\bibfnamefont {W.~H.}\ \bibnamefont {{Press}}}, \ and\
  \bibinfo {author} {\bibfnamefont {R.~H.}\ \bibnamefont {{Price}}},\ }\href
  {\doibase 10.1103/PhysRevLett.27.1466} {\bibfield  {journal} {\bibinfo
  {journal} {\prl}\ }\textbf {\bibinfo {volume} {27}},\ \bibinfo {pages} {1466}
  (\bibinfo {year} {1971})}\BibitemShut {NoStop}%
\bibitem [{\citenamefont {{Goebel}}(1972)}]{1972ApJ...172L..95G}%
  \BibitemOpen
  \bibfield  {author} {\bibinfo {author} {\bibfnamefont {C.~J.}\ \bibnamefont
  {{Goebel}}},\ }\href {\doibase 10.1086/180898} {\bibfield  {journal}
  {\bibinfo  {journal} {\apjl}\ }\textbf {\bibinfo {volume} {172}},\ \bibinfo
  {pages} {L95} (\bibinfo {year} {1972})}\BibitemShut {NoStop}%
\bibitem [{\citenamefont {{Cardoso}}\ \emph {et~al.}(2009)\citenamefont
  {{Cardoso}}, \citenamefont {{Miranda}}, \citenamefont {{Berti}},
  \citenamefont {{Witek}},\ and\ \citenamefont
  {{Zanchin}}}]{2009PhRvD..79f4016C}%
  \BibitemOpen
  \bibfield  {author} {\bibinfo {author} {\bibfnamefont {V.}~\bibnamefont
  {{Cardoso}}}, \bibinfo {author} {\bibfnamefont {A.~S.}\ \bibnamefont
  {{Miranda}}}, \bibinfo {author} {\bibfnamefont {E.}~\bibnamefont {{Berti}}},
  \bibinfo {author} {\bibfnamefont {H.}~\bibnamefont {{Witek}}}, \ and\
  \bibinfo {author} {\bibfnamefont {V.~T.}\ \bibnamefont {{Zanchin}}},\ }\href
  {\doibase 10.1103/PhysRevD.79.064016} {\bibfield  {journal} {\bibinfo
  {journal} {\prd}\ }\textbf {\bibinfo {volume} {79}},\ \bibinfo {eid} {064016}
  (\bibinfo {year} {2009})},\ \Eprint {http://arxiv.org/abs/0812.1806}
  {arXiv:0812.1806 [hep-th]} \BibitemShut {NoStop}%
\bibitem [{\citenamefont {{Teukolsky}}(1973)}]{1973ApJ...185..635T}%
  \BibitemOpen
  \bibfield  {author} {\bibinfo {author} {\bibfnamefont {S.~A.}\ \bibnamefont
  {{Teukolsky}}},\ }\href {\doibase 10.1086/152444} {\bibfield  {journal}
  {\bibinfo  {journal} {\apj}\ }\textbf {\bibinfo {volume} {185}},\ \bibinfo
  {pages} {635} (\bibinfo {year} {1973})}\BibitemShut {NoStop}%
\bibitem [{\citenamefont {{Press}}\ and\ \citenamefont
  {{Teukolsky}}(1973)}]{1973ApJ...185..649P}%
  \BibitemOpen
  \bibfield  {author} {\bibinfo {author} {\bibfnamefont {W.~H.}\ \bibnamefont
  {{Press}}}\ and\ \bibinfo {author} {\bibfnamefont {S.~A.}\ \bibnamefont
  {{Teukolsky}}},\ }\href {\doibase 10.1086/152445} {\bibfield  {journal}
  {\bibinfo  {journal} {\apj}\ }\textbf {\bibinfo {volume} {185}},\ \bibinfo
  {pages} {649} (\bibinfo {year} {1973})}\BibitemShut {NoStop}%
\bibitem [{\citenamefont {{Maggiore}}(2018)}]{2018gwv..book.....M}%
  \BibitemOpen
  \bibfield  {author} {\bibinfo {author} {\bibfnamefont {M.}~\bibnamefont
  {{Maggiore}}},\ }\href {\doibase 10.1093/oso/9780198570899.001.0001} {\emph
  {\bibinfo {title} {{Gravitational Waves: Volume 2: Astrophysics and
  Cosmology}}}}\ (\bibinfo {year} {2018})\BibitemShut {NoStop}%
\bibitem [{\citenamefont {{Berti}}\ \emph {et~al.}(2018)\citenamefont
  {{Berti}}, \citenamefont {{Yagi}}, \citenamefont {{Yang}},\ and\
  \citenamefont {{Yunes}}}]{2018GReGr..50...49B}%
  \BibitemOpen
  \bibfield  {author} {\bibinfo {author} {\bibfnamefont {E.}~\bibnamefont
  {{Berti}}}, \bibinfo {author} {\bibfnamefont {K.}~\bibnamefont {{Yagi}}},
  \bibinfo {author} {\bibfnamefont {H.}~\bibnamefont {{Yang}}}, \ and\ \bibinfo
  {author} {\bibfnamefont {N.}~\bibnamefont {{Yunes}}},\ }\href {\doibase
  10.1007/s10714-018-2372-6} {\bibfield  {journal} {\bibinfo  {journal}
  {General Relativity and Gravitation}\ }\textbf {\bibinfo {volume} {50}},\
  \bibinfo {eid} {49} (\bibinfo {year} {2018})},\ \Eprint
  {http://arxiv.org/abs/1801.03587} {arXiv:1801.03587 [gr-qc]} \BibitemShut
  {NoStop}%
\bibitem [{\citenamefont {Abbott}\ \emph
  {et~al.}(2016{\natexlab{a}})\citenamefont {Abbott} \emph
  {et~al.}}]{LIGOScientific:2016aoc}%
  \BibitemOpen
  \bibfield  {author} {\bibinfo {author} {\bibfnamefont {B.~P.}\ \bibnamefont
  {Abbott}} \emph {et~al.} (\bibinfo {collaboration} {LIGO Scientific,
  Virgo}),\ }\href {\doibase 10.1103/PhysRevLett.116.061102} {\bibfield
  {journal} {\bibinfo  {journal} {Phys. Rev. Lett.}\ }\textbf {\bibinfo
  {volume} {116}},\ \bibinfo {pages} {061102} (\bibinfo {year}
  {2016}{\natexlab{a}})},\ \Eprint {http://arxiv.org/abs/1602.03837}
  {arXiv:1602.03837 [gr-qc]} \BibitemShut {NoStop}%
\bibitem [{\citenamefont {Abbott}\ \emph
  {et~al.}(2016{\natexlab{b}})\citenamefont {Abbott} \emph
  {et~al.}}]{LIGOScientific:2016lio}%
  \BibitemOpen
  \bibfield  {author} {\bibinfo {author} {\bibfnamefont {B.~P.}\ \bibnamefont
  {Abbott}} \emph {et~al.} (\bibinfo {collaboration} {LIGO Scientific,
  Virgo}),\ }\href {\doibase 10.1103/PhysRevLett.116.221101} {\bibfield
  {journal} {\bibinfo  {journal} {Phys. Rev. Lett.}\ }\textbf {\bibinfo
  {volume} {116}},\ \bibinfo {pages} {221101} (\bibinfo {year}
  {2016}{\natexlab{b}})},\ \bibinfo {note} {[Erratum: Phys.Rev.Lett. 121,
  129902 (2018)]},\ \Eprint {http://arxiv.org/abs/1602.03841} {arXiv:1602.03841
  [gr-qc]} \BibitemShut {NoStop}%
\bibitem [{\citenamefont {Abbott}\ \emph {et~al.}(2021)\citenamefont {Abbott}
  \emph {et~al.}}]{LIGOScientific:2021sio}%
  \BibitemOpen
  \bibfield  {author} {\bibinfo {author} {\bibfnamefont {R.}~\bibnamefont
  {Abbott}} \emph {et~al.} (\bibinfo {collaboration} {LIGO Scientific, VIRGO,
  KAGRA}),\ }\href@noop {} {\  (\bibinfo {year} {2021})},\ \Eprint
  {http://arxiv.org/abs/2112.06861} {arXiv:2112.06861 [gr-qc]} \BibitemShut
  {NoStop}%
\bibitem [{\citenamefont {Echeverria}(1989)}]{Echeverria:1989hg}%
  \BibitemOpen
  \bibfield  {author} {\bibinfo {author} {\bibfnamefont {F.}~\bibnamefont
  {Echeverria}},\ }\href {\doibase 10.1103/PhysRevD.40.3194} {\bibfield
  {journal} {\bibinfo  {journal} {Phys. Rev. D}\ }\textbf {\bibinfo {volume}
  {40}},\ \bibinfo {pages} {3194} (\bibinfo {year} {1989})}\BibitemShut
  {NoStop}%
\bibitem [{\citenamefont {{Berti}}\ \emph {et~al.}(2007)\citenamefont
  {{Berti}}, \citenamefont {{Cardoso}}, \citenamefont {{Cardoso}},\ and\
  \citenamefont {{Cavagli{\`a}}}}]{2007PhRvD..76j4044B}%
  \BibitemOpen
  \bibfield  {author} {\bibinfo {author} {\bibfnamefont {E.}~\bibnamefont
  {{Berti}}}, \bibinfo {author} {\bibfnamefont {J.}~\bibnamefont {{Cardoso}}},
  \bibinfo {author} {\bibfnamefont {V.}~\bibnamefont {{Cardoso}}}, \ and\
  \bibinfo {author} {\bibfnamefont {M.}~\bibnamefont {{Cavagli{\`a}}}},\ }\href
  {\doibase 10.1103/PhysRevD.76.104044} {\bibfield  {journal} {\bibinfo
  {journal} {\prd}\ }\textbf {\bibinfo {volume} {76}},\ \bibinfo {eid} {104044}
  (\bibinfo {year} {2007})},\ \Eprint {http://arxiv.org/abs/0707.1202}
  {arXiv:0707.1202 [gr-qc]} \BibitemShut {NoStop}%
\bibitem [{\citenamefont {{Detweiler}}(1980)}]{1980ApJ...239..292D}%
  \BibitemOpen
  \bibfield  {author} {\bibinfo {author} {\bibfnamefont {S.}~\bibnamefont
  {{Detweiler}}},\ }\href {\doibase 10.1086/158109} {\bibfield  {journal}
  {\bibinfo  {journal} {\apj}\ }\textbf {\bibinfo {volume} {239}},\ \bibinfo
  {pages} {292} (\bibinfo {year} {1980})}\BibitemShut {NoStop}%
\bibitem [{\citenamefont {{Dreyer}}\ \emph {et~al.}(2004)\citenamefont
  {{Dreyer}}, \citenamefont {{Kelly}}, \citenamefont {{Krishnan}},
  \citenamefont {{Finn}}, \citenamefont {{Garrison}},\ and\ \citenamefont
  {{Lopez-Aleman}}}]{2004CQGra..21..787D}%
  \BibitemOpen
  \bibfield  {author} {\bibinfo {author} {\bibfnamefont {O.}~\bibnamefont
  {{Dreyer}}}, \bibinfo {author} {\bibfnamefont {B.}~\bibnamefont {{Kelly}}},
  \bibinfo {author} {\bibfnamefont {B.}~\bibnamefont {{Krishnan}}}, \bibinfo
  {author} {\bibfnamefont {L.~S.}\ \bibnamefont {{Finn}}}, \bibinfo {author}
  {\bibfnamefont {D.}~\bibnamefont {{Garrison}}}, \ and\ \bibinfo {author}
  {\bibfnamefont {R.}~\bibnamefont {{Lopez-Aleman}}},\ }\href {\doibase
  10.1088/0264-9381/21/4/003} {\bibfield  {journal} {\bibinfo  {journal}
  {Classical and Quantum Gravity}\ }\textbf {\bibinfo {volume} {21}},\ \bibinfo
  {pages} {787} (\bibinfo {year} {2004})},\ \Eprint
  {http://arxiv.org/abs/gr-qc/0309007} {arXiv:gr-qc/0309007 [gr-qc]}
  \BibitemShut {NoStop}%
\bibitem [{\citenamefont {{Berti}}\ \emph {et~al.}(2006)\citenamefont
  {{Berti}}, \citenamefont {{Cardoso}},\ and\ \citenamefont
  {{Will}}}]{2006PhRvD..73f4030B}%
  \BibitemOpen
  \bibfield  {author} {\bibinfo {author} {\bibfnamefont {E.}~\bibnamefont
  {{Berti}}}, \bibinfo {author} {\bibfnamefont {V.}~\bibnamefont {{Cardoso}}},
  \ and\ \bibinfo {author} {\bibfnamefont {C.~M.}\ \bibnamefont {{Will}}},\
  }\href {\doibase 10.1103/PhysRevD.73.064030} {\bibfield  {journal} {\bibinfo
  {journal} {\prd}\ }\textbf {\bibinfo {volume} {73}},\ \bibinfo {eid} {064030}
  (\bibinfo {year} {2006})},\ \Eprint {http://arxiv.org/abs/gr-qc/0512160}
  {arXiv:gr-qc/0512160 [gr-qc]} \BibitemShut {NoStop}%
\bibitem [{\citenamefont {{Berti}}\ \emph {et~al.}(2015)\citenamefont {{Berti}}
  \emph {et~al.}}]{2015CQGra..32x3001B}%
  \BibitemOpen
  \bibfield  {author} {\bibinfo {author} {\bibfnamefont {E.}~\bibnamefont
  {{Berti}}} \emph {et~al.},\ }\href {\doibase 10.1088/0264-9381/32/24/243001}
  {\bibfield  {journal} {\bibinfo  {journal} {Classical and Quantum Gravity}\
  }\textbf {\bibinfo {volume} {32}},\ \bibinfo {eid} {243001} (\bibinfo {year}
  {2015})},\ \Eprint {http://arxiv.org/abs/1501.07274} {arXiv:1501.07274
  [gr-qc]} \BibitemShut {NoStop}%
\bibitem [{\citenamefont {{Isi}}\ \emph {et~al.}(2019)\citenamefont {{Isi}},
  \citenamefont {{Giesler}}, \citenamefont {{Farr}}, \citenamefont {{Scheel}},\
  and\ \citenamefont {{Teukolsky}}}]{2019PhRvL.123k1102I}%
  \BibitemOpen
  \bibfield  {author} {\bibinfo {author} {\bibfnamefont {M.}~\bibnamefont
  {{Isi}}}, \bibinfo {author} {\bibfnamefont {M.}~\bibnamefont {{Giesler}}},
  \bibinfo {author} {\bibfnamefont {W.~M.}\ \bibnamefont {{Farr}}}, \bibinfo
  {author} {\bibfnamefont {M.~A.}\ \bibnamefont {{Scheel}}}, \ and\ \bibinfo
  {author} {\bibfnamefont {S.~A.}\ \bibnamefont {{Teukolsky}}},\ }\href
  {\doibase 10.1103/PhysRevLett.123.111102} {\bibfield  {journal} {\bibinfo
  {journal} {\prl}\ }\textbf {\bibinfo {volume} {123}},\ \bibinfo {eid}
  {111102} (\bibinfo {year} {2019})},\ \Eprint
  {http://arxiv.org/abs/1905.00869} {arXiv:1905.00869 [gr-qc]} \BibitemShut
  {NoStop}%
\bibitem [{\citenamefont {{Cotesta}}\ \emph {et~al.}(2022)\citenamefont
  {{Cotesta}}, \citenamefont {{Carullo}}, \citenamefont {{Berti}},\ and\
  \citenamefont {{Cardoso}}}]{2022PhRvL.129k1102C}%
  \BibitemOpen
  \bibfield  {author} {\bibinfo {author} {\bibfnamefont {R.}~\bibnamefont
  {{Cotesta}}}, \bibinfo {author} {\bibfnamefont {G.}~\bibnamefont
  {{Carullo}}}, \bibinfo {author} {\bibfnamefont {E.}~\bibnamefont {{Berti}}},
  \ and\ \bibinfo {author} {\bibfnamefont {V.}~\bibnamefont {{Cardoso}}},\
  }\href {\doibase 10.1103/PhysRevLett.129.111102} {\bibfield  {journal}
  {\bibinfo  {journal} {\prl}\ }\textbf {\bibinfo {volume} {129}},\ \bibinfo
  {eid} {111102} (\bibinfo {year} {2022})},\ \Eprint
  {http://arxiv.org/abs/2201.00822} {arXiv:2201.00822 [gr-qc]} \BibitemShut
  {NoStop}%
\bibitem [{\citenamefont {{Capano}}\ \emph {et~al.}(2023)\citenamefont
  {{Capano}}, \citenamefont {{Cabero}}, \citenamefont {{Westerweck}},
  \citenamefont {{Abedi}}, \citenamefont {{Kastha}}, \citenamefont {{Nitz}},
  \citenamefont {{Wang}}, \citenamefont {{Nielsen}},\ and\ \citenamefont
  {{Krishnan}}}]{2023PhRvL.131v1402C}%
  \BibitemOpen
  \bibfield  {author} {\bibinfo {author} {\bibfnamefont {C.~D.}\ \bibnamefont
  {{Capano}}}, \bibinfo {author} {\bibfnamefont {M.}~\bibnamefont {{Cabero}}},
  \bibinfo {author} {\bibfnamefont {J.}~\bibnamefont {{Westerweck}}}, \bibinfo
  {author} {\bibfnamefont {J.}~\bibnamefont {{Abedi}}}, \bibinfo {author}
  {\bibfnamefont {S.}~\bibnamefont {{Kastha}}}, \bibinfo {author}
  {\bibfnamefont {A.~H.}\ \bibnamefont {{Nitz}}}, \bibinfo {author}
  {\bibfnamefont {Y.-F.}\ \bibnamefont {{Wang}}}, \bibinfo {author}
  {\bibfnamefont {A.~B.}\ \bibnamefont {{Nielsen}}}, \ and\ \bibinfo {author}
  {\bibfnamefont {B.}~\bibnamefont {{Krishnan}}},\ }\href {\doibase
  10.1103/PhysRevLett.131.221402} {\bibfield  {journal} {\bibinfo  {journal}
  {\prl}\ }\textbf {\bibinfo {volume} {131}},\ \bibinfo {eid} {221402}
  (\bibinfo {year} {2023})},\ \Eprint {http://arxiv.org/abs/2105.05238}
  {arXiv:2105.05238 [gr-qc]} \BibitemShut {NoStop}%
\bibitem [{\citenamefont {{Gu}}\ \emph {et~al.}(2024)\citenamefont {{Gu}},
  \citenamefont {{Wang}},\ and\ \citenamefont {{Shao}}}]{2024PhRvD.109b4058G}%
  \BibitemOpen
  \bibfield  {author} {\bibinfo {author} {\bibfnamefont {H.-P.}\ \bibnamefont
  {{Gu}}}, \bibinfo {author} {\bibfnamefont {H.-T.}\ \bibnamefont {{Wang}}}, \
  and\ \bibinfo {author} {\bibfnamefont {L.}~\bibnamefont {{Shao}}},\ }\href
  {\doibase 10.1103/PhysRevD.109.024058} {\bibfield  {journal} {\bibinfo
  {journal} {\prd}\ }\textbf {\bibinfo {volume} {109}},\ \bibinfo {eid}
  {024058} (\bibinfo {year} {2024})},\ \Eprint
  {http://arxiv.org/abs/2310.10447} {arXiv:2310.10447 [gr-qc]} \BibitemShut
  {NoStop}%
\bibitem [{\citenamefont {{Wang}}\ and\ \citenamefont
  {{Shao}}(2024)}]{2024PhRvD.109d3027W}%
  \BibitemOpen
  \bibfield  {author} {\bibinfo {author} {\bibfnamefont {H.-T.}\ \bibnamefont
  {{Wang}}}\ and\ \bibinfo {author} {\bibfnamefont {L.}~\bibnamefont
  {{Shao}}},\ }\href {\doibase 10.1103/PhysRevD.109.043027} {\bibfield
  {journal} {\bibinfo  {journal} {\prd}\ }\textbf {\bibinfo {volume} {109}},\
  \bibinfo {eid} {043027} (\bibinfo {year} {2024})},\ \Eprint
  {http://arxiv.org/abs/2401.13997} {arXiv:2401.13997 [gr-qc]} \BibitemShut
  {NoStop}%
\bibitem [{\citenamefont {{Baibhav}}\ \emph {et~al.}(2018)\citenamefont
  {{Baibhav}}, \citenamefont {{Berti}}, \citenamefont {{Cardoso}},\ and\
  \citenamefont {{Khanna}}}]{2018PhRvD..97d4048B}%
  \BibitemOpen
  \bibfield  {author} {\bibinfo {author} {\bibfnamefont {V.}~\bibnamefont
  {{Baibhav}}}, \bibinfo {author} {\bibfnamefont {E.}~\bibnamefont {{Berti}}},
  \bibinfo {author} {\bibfnamefont {V.}~\bibnamefont {{Cardoso}}}, \ and\
  \bibinfo {author} {\bibfnamefont {G.}~\bibnamefont {{Khanna}}},\ }\href
  {\doibase 10.1103/PhysRevD.97.044048} {\bibfield  {journal} {\bibinfo
  {journal} {\prd}\ }\textbf {\bibinfo {volume} {97}},\ \bibinfo {eid} {044048}
  (\bibinfo {year} {2018})},\ \Eprint {http://arxiv.org/abs/1710.02156}
  {arXiv:1710.02156 [gr-qc]} \BibitemShut {NoStop}%
\bibitem [{\citenamefont {{Giesler}}\ \emph {et~al.}(2019)\citenamefont
  {{Giesler}}, \citenamefont {{Isi}}, \citenamefont {{Scheel}},\ and\
  \citenamefont {{Teukolsky}}}]{2019PhRvX...9d1060G}%
  \BibitemOpen
  \bibfield  {author} {\bibinfo {author} {\bibfnamefont {M.}~\bibnamefont
  {{Giesler}}}, \bibinfo {author} {\bibfnamefont {M.}~\bibnamefont {{Isi}}},
  \bibinfo {author} {\bibfnamefont {M.~A.}\ \bibnamefont {{Scheel}}}, \ and\
  \bibinfo {author} {\bibfnamefont {S.~A.}\ \bibnamefont {{Teukolsky}}},\
  }\href {\doibase 10.1103/PhysRevX.9.041060} {\bibfield  {journal} {\bibinfo
  {journal} {Physical Review X}\ }\textbf {\bibinfo {volume} {9}},\ \bibinfo
  {eid} {041060} (\bibinfo {year} {2019})},\ \Eprint
  {http://arxiv.org/abs/1903.08284} {arXiv:1903.08284 [gr-qc]} \BibitemShut
  {NoStop}%
\bibitem [{\citenamefont {{Baibhav}}\ \emph {et~al.}(2023)\citenamefont
  {{Baibhav}}, \citenamefont {{Cheung}}, \citenamefont {{Berti}}, \citenamefont
  {{Cardoso}}, \citenamefont {{Carullo}}, \citenamefont {{Cotesta}},
  \citenamefont {{Del Pozzo}},\ and\ \citenamefont
  {{Duque}}}]{2023PhRvD.108j4020B}%
  \BibitemOpen
  \bibfield  {author} {\bibinfo {author} {\bibfnamefont {V.}~\bibnamefont
  {{Baibhav}}}, \bibinfo {author} {\bibfnamefont {M.~H.-Y.}\ \bibnamefont
  {{Cheung}}}, \bibinfo {author} {\bibfnamefont {E.}~\bibnamefont {{Berti}}},
  \bibinfo {author} {\bibfnamefont {V.}~\bibnamefont {{Cardoso}}}, \bibinfo
  {author} {\bibfnamefont {G.}~\bibnamefont {{Carullo}}}, \bibinfo {author}
  {\bibfnamefont {R.}~\bibnamefont {{Cotesta}}}, \bibinfo {author}
  {\bibfnamefont {W.}~\bibnamefont {{Del Pozzo}}}, \ and\ \bibinfo {author}
  {\bibfnamefont {F.}~\bibnamefont {{Duque}}},\ }\href {\doibase
  10.1103/PhysRevD.108.104020} {\bibfield  {journal} {\bibinfo  {journal}
  {\prd}\ }\textbf {\bibinfo {volume} {108}},\ \bibinfo {eid} {104020}
  (\bibinfo {year} {2023})},\ \Eprint {http://arxiv.org/abs/2302.03050}
  {arXiv:2302.03050 [gr-qc]} \BibitemShut {NoStop}%
\bibitem [{\citenamefont {{Nee}}\ \emph {et~al.}(2023)\citenamefont {{Nee}},
  \citenamefont {{V{\"o}lkel}},\ and\ \citenamefont
  {{Pfeiffer}}}]{2023PhRvD.108d4032N}%
  \BibitemOpen
  \bibfield  {author} {\bibinfo {author} {\bibfnamefont {P.~J.}\ \bibnamefont
  {{Nee}}}, \bibinfo {author} {\bibfnamefont {S.~H.}\ \bibnamefont
  {{V{\"o}lkel}}}, \ and\ \bibinfo {author} {\bibfnamefont {H.~P.}\
  \bibnamefont {{Pfeiffer}}},\ }\href {\doibase 10.1103/PhysRevD.108.044032}
  {\bibfield  {journal} {\bibinfo  {journal} {\prd}\ }\textbf {\bibinfo
  {volume} {108}},\ \bibinfo {eid} {044032} (\bibinfo {year} {2023})},\ \Eprint
  {http://arxiv.org/abs/2302.06634} {arXiv:2302.06634 [gr-qc]} \BibitemShut
  {NoStop}%
\bibitem [{\citenamefont {{Wang}}\ and\ \citenamefont
  {{Shao}}(2023)}]{2023PhRvD.108l3018W}%
  \BibitemOpen
  \bibfield  {author} {\bibinfo {author} {\bibfnamefont {H.-T.}\ \bibnamefont
  {{Wang}}}\ and\ \bibinfo {author} {\bibfnamefont {L.}~\bibnamefont
  {{Shao}}},\ }\href {\doibase 10.1103/PhysRevD.108.123018} {\bibfield
  {journal} {\bibinfo  {journal} {\prd}\ }\textbf {\bibinfo {volume} {108}},\
  \bibinfo {eid} {123018} (\bibinfo {year} {2023})},\ \Eprint
  {http://arxiv.org/abs/2311.13300} {arXiv:2311.13300 [gr-qc]} \BibitemShut
  {NoStop}%
\bibitem [{\citenamefont {{Nollert}}(1996)}]{1996PhRvD..53.4397N}%
  \BibitemOpen
  \bibfield  {author} {\bibinfo {author} {\bibfnamefont {H.-P.}\ \bibnamefont
  {{Nollert}}},\ }\href {\doibase 10.1103/PhysRevD.53.4397} {\bibfield
  {journal} {\bibinfo  {journal} {\prd}\ }\textbf {\bibinfo {volume} {53}},\
  \bibinfo {pages} {4397} (\bibinfo {year} {1996})},\ \Eprint
  {http://arxiv.org/abs/gr-qc/9602032} {arXiv:gr-qc/9602032 [gr-qc]}
  \BibitemShut {NoStop}%
\bibitem [{\citenamefont {{Nollert}}\ and\ \citenamefont
  {{Price}}(1999)}]{1999JMP....40..980N}%
  \BibitemOpen
  \bibfield  {author} {\bibinfo {author} {\bibfnamefont {H.-P.}\ \bibnamefont
  {{Nollert}}}\ and\ \bibinfo {author} {\bibfnamefont {R.~H.}\ \bibnamefont
  {{Price}}},\ }\href {\doibase 10.1063/1.532698} {\bibfield  {journal}
  {\bibinfo  {journal} {Journal of Mathematical Physics}\ }\textbf {\bibinfo
  {volume} {40}},\ \bibinfo {pages} {980} (\bibinfo {year} {1999})},\ \Eprint
  {http://arxiv.org/abs/gr-qc/9810074} {arXiv:gr-qc/9810074 [gr-qc]}
  \BibitemShut {NoStop}%
\bibitem [{\citenamefont {{Daghigh}}\ \emph {et~al.}(2020)\citenamefont
  {{Daghigh}}, \citenamefont {{Green}},\ and\ \citenamefont
  {{Morey}}}]{2020PhRvD.101j4009D}%
  \BibitemOpen
  \bibfield  {author} {\bibinfo {author} {\bibfnamefont {R.~G.}\ \bibnamefont
  {{Daghigh}}}, \bibinfo {author} {\bibfnamefont {M.~D.}\ \bibnamefont
  {{Green}}}, \ and\ \bibinfo {author} {\bibfnamefont {J.~C.}\ \bibnamefont
  {{Morey}}},\ }\href {\doibase 10.1103/PhysRevD.101.104009} {\bibfield
  {journal} {\bibinfo  {journal} {\prd}\ }\textbf {\bibinfo {volume} {101}},\
  \bibinfo {eid} {104009} (\bibinfo {year} {2020})},\ \Eprint
  {http://arxiv.org/abs/2002.07251} {arXiv:2002.07251 [gr-qc]} \BibitemShut
  {NoStop}%
\bibitem [{\citenamefont {{Leung}}\ \emph {et~al.}(1997)\citenamefont
  {{Leung}}, \citenamefont {{Liu}}, \citenamefont {{Suen}}, \citenamefont
  {{Tam}},\ and\ \citenamefont {{Young}}}]{1997PhRvL..78.2894L}%
  \BibitemOpen
  \bibfield  {author} {\bibinfo {author} {\bibfnamefont {P.~T.}\ \bibnamefont
  {{Leung}}}, \bibinfo {author} {\bibfnamefont {Y.~T.}\ \bibnamefont {{Liu}}},
  \bibinfo {author} {\bibfnamefont {W.~M.}\ \bibnamefont {{Suen}}}, \bibinfo
  {author} {\bibfnamefont {C.~Y.}\ \bibnamefont {{Tam}}}, \ and\ \bibinfo
  {author} {\bibfnamefont {K.}~\bibnamefont {{Young}}},\ }\href {\doibase
  10.1103/PhysRevLett.78.2894} {\bibfield  {journal} {\bibinfo  {journal}
  {\prl}\ }\textbf {\bibinfo {volume} {78}},\ \bibinfo {pages} {2894} (\bibinfo
  {year} {1997})},\ \Eprint {http://arxiv.org/abs/gr-qc/9903031}
  {arXiv:gr-qc/9903031 [gr-qc]} \BibitemShut {NoStop}%
\bibitem [{\citenamefont {{Barausse}}\ \emph {et~al.}(2014)\citenamefont
  {{Barausse}}, \citenamefont {{Cardoso}},\ and\ \citenamefont
  {{Pani}}}]{2014PhRvD..89j4059B}%
  \BibitemOpen
  \bibfield  {author} {\bibinfo {author} {\bibfnamefont {E.}~\bibnamefont
  {{Barausse}}}, \bibinfo {author} {\bibfnamefont {V.}~\bibnamefont
  {{Cardoso}}}, \ and\ \bibinfo {author} {\bibfnamefont {P.}~\bibnamefont
  {{Pani}}},\ }\href {\doibase 10.1103/PhysRevD.89.104059} {\bibfield
  {journal} {\bibinfo  {journal} {\prd}\ }\textbf {\bibinfo {volume} {89}},\
  \bibinfo {eid} {104059} (\bibinfo {year} {2014})},\ \Eprint
  {http://arxiv.org/abs/1404.7149} {arXiv:1404.7149 [gr-qc]} \BibitemShut
  {NoStop}%
\bibitem [{\citenamefont {{Bamber}}\ \emph {et~al.}(2021)\citenamefont
  {{Bamber}}, \citenamefont {{Tattersall}}, \citenamefont {{Clough}},\ and\
  \citenamefont {{Ferreira}}}]{2021PhRvD.103l4013B}%
  \BibitemOpen
  \bibfield  {author} {\bibinfo {author} {\bibfnamefont {J.}~\bibnamefont
  {{Bamber}}}, \bibinfo {author} {\bibfnamefont {O.~J.}\ \bibnamefont
  {{Tattersall}}}, \bibinfo {author} {\bibfnamefont {K.}~\bibnamefont
  {{Clough}}}, \ and\ \bibinfo {author} {\bibfnamefont {P.~G.}\ \bibnamefont
  {{Ferreira}}},\ }\href {\doibase 10.1103/PhysRevD.103.124013} {\bibfield
  {journal} {\bibinfo  {journal} {\prd}\ }\textbf {\bibinfo {volume} {103}},\
  \bibinfo {eid} {124013} (\bibinfo {year} {2021})},\ \Eprint
  {http://arxiv.org/abs/2103.00026} {arXiv:2103.00026 [gr-qc]} \BibitemShut
  {NoStop}%
\bibitem [{\citenamefont {{Cardoso}}\ \emph
  {et~al.}(2022{\natexlab{a}})\citenamefont {{Cardoso}}, \citenamefont
  {{Destounis}}, \citenamefont {{Duque}}, \citenamefont {{Macedo}},\ and\
  \citenamefont {{Maselli}}}]{2022PhRvD.105f1501C}%
  \BibitemOpen
  \bibfield  {author} {\bibinfo {author} {\bibfnamefont {V.}~\bibnamefont
  {{Cardoso}}}, \bibinfo {author} {\bibfnamefont {K.}~\bibnamefont
  {{Destounis}}}, \bibinfo {author} {\bibfnamefont {F.}~\bibnamefont
  {{Duque}}}, \bibinfo {author} {\bibfnamefont {R.~P.}\ \bibnamefont
  {{Macedo}}}, \ and\ \bibinfo {author} {\bibfnamefont {A.}~\bibnamefont
  {{Maselli}}},\ }\href {\doibase 10.1103/PhysRevD.105.L061501} {\bibfield
  {journal} {\bibinfo  {journal} {\prd}\ }\textbf {\bibinfo {volume} {105}},\
  \bibinfo {eid} {L061501} (\bibinfo {year} {2022}{\natexlab{a}})},\ \Eprint
  {http://arxiv.org/abs/2109.00005} {arXiv:2109.00005 [gr-qc]} \BibitemShut
  {NoStop}%
\bibitem [{\citenamefont {{Cardoso}}\ \emph
  {et~al.}(2022{\natexlab{b}})\citenamefont {{Cardoso}}, \citenamefont
  {{Destounis}}, \citenamefont {{Duque}}, \citenamefont {{Macedo}},\ and\
  \citenamefont {{Maselli}}}]{2022PhRvL.129x1103C}%
  \BibitemOpen
  \bibfield  {author} {\bibinfo {author} {\bibfnamefont {V.}~\bibnamefont
  {{Cardoso}}}, \bibinfo {author} {\bibfnamefont {K.}~\bibnamefont
  {{Destounis}}}, \bibinfo {author} {\bibfnamefont {F.}~\bibnamefont
  {{Duque}}}, \bibinfo {author} {\bibfnamefont {R.~P.}\ \bibnamefont
  {{Macedo}}}, \ and\ \bibinfo {author} {\bibfnamefont {A.}~\bibnamefont
  {{Maselli}}},\ }\href {\doibase 10.1103/PhysRevLett.129.241103} {\bibfield
  {journal} {\bibinfo  {journal} {\prl}\ }\textbf {\bibinfo {volume} {129}},\
  \bibinfo {eid} {241103} (\bibinfo {year} {2022}{\natexlab{b}})},\ \Eprint
  {http://arxiv.org/abs/2210.01133} {arXiv:2210.01133 [gr-qc]} \BibitemShut
  {NoStop}%
\bibitem [{\citenamefont {{Cardoso}}\ and\ \citenamefont
  {{Pani}}(2019)}]{2019LRR....22....4C}%
  \BibitemOpen
  \bibfield  {author} {\bibinfo {author} {\bibfnamefont {V.}~\bibnamefont
  {{Cardoso}}}\ and\ \bibinfo {author} {\bibfnamefont {P.}~\bibnamefont
  {{Pani}}},\ }\href {\doibase 10.1007/s41114-019-0020-4} {\bibfield  {journal}
  {\bibinfo  {journal} {Living Reviews in Relativity}\ }\textbf {\bibinfo
  {volume} {22}},\ \bibinfo {eid} {4} (\bibinfo {year} {2019})},\ \Eprint
  {http://arxiv.org/abs/1904.05363} {arXiv:1904.05363 [gr-qc]} \BibitemShut
  {NoStop}%
\bibitem [{\citenamefont {{Cardoso}}\ \emph {et~al.}(2016)\citenamefont
  {{Cardoso}}, \citenamefont {{Hopper}}, \citenamefont {{Macedo}},
  \citenamefont {{Palenzuela}},\ and\ \citenamefont
  {{Pani}}}]{2016PhRvD..94h4031C}%
  \BibitemOpen
  \bibfield  {author} {\bibinfo {author} {\bibfnamefont {V.}~\bibnamefont
  {{Cardoso}}}, \bibinfo {author} {\bibfnamefont {S.}~\bibnamefont {{Hopper}}},
  \bibinfo {author} {\bibfnamefont {C.~F.~B.}\ \bibnamefont {{Macedo}}},
  \bibinfo {author} {\bibfnamefont {C.}~\bibnamefont {{Palenzuela}}}, \ and\
  \bibinfo {author} {\bibfnamefont {P.}~\bibnamefont {{Pani}}},\ }\href
  {\doibase 10.1103/PhysRevD.94.084031} {\bibfield  {journal} {\bibinfo
  {journal} {\prd}\ }\textbf {\bibinfo {volume} {94}},\ \bibinfo {eid} {084031}
  (\bibinfo {year} {2016})},\ \Eprint {http://arxiv.org/abs/1608.08637}
  {arXiv:1608.08637 [gr-qc]} \BibitemShut {NoStop}%
\bibitem [{\citenamefont {{Jaramillo}}\ \emph {et~al.}(2021)\citenamefont
  {{Jaramillo}}, \citenamefont {{Macedo}},\ and\ \citenamefont
  {{Sheikh}}}]{2021PhRvX..11c1003J}%
  \BibitemOpen
  \bibfield  {author} {\bibinfo {author} {\bibfnamefont {J.~L.}\ \bibnamefont
  {{Jaramillo}}}, \bibinfo {author} {\bibfnamefont {R.~P.}\ \bibnamefont
  {{Macedo}}}, \ and\ \bibinfo {author} {\bibfnamefont {L.~A.}\ \bibnamefont
  {{Sheikh}}},\ }\href {\doibase 10.1103/PhysRevX.11.031003} {\bibfield
  {journal} {\bibinfo  {journal} {Physical Review X}\ }\textbf {\bibinfo
  {volume} {11}},\ \bibinfo {eid} {031003} (\bibinfo {year} {2021})},\ \Eprint
  {http://arxiv.org/abs/2004.06434} {arXiv:2004.06434 [gr-qc]} \BibitemShut
  {NoStop}%
\bibitem [{\citenamefont {{Destounis}}\ \emph {et~al.}(2021)\citenamefont
  {{Destounis}}, \citenamefont {{Macedo}}, \citenamefont {{Berti}},
  \citenamefont {{Cardoso}},\ and\ \citenamefont
  {{Jaramillo}}}]{2021PhRvD.104h4091D}%
  \BibitemOpen
  \bibfield  {author} {\bibinfo {author} {\bibfnamefont {K.}~\bibnamefont
  {{Destounis}}}, \bibinfo {author} {\bibfnamefont {R.~P.}\ \bibnamefont
  {{Macedo}}}, \bibinfo {author} {\bibfnamefont {E.}~\bibnamefont {{Berti}}},
  \bibinfo {author} {\bibfnamefont {V.}~\bibnamefont {{Cardoso}}}, \ and\
  \bibinfo {author} {\bibfnamefont {J.~L.}\ \bibnamefont {{Jaramillo}}},\
  }\href {\doibase 10.1103/PhysRevD.104.084091} {\bibfield  {journal} {\bibinfo
   {journal} {\prd}\ }\textbf {\bibinfo {volume} {104}},\ \bibinfo {eid}
  {084091} (\bibinfo {year} {2021})},\ \Eprint
  {http://arxiv.org/abs/2107.09673} {arXiv:2107.09673 [gr-qc]} \BibitemShut
  {NoStop}%
\bibitem [{\citenamefont {{Jaramillo}}\ \emph {et~al.}(2022)\citenamefont
  {{Jaramillo}}, \citenamefont {{Macedo}},\ and\ \citenamefont
  {{Sheikh}}}]{2022PhRvL.128u1102J}%
  \BibitemOpen
  \bibfield  {author} {\bibinfo {author} {\bibfnamefont {J.~L.}\ \bibnamefont
  {{Jaramillo}}}, \bibinfo {author} {\bibfnamefont {R.~P.}\ \bibnamefont
  {{Macedo}}}, \ and\ \bibinfo {author} {\bibfnamefont {L.~A.}\ \bibnamefont
  {{Sheikh}}},\ }\href {\doibase 10.1103/PhysRevLett.128.211102} {\bibfield
  {journal} {\bibinfo  {journal} {\prl}\ }\textbf {\bibinfo {volume} {128}},\
  \bibinfo {eid} {211102} (\bibinfo {year} {2022})},\ \Eprint
  {http://arxiv.org/abs/2105.03451} {arXiv:2105.03451 [gr-qc]} \BibitemShut
  {NoStop}%
\bibitem [{\citenamefont {{Yang}}\ and\ \citenamefont
  {{Zhang}}(2023)}]{2023PhRvD.107f4045Y}%
  \BibitemOpen
  \bibfield  {author} {\bibinfo {author} {\bibfnamefont {H.}~\bibnamefont
  {{Yang}}}\ and\ \bibinfo {author} {\bibfnamefont {J.}~\bibnamefont
  {{Zhang}}},\ }\href {\doibase 10.1103/PhysRevD.107.064045} {\bibfield
  {journal} {\bibinfo  {journal} {\prd}\ }\textbf {\bibinfo {volume} {107}},\
  \bibinfo {eid} {064045} (\bibinfo {year} {2023})},\ \Eprint
  {http://arxiv.org/abs/2210.01724} {arXiv:2210.01724 [gr-qc]} \BibitemShut
  {NoStop}%
\bibitem [{\citenamefont {{Boyanov}}\ \emph {et~al.}(2023)\citenamefont
  {{Boyanov}}, \citenamefont {{Destounis}}, \citenamefont {{Panosso Macedo}},
  \citenamefont {{Cardoso}},\ and\ \citenamefont
  {{Jaramillo}}}]{2023PhRvD.107f4012B}%
  \BibitemOpen
  \bibfield  {author} {\bibinfo {author} {\bibfnamefont {V.}~\bibnamefont
  {{Boyanov}}}, \bibinfo {author} {\bibfnamefont {K.}~\bibnamefont
  {{Destounis}}}, \bibinfo {author} {\bibfnamefont {R.}~\bibnamefont {{Panosso
  Macedo}}}, \bibinfo {author} {\bibfnamefont {V.}~\bibnamefont {{Cardoso}}}, \
  and\ \bibinfo {author} {\bibfnamefont {J.~L.}\ \bibnamefont {{Jaramillo}}},\
  }\href {\doibase 10.1103/PhysRevD.107.064012} {\bibfield  {journal} {\bibinfo
   {journal} {\prd}\ }\textbf {\bibinfo {volume} {107}},\ \bibinfo {eid}
  {064012} (\bibinfo {year} {2023})},\ \Eprint
  {http://arxiv.org/abs/2209.12950} {arXiv:2209.12950 [gr-qc]} \BibitemShut
  {NoStop}%
\bibitem [{\citenamefont {{Sarkar}}\ \emph {et~al.}(2023)\citenamefont
  {{Sarkar}}, \citenamefont {{Rahman}},\ and\ \citenamefont
  {{Chakraborty}}}]{2023PhRvD.108j4002S}%
  \BibitemOpen
  \bibfield  {author} {\bibinfo {author} {\bibfnamefont {S.}~\bibnamefont
  {{Sarkar}}}, \bibinfo {author} {\bibfnamefont {M.}~\bibnamefont {{Rahman}}},
  \ and\ \bibinfo {author} {\bibfnamefont {S.}~\bibnamefont {{Chakraborty}}},\
  }\href {\doibase 10.1103/PhysRevD.108.104002} {\bibfield  {journal} {\bibinfo
   {journal} {\prd}\ }\textbf {\bibinfo {volume} {108}},\ \bibinfo {eid}
  {104002} (\bibinfo {year} {2023})},\ \Eprint
  {http://arxiv.org/abs/2304.06829} {arXiv:2304.06829 [gr-qc]} \BibitemShut
  {NoStop}%
\bibitem [{\citenamefont {{Courty}}\ \emph {et~al.}(2023)\citenamefont
  {{Courty}}, \citenamefont {{Destounis}},\ and\ \citenamefont
  {{Pani}}}]{2023PhRvD.108j4027C}%
  \BibitemOpen
  \bibfield  {author} {\bibinfo {author} {\bibfnamefont {A.}~\bibnamefont
  {{Courty}}}, \bibinfo {author} {\bibfnamefont {K.}~\bibnamefont
  {{Destounis}}}, \ and\ \bibinfo {author} {\bibfnamefont {P.}~\bibnamefont
  {{Pani}}},\ }\href {\doibase 10.1103/PhysRevD.108.104027} {\bibfield
  {journal} {\bibinfo  {journal} {\prd}\ }\textbf {\bibinfo {volume} {108}},\
  \bibinfo {eid} {104027} (\bibinfo {year} {2023})},\ \Eprint
  {http://arxiv.org/abs/2307.11155} {arXiv:2307.11155 [gr-qc]} \BibitemShut
  {NoStop}%
\bibitem [{\citenamefont {{Cownden}}\ \emph {et~al.}(2024)\citenamefont
  {{Cownden}}, \citenamefont {{Pantelidou}},\ and\ \citenamefont
  {{Zilh{\~a}o}}}]{2024JHEP...05..202C}%
  \BibitemOpen
  \bibfield  {author} {\bibinfo {author} {\bibfnamefont {B.}~\bibnamefont
  {{Cownden}}}, \bibinfo {author} {\bibfnamefont {C.}~\bibnamefont
  {{Pantelidou}}}, \ and\ \bibinfo {author} {\bibfnamefont {M.}~\bibnamefont
  {{Zilh{\~a}o}}},\ }\href {\doibase 10.1007/JHEP05(2024)202} {\bibfield
  {journal} {\bibinfo  {journal} {Journal of High Energy Physics}\ }\textbf
  {\bibinfo {volume} {2024}},\ \bibinfo {eid} {202} (\bibinfo {year} {2024})},\
  \Eprint {http://arxiv.org/abs/2312.08352} {arXiv:2312.08352 [gr-qc]}
  \BibitemShut {NoStop}%
\bibitem [{\citenamefont {{Boyanov}}\ \emph {et~al.}(2024)\citenamefont
  {{Boyanov}}, \citenamefont {{Cardoso}}, \citenamefont {{Destounis}},
  \citenamefont {{Jaramillo}},\ and\ \citenamefont
  {{Macedo}}}]{2024PhRvD.109f4068B}%
  \BibitemOpen
  \bibfield  {author} {\bibinfo {author} {\bibfnamefont {V.}~\bibnamefont
  {{Boyanov}}}, \bibinfo {author} {\bibfnamefont {V.}~\bibnamefont
  {{Cardoso}}}, \bibinfo {author} {\bibfnamefont {K.}~\bibnamefont
  {{Destounis}}}, \bibinfo {author} {\bibfnamefont {J.~L.}\ \bibnamefont
  {{Jaramillo}}}, \ and\ \bibinfo {author} {\bibfnamefont {R.~P.}\ \bibnamefont
  {{Macedo}}},\ }\href {\doibase 10.1103/PhysRevD.109.064068} {\bibfield
  {journal} {\bibinfo  {journal} {\prd}\ }\textbf {\bibinfo {volume} {109}},\
  \bibinfo {eid} {064068} (\bibinfo {year} {2024})},\ \Eprint
  {http://arxiv.org/abs/2312.11998} {arXiv:2312.11998 [gr-qc]} \BibitemShut
  {NoStop}%
\bibitem [{\citenamefont {{Destounis}}\ \emph {et~al.}(2024)\citenamefont
  {{Destounis}}, \citenamefont {{Boyanov}},\ and\ \citenamefont
  {{Macedo}}}]{2024PhRvD.109d4023D}%
  \BibitemOpen
  \bibfield  {author} {\bibinfo {author} {\bibfnamefont {K.}~\bibnamefont
  {{Destounis}}}, \bibinfo {author} {\bibfnamefont {V.}~\bibnamefont
  {{Boyanov}}}, \ and\ \bibinfo {author} {\bibfnamefont {R.~P.}\ \bibnamefont
  {{Macedo}}},\ }\href {\doibase 10.1103/PhysRevD.109.044023} {\bibfield
  {journal} {\bibinfo  {journal} {\prd}\ }\textbf {\bibinfo {volume} {109}},\
  \bibinfo {eid} {044023} (\bibinfo {year} {2024})},\ \Eprint
  {http://arxiv.org/abs/2312.11630} {arXiv:2312.11630 [gr-qc]} \BibitemShut
  {NoStop}%
\bibitem [{\citenamefont {Destounis}\ and\ \citenamefont
  {Duque}(2024)}]{2023arXiv230816227D}%
  \BibitemOpen
  \bibfield  {author} {\bibinfo {author} {\bibfnamefont {K.}~\bibnamefont
  {Destounis}}\ and\ \bibinfo {author} {\bibfnamefont {F.}~\bibnamefont
  {Duque}},\ }in\ \href {\doibase 10.1007/978-3-031-55098-0_6} {\emph {\bibinfo
  {booktitle} {Compact Objects in the Universe}}}\ (\bibinfo  {publisher}
  {Springer Nature, Switzerland},\ \bibinfo {year} {2024})\ \Eprint
  {http://arxiv.org/abs/2308.16227} {arXiv:2308.16227 [gr-qc]} \BibitemShut
  {NoStop}%
\bibitem [{\citenamefont {{Cheung}}\ \emph {et~al.}(2022)\citenamefont
  {{Cheung}}, \citenamefont {{Destounis}}, \citenamefont {{Macedo}},
  \citenamefont {{Berti}},\ and\ \citenamefont
  {{Cardoso}}}]{2022PhRvL.128k1103C}%
  \BibitemOpen
  \bibfield  {author} {\bibinfo {author} {\bibfnamefont {M.~H.-Y.}\
  \bibnamefont {{Cheung}}}, \bibinfo {author} {\bibfnamefont {K.}~\bibnamefont
  {{Destounis}}}, \bibinfo {author} {\bibfnamefont {R.~P.}\ \bibnamefont
  {{Macedo}}}, \bibinfo {author} {\bibfnamefont {E.}~\bibnamefont {{Berti}}}, \
  and\ \bibinfo {author} {\bibfnamefont {V.}~\bibnamefont {{Cardoso}}},\ }\href
  {\doibase 10.1103/PhysRevLett.128.111103} {\bibfield  {journal} {\bibinfo
  {journal} {\prl}\ }\textbf {\bibinfo {volume} {128}},\ \bibinfo {eid}
  {111103} (\bibinfo {year} {2022})},\ \Eprint
  {http://arxiv.org/abs/2111.05415} {arXiv:2111.05415 [gr-qc]} \BibitemShut
  {NoStop}%
\bibitem [{\citenamefont {{Berti}}\ \emph {et~al.}(2022)\citenamefont
  {{Berti}}, \citenamefont {{Cardoso}}, \citenamefont {{Cheung}}, \citenamefont
  {{Di Filippo}}, \citenamefont {{Duque}}, \citenamefont {{Martens}},\ and\
  \citenamefont {{Mukohyama}}}]{2022PhRvD.106h4011B}%
  \BibitemOpen
  \bibfield  {author} {\bibinfo {author} {\bibfnamefont {E.}~\bibnamefont
  {{Berti}}}, \bibinfo {author} {\bibfnamefont {V.}~\bibnamefont {{Cardoso}}},
  \bibinfo {author} {\bibfnamefont {M.~H.-Y.}\ \bibnamefont {{Cheung}}},
  \bibinfo {author} {\bibfnamefont {F.}~\bibnamefont {{Di Filippo}}}, \bibinfo
  {author} {\bibfnamefont {F.}~\bibnamefont {{Duque}}}, \bibinfo {author}
  {\bibfnamefont {P.}~\bibnamefont {{Martens}}}, \ and\ \bibinfo {author}
  {\bibfnamefont {S.}~\bibnamefont {{Mukohyama}}},\ }\href {\doibase
  10.1103/PhysRevD.106.084011} {\bibfield  {journal} {\bibinfo  {journal}
  {\prd}\ }\textbf {\bibinfo {volume} {106}},\ \bibinfo {eid} {084011}
  (\bibinfo {year} {2022})},\ \Eprint {http://arxiv.org/abs/2205.08547}
  {arXiv:2205.08547 [gr-qc]} \BibitemShut {NoStop}%
\bibitem [{\citenamefont {{Kyutoku}}\ \emph {et~al.}(2023)\citenamefont
  {{Kyutoku}}, \citenamefont {{Motohashi}},\ and\ \citenamefont
  {{Tanaka}}}]{2023PhRvD.107d4012K}%
  \BibitemOpen
  \bibfield  {author} {\bibinfo {author} {\bibfnamefont {K.}~\bibnamefont
  {{Kyutoku}}}, \bibinfo {author} {\bibfnamefont {H.}~\bibnamefont
  {{Motohashi}}}, \ and\ \bibinfo {author} {\bibfnamefont {T.}~\bibnamefont
  {{Tanaka}}},\ }\href {\doibase 10.1103/PhysRevD.107.044012} {\bibfield
  {journal} {\bibinfo  {journal} {\prd}\ }\textbf {\bibinfo {volume} {107}},\
  \bibinfo {eid} {044012} (\bibinfo {year} {2023})},\ \Eprint
  {http://arxiv.org/abs/2206.00671} {arXiv:2206.00671 [gr-qc]} \BibitemShut
  {NoStop}%
\bibitem [{\citenamefont {{Torres}}(2023)}]{2023PhRvL.131k1401T}%
  \BibitemOpen
  \bibfield  {author} {\bibinfo {author} {\bibfnamefont {T.}~\bibnamefont
  {{Torres}}},\ }\href {\doibase 10.1103/PhysRevLett.131.111401} {\bibfield
  {journal} {\bibinfo  {journal} {\prl}\ }\textbf {\bibinfo {volume} {131}},\
  \bibinfo {eid} {111401} (\bibinfo {year} {2023})},\ \Eprint
  {http://arxiv.org/abs/2304.10252} {arXiv:2304.10252 [gr-qc]} \BibitemShut
  {NoStop}%
\bibitem [{\citenamefont {Rosato}\ \emph {et~al.}(2024)\citenamefont {Rosato},
  \citenamefont {Destounis},\ and\ \citenamefont {Pani}}]{2024arXiv240601692R}%
  \BibitemOpen
  \bibfield  {author} {\bibinfo {author} {\bibfnamefont {R.~F.}\ \bibnamefont
  {Rosato}}, \bibinfo {author} {\bibfnamefont {K.}~\bibnamefont {Destounis}}, \
  and\ \bibinfo {author} {\bibfnamefont {P.}~\bibnamefont {Pani}},\ }\href@noop
  {} {\  (\bibinfo {year} {2024})},\ \Eprint {http://arxiv.org/abs/2406.01692}
  {arXiv:2406.01692 [gr-qc]} \BibitemShut {NoStop}%
\bibitem [{\citenamefont {Oshita}\ \emph {et~al.}(2024)\citenamefont {Oshita},
  \citenamefont {Takahashi},\ and\ \citenamefont
  {Mukohyama}}]{2024arXiv240604525O}%
  \BibitemOpen
  \bibfield  {author} {\bibinfo {author} {\bibfnamefont {N.}~\bibnamefont
  {Oshita}}, \bibinfo {author} {\bibfnamefont {K.}~\bibnamefont {Takahashi}}, \
  and\ \bibinfo {author} {\bibfnamefont {S.}~\bibnamefont {Mukohyama}},\
  }\href@noop {} {\  (\bibinfo {year} {2024})},\ \Eprint
  {http://arxiv.org/abs/2406.04525} {arXiv:2406.04525 [gr-qc]} \BibitemShut
  {NoStop}%
\bibitem [{\citenamefont {{Leaver}}(1986)}]{1986PhRvD..34..384L}%
  \BibitemOpen
  \bibfield  {author} {\bibinfo {author} {\bibfnamefont {E.~W.}\ \bibnamefont
  {{Leaver}}},\ }\href {\doibase 10.1103/PhysRevD.34.384} {\bibfield  {journal}
  {\bibinfo  {journal} {\prd}\ }\textbf {\bibinfo {volume} {34}},\ \bibinfo
  {pages} {384} (\bibinfo {year} {1986})}\BibitemShut {NoStop}%
\bibitem [{\citenamefont {{Casals}}\ and\ \citenamefont
  {{Ottewill}}(2013)}]{2013PhRvD..87f4010C}%
  \BibitemOpen
  \bibfield  {author} {\bibinfo {author} {\bibfnamefont {M.}~\bibnamefont
  {{Casals}}}\ and\ \bibinfo {author} {\bibfnamefont {A.}~\bibnamefont
  {{Ottewill}}},\ }\href {\doibase 10.1103/PhysRevD.87.064010} {\bibfield
  {journal} {\bibinfo  {journal} {\prd}\ }\textbf {\bibinfo {volume} {87}},\
  \bibinfo {eid} {064010} (\bibinfo {year} {2013})},\ \Eprint
  {http://arxiv.org/abs/1210.0519} {arXiv:1210.0519 [gr-qc]} \BibitemShut
  {NoStop}%
\bibitem [{\citenamefont {{Casals}}\ \emph {et~al.}(2019)\citenamefont
  {{Casals}}, \citenamefont {{Nolan}}, \citenamefont {{Ottewill}},\ and\
  \citenamefont {{Wardell}}}]{2019PhRvD.100j4037C}%
  \BibitemOpen
  \bibfield  {author} {\bibinfo {author} {\bibfnamefont {M.}~\bibnamefont
  {{Casals}}}, \bibinfo {author} {\bibfnamefont {B.~C.}\ \bibnamefont
  {{Nolan}}}, \bibinfo {author} {\bibfnamefont {A.~C.}\ \bibnamefont
  {{Ottewill}}}, \ and\ \bibinfo {author} {\bibfnamefont {B.}~\bibnamefont
  {{Wardell}}},\ }\href {\doibase 10.1103/PhysRevD.100.104037} {\bibfield
  {journal} {\bibinfo  {journal} {\prd}\ }\textbf {\bibinfo {volume} {100}},\
  \bibinfo {eid} {104037} (\bibinfo {year} {2019})},\ \Eprint
  {http://arxiv.org/abs/1910.02567} {arXiv:1910.02567 [gr-qc]} \BibitemShut
  {NoStop}%
\bibitem [{\citenamefont {{Cardoso}}\ \emph {et~al.}(2024)\citenamefont
  {{Cardoso}}, \citenamefont {{Kastha}},\ and\ \citenamefont
  {{Macedo}}}]{2024PhRvD.110b4016C}%
  \BibitemOpen
  \bibfield  {author} {\bibinfo {author} {\bibfnamefont {V.}~\bibnamefont
  {{Cardoso}}}, \bibinfo {author} {\bibfnamefont {S.}~\bibnamefont {{Kastha}}},
  \ and\ \bibinfo {author} {\bibfnamefont {R.~P.}\ \bibnamefont {{Macedo}}},\
  }\href {\doibase 10.1103/PhysRevD.110.024016} {\bibfield  {journal} {\bibinfo
   {journal} {\prd}\ }\textbf {\bibinfo {volume} {110}},\ \bibinfo {eid}
  {024016} (\bibinfo {year} {2024})},\ \Eprint
  {http://arxiv.org/abs/2404.01374} {arXiv:2404.01374 [gr-qc]} \BibitemShut
  {NoStop}%
\bibitem [{\citenamefont {{Regge}}\ and\ \citenamefont
  {{Wheeler}}(1957)}]{1957PhRv..108.1063R}%
  \BibitemOpen
  \bibfield  {author} {\bibinfo {author} {\bibfnamefont {T.}~\bibnamefont
  {{Regge}}}\ and\ \bibinfo {author} {\bibfnamefont {J.~A.}\ \bibnamefont
  {{Wheeler}}},\ }\href {\doibase 10.1103/PhysRev.108.1063} {\bibfield
  {journal} {\bibinfo  {journal} {Physical Review}\ }\textbf {\bibinfo {volume}
  {108}},\ \bibinfo {pages} {1063} (\bibinfo {year} {1957})}\BibitemShut
  {NoStop}%
\bibitem [{\citenamefont {{Zerilli}}(1970)}]{1970PhRvL..24..737Z}%
  \BibitemOpen
  \bibfield  {author} {\bibinfo {author} {\bibfnamefont {F.~J.}\ \bibnamefont
  {{Zerilli}}},\ }\href {\doibase 10.1103/PhysRevLett.24.737} {\bibfield
  {journal} {\bibinfo  {journal} {\prl}\ }\textbf {\bibinfo {volume} {24}},\
  \bibinfo {pages} {737} (\bibinfo {year} {1970})}\BibitemShut {NoStop}%
\bibitem [{\citenamefont {{Zerilli}}(1969)}]{1969PhDT........13Z}%
  \BibitemOpen
  \bibfield  {author} {\bibinfo {author} {\bibfnamefont {F.~J.}\ \bibnamefont
  {{Zerilli}}},\ }\emph {\bibinfo {title} {{The Gravitational Field of a
  Particle Falling in a Schwarzschild Geometry Analyzed in Tensor
  Harmonics.}}},\ \href@noop {} {Ph.D. thesis},\ \bibinfo  {school} {Princeton
  University, New Jersey} (\bibinfo {year} {1969})\BibitemShut {NoStop}%
\bibitem [{\citenamefont {{Teukolsky}}(1972)}]{1972PhRvL..29.1114T}%
  \BibitemOpen
  \bibfield  {author} {\bibinfo {author} {\bibfnamefont {S.~A.}\ \bibnamefont
  {{Teukolsky}}},\ }\href {\doibase 10.1103/PhysRevLett.29.1114} {\bibfield
  {journal} {\bibinfo  {journal} {\prl}\ }\textbf {\bibinfo {volume} {29}},\
  \bibinfo {pages} {1114} (\bibinfo {year} {1972})}\BibitemShut {NoStop}%
\bibitem [{\citenamefont {{Mark}}\ \emph {et~al.}(2017)\citenamefont {{Mark}},
  \citenamefont {{Zimmerman}}, \citenamefont {{Du}},\ and\ \citenamefont
  {{Chen}}}]{2017PhRvD..96h4002M}%
  \BibitemOpen
  \bibfield  {author} {\bibinfo {author} {\bibfnamefont {Z.}~\bibnamefont
  {{Mark}}}, \bibinfo {author} {\bibfnamefont {A.}~\bibnamefont {{Zimmerman}}},
  \bibinfo {author} {\bibfnamefont {S.~M.}\ \bibnamefont {{Du}}}, \ and\
  \bibinfo {author} {\bibfnamefont {Y.}~\bibnamefont {{Chen}}},\ }\href
  {\doibase 10.1103/PhysRevD.96.084002} {\bibfield  {journal} {\bibinfo
  {journal} {\prd}\ }\textbf {\bibinfo {volume} {96}},\ \bibinfo {eid} {084002}
  (\bibinfo {year} {2017})},\ \Eprint {http://arxiv.org/abs/1706.06155}
  {arXiv:1706.06155 [gr-qc]} \BibitemShut {NoStop}%
\bibitem [{\citenamefont {{Hui}}\ \emph {et~al.}(2019)\citenamefont {{Hui}},
  \citenamefont {{Kabat}},\ and\ \citenamefont {{Wong}}}]{2019JCAP...12..020H}%
  \BibitemOpen
  \bibfield  {author} {\bibinfo {author} {\bibfnamefont {L.}~\bibnamefont
  {{Hui}}}, \bibinfo {author} {\bibfnamefont {D.}~\bibnamefont {{Kabat}}}, \
  and\ \bibinfo {author} {\bibfnamefont {S.~S.~C.}\ \bibnamefont {{Wong}}},\
  }\href {\doibase 10.1088/1475-7516/2019/12/020} {\bibfield  {journal}
  {\bibinfo  {journal} {\jcap}\ }\textbf {\bibinfo {volume} {2019}},\ \bibinfo
  {eid} {020} (\bibinfo {year} {2019})},\ \Eprint
  {http://arxiv.org/abs/1909.10382} {arXiv:1909.10382 [gr-qc]} \BibitemShut
  {NoStop}%
\bibitem [{\citenamefont {{Leung}}\ \emph {et~al.}(1999)\citenamefont
  {{Leung}}, \citenamefont {{Liu}}, \citenamefont {{Suen}}, \citenamefont
  {{Tam}},\ and\ \citenamefont {{Young}}}]{1999PhRvD..59d4034L}%
  \BibitemOpen
  \bibfield  {author} {\bibinfo {author} {\bibfnamefont {P.~T.}\ \bibnamefont
  {{Leung}}}, \bibinfo {author} {\bibfnamefont {Y.~T.}\ \bibnamefont {{Liu}}},
  \bibinfo {author} {\bibfnamefont {W.~M.}\ \bibnamefont {{Suen}}}, \bibinfo
  {author} {\bibfnamefont {C.~Y.}\ \bibnamefont {{Tam}}}, \ and\ \bibinfo
  {author} {\bibfnamefont {K.}~\bibnamefont {{Young}}},\ }\href {\doibase
  10.1103/PhysRevD.59.044034} {\bibfield  {journal} {\bibinfo  {journal}
  {\prd}\ }\textbf {\bibinfo {volume} {59}},\ \bibinfo {eid} {044034} (\bibinfo
  {year} {1999})},\ \Eprint {http://arxiv.org/abs/gr-qc/9903032}
  {arXiv:gr-qc/9903032 [gr-qc]} \BibitemShut {NoStop}%
\bibitem [{\citenamefont {{Sasaki}}\ and\ \citenamefont
  {{Nakamura}}(1982)}]{1982PThPh..67.1788S}%
  \BibitemOpen
  \bibfield  {author} {\bibinfo {author} {\bibfnamefont {M.}~\bibnamefont
  {{Sasaki}}}\ and\ \bibinfo {author} {\bibfnamefont {T.}~\bibnamefont
  {{Nakamura}}},\ }\href {\doibase 10.1143/PTP.67.1788} {\bibfield  {journal}
  {\bibinfo  {journal} {Progress of Theoretical Physics}\ }\textbf {\bibinfo
  {volume} {67}},\ \bibinfo {pages} {1788} (\bibinfo {year}
  {1982})}\BibitemShut {NoStop}%
\bibitem [{\citenamefont {{Zhang}}\ \emph {et~al.}(2013)\citenamefont
  {{Zhang}}, \citenamefont {{Berti}},\ and\ \citenamefont
  {{Cardoso}}}]{2013PhRvD..88d4018Z}%
  \BibitemOpen
  \bibfield  {author} {\bibinfo {author} {\bibfnamefont {Z.}~\bibnamefont
  {{Zhang}}}, \bibinfo {author} {\bibfnamefont {E.}~\bibnamefont {{Berti}}}, \
  and\ \bibinfo {author} {\bibfnamefont {V.}~\bibnamefont {{Cardoso}}},\ }\href
  {\doibase 10.1103/PhysRevD.88.044018} {\bibfield  {journal} {\bibinfo
  {journal} {\prd}\ }\textbf {\bibinfo {volume} {88}},\ \bibinfo {eid} {044018}
  (\bibinfo {year} {2013})},\ \Eprint {http://arxiv.org/abs/1305.4306}
  {arXiv:1305.4306 [gr-qc]} \BibitemShut {NoStop}%
\bibitem [{\citenamefont {Andersson}(1995)}]{Andersson:1995zk}%
  \BibitemOpen
  \bibfield  {author} {\bibinfo {author} {\bibfnamefont {N.}~\bibnamefont
  {Andersson}},\ }\href {\doibase 10.1103/PhysRevD.51.353} {\bibfield
  {journal} {\bibinfo  {journal} {Phys. Rev. D}\ }\textbf {\bibinfo {volume}
  {51}},\ \bibinfo {pages} {353} (\bibinfo {year} {1995})}\BibitemShut
  {NoStop}%
\bibitem [{\citenamefont {{Casals}}\ and\ \citenamefont
  {{Ottewill}}(2012)}]{2012PhRvD..86b4021C}%
  \BibitemOpen
  \bibfield  {author} {\bibinfo {author} {\bibfnamefont {M.}~\bibnamefont
  {{Casals}}}\ and\ \bibinfo {author} {\bibfnamefont {A.}~\bibnamefont
  {{Ottewill}}},\ }\href {\doibase 10.1103/PhysRevD.86.024021} {\bibfield
  {journal} {\bibinfo  {journal} {\prd}\ }\textbf {\bibinfo {volume} {86}},\
  \bibinfo {eid} {024021} (\bibinfo {year} {2012})},\ \Eprint
  {http://arxiv.org/abs/1112.2695} {arXiv:1112.2695 [gr-qc]} \BibitemShut
  {NoStop}%
\bibitem [{\citenamefont {Konoplya}\ and\ \citenamefont
  {Zhidenko}(2022)}]{2022arXiv220900679K}%
  \BibitemOpen
  \bibfield  {author} {\bibinfo {author} {\bibfnamefont {R.~A.}\ \bibnamefont
  {Konoplya}}\ and\ \bibinfo {author} {\bibfnamefont {A.}~\bibnamefont
  {Zhidenko}},\ }\href@noop {} {\  (\bibinfo {year} {2022})},\ \Eprint
  {http://arxiv.org/abs/2209.00679} {arXiv:2209.00679 [gr-qc]} \BibitemShut
  {NoStop}%
\bibitem [{\citenamefont {Motohashi}(2024)}]{2024arXiv240715191M}%
  \BibitemOpen
  \bibfield  {author} {\bibinfo {author} {\bibfnamefont {H.}~\bibnamefont
  {Motohashi}},\ }\href@noop {} {\  (\bibinfo {year} {2024})},\ \Eprint
  {http://arxiv.org/abs/2407.15191} {arXiv:2407.15191 [gr-qc]} \BibitemShut
  {NoStop}%
\bibitem [{\citenamefont {Stanzione}\ \emph {et~al.}(2020)\citenamefont
  {Stanzione}, \citenamefont {West}, \citenamefont {Evans}, \citenamefont
  {Minyard}, \citenamefont {Ghattas},\ and\ \citenamefont
  {Panda}}]{10.1145/3311790.3396656}%
  \BibitemOpen
  \bibfield  {author} {\bibinfo {author} {\bibfnamefont {D.}~\bibnamefont
  {Stanzione}}, \bibinfo {author} {\bibfnamefont {J.}~\bibnamefont {West}},
  \bibinfo {author} {\bibfnamefont {R.~T.}\ \bibnamefont {Evans}}, \bibinfo
  {author} {\bibfnamefont {T.}~\bibnamefont {Minyard}}, \bibinfo {author}
  {\bibfnamefont {O.}~\bibnamefont {Ghattas}}, \ and\ \bibinfo {author}
  {\bibfnamefont {D.~K.}\ \bibnamefont {Panda}},\ }in\ \href {\doibase
  10.1145/3311790.3396656} {\emph {\bibinfo {booktitle} {Practice and
  Experience in Advanced Research Computing}}},\ \bibinfo {series and number}
  {PEARC '20}\ (\bibinfo  {publisher} {Association for Computing Machinery},\
  \bibinfo {address} {New York, NY, USA},\ \bibinfo {year} {2020})\ p.\
  \bibinfo {pages} {106–111}\BibitemShut {NoStop}%
\end{thebibliography}%
	
\end{document}